\documentclass[11pt,letterpaper]{article}

\usepackage{arxiv} %

\usepackage{amsmath}
\usepackage{amssymb}
\usepackage{booktabs}
\usepackage{array}
\usepackage{placeins}
\usepackage{float}
\usepackage[authoryear,round]{natbib}

\usepackage{graphicx}
\usepackage{xcolor}
\usepackage{pgfplots}
\usetikzlibrary{arrows.meta,positioning,calc,fit,backgrounds,plotmarks}
\pgfplotsset{compat=1.18}
\usepackage{url}
\usepackage[hidelinks]{hyperref}

\graphicspath{{figures/}{./}}

\hypersetup{
  pdftitle={E2Former-V3: Atom-Role-Aware Equivariant Interatomic Potentials},
  pdfauthor={Frank Peng, Lin Huang, Arthur Jiang, Jia Zhang},
  pdfsubject={Machine-learned interatomic potentials; equivariant neural networks; molecular dynamics},
  pdfkeywords={equivariant interatomic potential, atom-role-aware, mixed resolution, molecular dynamics}
}

\title{E2Former-V3: Atom-Role-Aware Equivariant Interatomic Potentials}

\author{%
  Frank Peng, Lin Huang, Arthur Jiang, Jia Zhang$^{\dagger}$\\[0.6em]
  \normalsize IQuest Research, UBio Team\\[0.35em]
  \normalsize $^{\dagger}$Correspondence: \href{mailto:jialrs.z@iquestlab.com}{\texttt{jialrs.z@iquestlab.com}}%
}
\date{}

\begin{document}

\maketitle

\begin{abstract}
Machine-learned interatomic potentials approach density-functional accuracy, but explicit-solvent
atom counts are dominated by hydrogens and water. Uniform equivariant architectures still
assign every atom the same latent width and the same number of expensive block updates.
Prior work lowers cost by shrinking the model uniformly, approximating the physics, conditioning
computation, or optimizing kernels. To our knowledge, none reallocates capacity by chemical role
while retaining every atom and its first-principles force. We introduce E2Former-V3, which treats
representation capacity as an atom-level resource: a static chemical role, assigned before the
backbone runs, sets each atom's latent width and update count; no atom is removed, no coordinate
is integrated out, and every force remains the gradient of one conservative energy.
\mbox{\textbf{Standard}} reduces only the update count, sending hydrogen through half as many
full-width block updates; on the OMol25 validation split it reaches a force mean absolute error
(MAE) of $8.146$\,meV/\AA{}; E2Former-V2's published value is $8.38$
\citep{huang2026formerv}, at $1.75$--$1.84\times$ V2's throughput with lower peak
memory on one H200. \mbox{\textbf{Flash}} also narrows the low-resolution stream to one eighth of the
high-resolution width (32 vs 256 channels per degree) and routes water oxygen with hydrogen, reaching $4.00$--$4.88\times$ on the three
solvated systems (up to $5.85\times$ in the pure-water all-low-resolution limit) at
$11.272$\,meV/\AA{}, trading force accuracy for speed.
Across 54 20\,ps MD legs, a 100\,ps NVE check, and a 1\,ns NVT run on a 32{,}437-atom solvated
protein, all runs complete; the MD legs show no non-finite values, clashes, or water fusion, and
the water O--O and Na--O RDF peak positions match experiment within 0.04\,\AA{}. E2Former-V3 therefore moves the accuracy--throughput frontier for all-atom, conservative
simulation, not a uniform accuracy win: \mbox{Standard} is worse on biomolecules and metals, better on electrolytes and
organics; energy MAE is higher than E2Former-V2's; and water densities remain about $7.5\%$ too high.
Source code is available at \url{https://github.com/IQuestLab/UBio-MolFM}.
\end{abstract}

\section{Introduction}
\label{sec:intro}

Machine-learning interatomic potentials (MLIPs) now approach density-functional-theory
accuracy on molecular and biomolecular systems, returning an energy whose gradient is a
conservative force on every atom \citep{schutt2017schnet, batzner2022equivariant,
batatia2022mace}. Yet most equivariant MLIPs make the same architectural choice by default:
every atom receives the same latent width and the same number of expensive block updates.
That uniformity is an inheritance from earlier architectures, not a chemical requirement.

\begin{figure}[!t]
\centering
\includegraphics[width=\linewidth,keepaspectratio]{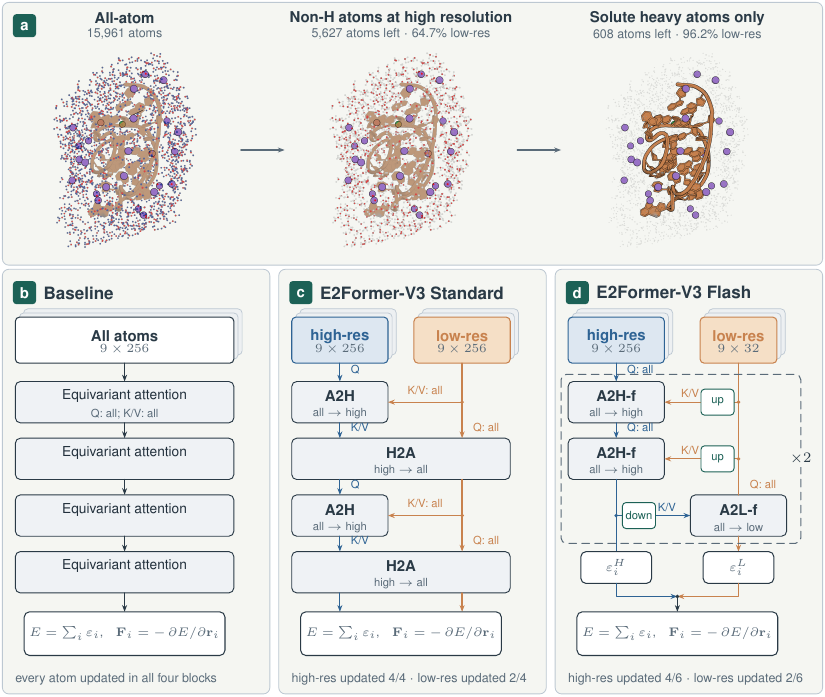}
\caption[E2Former-V3 under uniform and atom-role-aware capacity allocations.]{%
\textbf{E2Former-V3 under three capacity allocations.}
\textbf{(a)} One solvated nucleic acid (15{,}961 atoms; water drawn within 8\,\AA{} of the
RNA, counts are the full system's), at the three levels of atomic detail the allocations
use: every atom; then hydrogen at low resolution, so 64.7\% of atoms are low-resolution and
5{,}627 stay high-resolution; then the oxygens of water at low resolution as well, so 96.2\%
are low-resolution and only the 608 solute heavy atoms stay high-resolution.
\textbf{(b)} The authors' uniform control (not E2Former-V2): one shared $9\times256$ state,
every row updated in all four blocks ($4/4$).
\textbf{(c)} Standard: $[\mathrm{A2H},\mathrm{H2A}]\times2$ on one shared 256-channel state,
so the role streams have equal width and differ only in update count, $4/4$ against $2/4$.
\textbf{(d)} Flash: $[\mathrm{A2H\text{-}f},\mathrm{A2H\text{-}f},\mathrm{A2L\text{-}f}]\times2$
on two persistent streams, 256 channels per degree at high resolution against 32 at low
resolution; updates are $4/6$ and $2/6$. In both routed allocations the role heads merge
before forces are differentiated, so no atom is removed and no coordinate is integrated out.
The block schedules are specified in Section~\ref{sec:method}.}
\label{fig:framework}
\end{figure}

Explicit-solvent systems are where the default costs the most, and where longer trajectories
and larger boxes are most wanted. Figure~\ref{fig:framework}(a) shows one such system: of its
15{,}961 atoms, 10{,}334 are hydrogen and 15{,}057 are water atoms, with water's hydrogens
counted in both figures; the solute contributes 608 heavy atoms. Per-atom work is uniform, so
the cost follows those counts: the equivariant updates, the stored activations and the
autograd graph behind the conservative forces all scale with atom number.

Computational chemistry has long handled this cost by treating atoms unequally: implicit
solvent, coarse graining and constrained dynamics all show that hydrogens and solvent need not
be resolved at full fidelity \citep{marrink2007martini, kmiecik2016coarsegrained, tan2006how,
katzberger2023implicit, ryckaert1977numerical, andersen1983rattle, miyamoto1992settle,
hess1997lincs, hopkins2015longtimestep}. Each saves the cost by changing the physical model: a
continuum replaces the solvent, pseudo-particles merge atoms, or bonds are frozen. The all-atom
description is lost along with the cost saving, so none of these methods returns a
first-principles force on every atom. Existing equivariant potentials, efficiency methods and
conditional-computation models (Section~\ref{sec:related}) reduce the cost as well; none of
them allocates capacity by atom role while keeping every atom and its first-principles forces.

\paragraph{Problem setting: capacity is a resource.}

Uniform capacity is an architectural default, not a physical requirement. The energy is a sum
of per-atom contributions, so a model may spend different amounts of representation capacity
on different atoms as long as every atom remains in the graph and the sum is differentiated.
The default gives one width, one update count and one kernel, and it spends compute on atoms
that do not need the full representation. In an explicit-solvent box those atoms are the
majority: only 608 of the 15{,}961 atoms in Figure~\ref{fig:framework}(a) are
solute heavy atoms. The total capacity cost therefore follows the solvent, not the solute: width squared times
updates times atom count.

Fused attention, algebraic restructuring of the tensor product, and tile-level GPU kernels all
make the same per-node operation cheaper; they reduce the constant. Dual cutoffs and neighbour
filtering remove edges that a short-range model does not need; they reduce the edge set. Neither changes how wide or how often a node is updated. A uniformly
smaller model reduces capacity for the solute too, which is where the angular detail is needed.
The alternative studied here is reallocation: keep the all-atom graph and the conservative
readout, but give the solvent and hydrogen less width and fewer updates.

\paragraph{Hypotheses and design rationale.}

The design rests on three chemical hypotheses. \textbf{H1}: hydrogen carries little angular
information, so it can lose width and updates without changing the energy surface.
\textbf{H2}: bulk water oxygen is homogeneous enough that a narrow stream can capture its
angular environment. \textbf{H3}: solute heavy atoms are harder to predict and may need the full angular stack. The measured
error follows this partition: the high-to-low-resolution ratio is 3.634 under Standard and
3.324 under Flash, with high-resolution atoms carrying 76--78\% of absolute error while making
up 49.1--49.6\% of force components (Appendix~\ref{sec:app_role}).

\paragraph{This work.}
We introduce an atom-role-aware MLIP architecture. As in the reduced-fidelity methods above,
hydrogens and solvent carry less of the model, but no atom or degree of freedom is
removed: every atom and every degree of freedom stay, the forces still come from one
conservative energy, and only the capacity an atom consumes is smaller. A static chemical role, assigned before the
backbone runs, sets an atom's latent width and update count. \textbf{Standard} sends hydrogen to
low resolution, in solvent and solute alike, and holds the width at 256 channels per degree for
every atom; low-resolution atoms take two of the four block updates, high-resolution atoms all
four, so the saving comes from participation alone. \textbf{Flash} sends hydrogen and the
oxygens of water to low resolution, narrows that stream to one eighth of the high-resolution width (32 vs 256 channels per
degree), and runs six blocks, of which low-resolution atoms take two and
high-resolution atoms four. In both configurations the two role streams exchange information in
both directions and their energy heads merge before forces are differentiated. In an A2H block
every high-resolution row is a query, but the key/value bank need only cover the union of those
queries' neighbour tables; for Flash, where the high-resolution query set is small, skipping the
rest of the bank cuts latency by 23.1--25.5\% on the non-degenerate systems, while Standard's
deltas remain within $\pm0.4\%$
(Section~\ref{sec:filtering}). The equivariant message-passing backbone, the per-order channel
projections, the cross-role attention and the energy-gradient readout are standard components;
what is new here is the role-conditioned block masking and its schedules (A2H, H2A and A2L),
the reachable-row construction that follows from the resulting query set, and the matched-cost
evidence that compares role-conditioned allocation with uniform reductions at similar latency
(Section~\ref{sec:matcheduniform}).

\paragraph{Contributions and results.}
\begin{enumerate}
\item We formulate static atom-role capacity allocation inside an equivariant MLIP, with
latent width and block-update count as independent capacity axes, and introduce Standard
(participation-only) and Flash (width and participation) together with their block schedules
and the reachable-row key/value construction.
\item On OMol25, the two configurations form non-dominated force-accuracy--throughput points
against E2Former-V2 on one H200: Standard reaches 8.146\,meV/\AA{} against the published 8.38
at $1.75$--$1.84\times$ the throughput, and Flash reaches 11.272\,meV/\AA{} at
$4.00$--$4.88\times$ on the three solvated systems ($4.00$--$5.85\times$ including the
pure-water limit). Same-recipe uniform-reduction controls and the role-stratified error are
reported as supporting evidence.
\item We validate both configurations in 20\,ps molecular dynamics on three systems,
100\,ps of NVE on water and 1\,ns on one solvated protein; all 54 production legs complete,
the thermostatted cell means hold 300\,K within 2.4\,K, RDF peak positions match experiment
within 0.04\,\AA{}, and the peptide's 196-bond graph stays intact.
\end{enumerate}

\paragraph{Organization.} Section~\ref{sec:related} reviews related work. Section~\ref{sec:method}
gives the architecture, the two schedules and the implementation.
Section~\ref{sec:experiments} reports accuracy, efficiency, role-stratified error, physical
validation and ablations. Section~\ref{sec:discussion} discusses what the results establish,
what they do not, and how to choose between Standard and Flash.
Section~\ref{sec:conclusion} concludes. Appendices~\ref{sec:app_arch}--\ref{sec:app_bounds}
hold the architecture, protocols, full tables, physical validation and boundary conditions.

\section{Related Work}
\label{sec:related}

\subsection{From descriptor-based to equivariant interatomic potentials}

The first generation of machine-learned potentials used fixed local descriptors and a
regression head: Behler--Parrinello high-dimensional neural network potentials
\citep{behler2007generalized}, Gaussian approximation potentials \citep{bartok2010gaussian},
the atomic cluster expansion \citep{drautz2019atomic}, and Deep Potential
\citep{zhang2018deep}. These models established that a local, symmetry-invariant
representation plus a learned map can reach near-DFT accuracy, but they fix the descriptor
and the per-atom capacity by construction.

Message passing replaced the fixed descriptor with a learned neighbourhood aggregation.
SchNet \citep{schutt2017schnet}, DimeNet \citep{gasteiger2020directional} and GemNet
\citep{gasteiger2021gemnet} are invariant models; tensor-field networks
\citep{thomas2018tensor} and e3nn \citep{geiger2022euclidean} introduced irreducible
representations and equivariant tensor products; Cormorant \citep{anderson2019cormorant},
PaiNN \citep{schutt2021equivariant}, EGNN \citep{satorras2021n}, NequIP
\citep{batzner2022equivariant}, Allegro \citep{musaelian2023learning}, MACE
\citep{batatia2022mace}, Equiformer \citep{liao2022equiformer}, EquiformerV2
\citep{liao2023equiformerv}, SE(3)-Transformers \citep{fuchs2020transformers} and
TorchMD-NET \citep{tholke2022torchmdnet} form the equivariant and attention-based line.
The design-space analysis of \citet{batatia2025design} compares many of these choices.

All of these models fix latent width, angular degree and block-update count globally,
identically for a solvent hydrogen and a solute heavy atom. That global choice is the target
of this paper. Kernel-level improvements to these models, reviewed next, make the uniform
operation cheaper; they do not change how wide or how often a node is updated.

\subsection{Universal and foundation potentials}

A parallel line scales one model across chemical space rather than across a single system.
MACE-MP-0 \citep{batatia2025foundation}, CHGNet \citep{deng2023chgnet}, M3GNet
\citep{chen2022m3gnet}, Orb \citep{neumann2024orb} and MatterSim \citep{yang2024mattersim}
target materials and the periodic table; UMA \citep{wood2025uma} and the OMol25 models
\citep{levine2025open} target broad molecular and biomolecular coverage; MACE-OFF23
\citep{kovacs2025maceoff} targets organic and biomolecular transferability; eSEN
\citep{fu2025learning} targets smooth and conservative potentials. These models are
uniform-capacity by construction: every atom receives the same width and the same update
count, regardless of whether it is a solvent hydrogen or a solute heavy atom. They are the
right comparison class for accuracy, and they motivate the question this paper asks---whether
capacity itself can be allocated without changing the all-atom, conservative contract.

\subsection{Efficiency for equivariant potentials at uniform capacity}

Four families reduce the cost of the uniform operation without changing its width or update
count. Algebraic restructuring cheapens the tensor product
\citep{passaro2023reducing, luo2024enabling, li2025former}. Fused and IO-aware attention
recomputes edge intermediates rather than materialising them
\citep{huang2026formerv, dao2022flashattention}, with tile-level GPU languages and graph
compilers \citep{tillet2019triton, ansel2024pytorch}. Dropping some symmetry constraints
reduces the representation cost \citep{qu2024importance}. Graph-level work reduces which edges
exist \citep{kahle2024dualcutoff, hu2025efficient}. Each changes a constant factor or the edge
set; none changes the per-node width or update count. Both configurations in this paper keep
the E2Attention operator and the same kernel family, so the efficiency gaps are not
kernel-level. The accuracy gaps are not a pure allocation effect either: Flash also changes
the attention parameterisation, which the ablation in Section~\ref{sec:ablation} prices
separately.

\subsection{Conditional computation and non-uniform per-atom compute}

Non-uniform per-input compute is well validated in language and vision. Conditional
computation \citep{bengio2013estimating} and sparse mixture-of-experts layers
\citep{shazeer2017outrageously, lepikhin2020gshard, fedus2021switch, du2022glam,
zhou2022mixtureofexperts, zoph2022stmoe, dai2024deepseekmoe} route tokens to a subset of
experts; adaptive-depth models learn or cap the number of blocks
\citep{graves2016adaptive, dehghani2019universal, elbayad2019depthadaptive,
banino2021pondernet, raposo2024mixtureofdepths, fan2020layerdrop}; token pruning and merging
discard or fuse tokens \citep{rao2021dynamicvit, bolya2022token}. The cross-attention
structure with distinct query and key/value sets \citep{vaswani2017attention,
jaegle2021perceiver} is the structure our cross-role blocks take.

Three schemes already vary compute per atom inside interatomic potentials. Mixed precision
gives each atom a precise or a fast potential, chosen from a local-structure descriptor
\citep{immel2025adaptive}. Element-indexed routing sends each atom to expert sub-networks by
atomic number \citep{liu2026moe}. Multifidelity domain decomposition splits the system in
space and runs the higher-capacity model on the reactive region
\citep{nascimento2026multifidelity}. The first and third are energy-conserving by
construction: both derive forces from a single mixed Hamiltonian, and the latter adds an
agreement loss at the interface. The difference is architectural, not a matter of
conservation. They combine two separately trained potentials and therefore need a switching
rule or an agreement loss where the two meet; V3 keeps one trained network and one energy,
uses a fixed chemical prior rather than a learned gate, and exchanges information through
learned up/down projections, so there is no seam. Element-indexed routing has been
demonstrated in invariant architectures, where a nonlinear expert can act on a feature
without breaking symmetry; the same experts inside an equivariant block would have to preserve
each angular degree. A router fixed by atom role learns no gate, so ours can be applied before
the equivariant stack without that constraint. The Flash router also separates two environments
of the same species: solute oxygens stay high-resolution while the oxygens of intact water
drop to low.

\subsection{Mixed resolution, multi-fidelity and heterogeneous capacity}

We use \emph{resolution} to mean the width and update count assigned to an atom, not a second
physical model and not a spatial partition. Every atom remains in one all-atom network. This
distinguishes the paper from QM/MM and ML/MM partitioning
\citep{warshel1976theoretical, senn2009methods, boselt2021machine}, which keep two
descriptions plus a boundary; from multi-fidelity modelling
\citep{kennedy2000predicting, meng2020composite}, which combines models of different cost
over the same input; and from spatial domain decomposition
\citep{nascimento2026multifidelity}, which assigns a higher-capacity model to a region rather
than to a chemical role. The shared premise is that different parts of a system can receive
different fidelity. The mechanism here is different: one network, one energy, every atom
retained, and the exchange across the role boundary is learned.

\subsection{Coarse-graining, implicit solvent and constrained dynamics}

Classical reduced-fidelity methods establish the premise that resolution should follow atom
role. Coarse-grained fields merge atoms into pseudo-particles
\citep{marrink2007martini, kmiecik2016coarsegrained}; learned coarse-graining constructs a
conservative force field on a reduced representation
\citep{noid2008multiscale, shell2008relative, wang2019machine, husic2020coarse}; implicit
solvation substitutes a continuum \citep{tan2006how, katzberger2023implicit}; constrained
dynamics and mass repartitioning freeze or slow the fastest motions
\citep{ryckaert1977numerical, andersen1983rattle, miyamoto1992settle, hess1997lincs,
hopkins2015longtimestep}. Each saves cost by deleting degrees of freedom, changing the
potential, or splitting the Hamiltonian. V3 borrows the premise---resolution can follow atom
role---but rejects the mechanism: no atom is merged, no coordinate is constrained, and the
force is the gradient of one all-atom energy.

\subsection{Benchmarking and physical validation}

Averaged force error does not rank dynamical usability. Benchmarking studies show
that force accuracy alone is insufficient for molecular simulation \citep{fu2022forces}
and that error metrics disagree \citep{liu2023discrepancies}. Other work addresses
stability-aware training \citep{raja2024stabilityaware}, reviews the aqueous setting
\citep{omranpour2024perspective}, and compares accuracy and cost across potentials
\citep{zuo2020performance}; Matbench Discovery
\citep{riebesell2025matbench} evaluates crystal stability at scale. The physical-chemistry
literature adds reference sensitivity \citep{kuryla2025accurate, beck2026moreconverged} and
the need for explicit physical-validity tests \citep{merz2018testing}. These works motivate
the 20\,ps matrix, the 1\,ns campaign, the RDF landmarks, the repulsion-wall gate and the
proton-transfer gate used here, and they motivate reporting failures rather than only
averages.

\subsection{Relation to E2Former-V2 and novelty}

E2Former-V1 \citep{li2025former} introduced linear-scaling tensor products. E2Former-V2
\citep{huang2026formerv} kept uniform capacity but fused E2Attention and reduced activation
memory. V3 keeps the E2Attention operator and changes the allocation: a static role router,
two capacity axes (width and update count), and reachable-row construction for the Flash
inference path. We do not claim a new kernel, a learned or dynamic router, a separable
state-layout ablation, or zero-shot transfer of a routing policy. The controlled evidence is
the single-variable ablation in Appendix~\ref{sec:app_ablation}, not the comparison between
the two production configurations, which differ on several axes at once.

\begin{table}[t]
\centering
\caption{Where V3 sits relative to the four literatures it draws on.}
\label{tab:related}
\small
\resizebox{\linewidth}{!}{%
\begin{tabular}{@{}>{\raggedright\arraybackslash}p{0.20\linewidth}>{\raggedright\arraybackslash}p{0.22\linewidth}>{\raggedright\arraybackslash}p{0.15\linewidth}>{\raggedright\arraybackslash}p{0.15\linewidth}>{\raggedright\arraybackslash}p{0.18\linewidth}@{}}
\toprule
Family & What varies & All atoms? & One energy? & Relation to V3 \\
\midrule
Equivariant MLIPs & nothing (uniform) & yes & yes & baseline \\
Efficiency methods & constant / edge set & yes & yes & orthogonal \\
Conditional computation & per-token compute & often no & not applicable & structure, not setting \\
Reduced fidelity & deletes DOF / splits model & not always & not always & premise, not mechanism \\
V3 & width + updates by role & yes & yes & this work \\
\bottomrule
\end{tabular}}
\end{table}

\section{Method}
\label{sec:method}

\subsection{The Uniform MLIP}
\label{sec:uniform_mlip}

An input is a set of atoms: atomic numbers $Z_i$, coordinates $\mathbf{r}_i$, charge, spin
multiplicity, a validity mask, and optionally a periodic cell. A radial cutoff on the
coordinates turns that into a neighbour graph. The model returns one total energy and one
force on every atom. Force accuracy is always a \emph{per-component} mean absolute error
(MAE), averaged over the three Cartesian components.

Between input and output the chain is three steps. \emph{Embedding} initialises a per-atom
intermediate representation from neighbour distances, neighbour directions and element
identity. \emph{Attention} then refines that representation by attending over the neighbours
the graph supplies. \emph{The head} reads a per-atom energy out of the refined representation;
the forces are obtained by differentiating the summed energy.

The representation is a set of $\mathrm{SO}(3)$ irreducible representations under the
$\mathrm{SO}(2)$ reduction of \citet{passaro2023reducing}, one component per angular degree up
to a maximum degree $l_{\max}=2$. Each $\ell$-component is the $(2\ell+1)$-dimensional
spherical-harmonic space; the tensor product combines features degree by degree so that a
rotation of the inputs rotates the output consistently. Each atom therefore carries an
intermediate representation of
shape $[(l_{\max}+1)^2,C]=[9,C]$, where $C$ is the number of channels. We call this tensor the
atom's \emph{state}; ``state'' and ``representation'' are used interchangeably. Two quantities
set an atom's capacity: $C$, its \emph{width}, and the number of attention updates it receives,
its \emph{update count}. These two axes define the capacity allocation.

The readout fixes what an allocation may and may not do. Every atom contributes one scalar
atomic energy and the model returns their sum, $E = \sum_i \varepsilon_i$, with
$\mathbf{F}_i = -\partial E/\partial\mathbf{r}_i$, so every force is read out of that one
scalar by differentiation.

A uniform MLIP gives every atom the same capacity: the same width and the same number of
attention updates, whatever the atom is. In a solvated biological system, hydrogens and the
oxygens of intact water are most of the atoms, so a uniform allocation spends most of its
compute on them. The question studied here is whether assigning those atoms less capacity
produces a more efficient MLIP without sacrificing accuracy.

\begin{table}[t]
\centering
\caption{Notation used in the method.}
\label{tab:notation}
\small
\begin{tabular}{@{}ll@{}}
\toprule
Symbol & Meaning \\
\midrule
$N$ & number of atoms \\
$l_{\max}=2$ & maximum angular degree \\
$C$ & channels per angular degree (width) \\
$9=1+3+5$ & scalar, vector and rank-2 components \\
$x\in\mathbb{R}^{N\times9\times256}$ & Standard shared state \\
$x^{H},x^{L}$ & Flash high-/low-resolution persistent states \\
$\tau(i)\in\{H,L\}$ & atom role (high or low resolution) \\
$D_s$ & channels per angular degree in role $s$ \\
$M^{(b)}_{ts}$ & block mask, source role $s$ to target role $t$ \\
$P_{b,s\to t}$ & per-order channel projection, role $s$ to role $t$ \\
$N_H,N_L$ & atom counts in the high-/low-resolution role \\
A2H, H2A & all-to-high, high-to-all blocks \\
A2H-f, A2L-f & Flash blocks \\
$P_{\uparrow},P_{\downarrow}$ & up/down degree-wise width projections (Flash) \\
$D^{\ell}$ & degree-$\ell$ channel map in the attention logit \\
$m_H,m_W$ & hydrogen and water-oxygen masks \\
$\mathcal Z_i$ & per-atom initialization normalizer \\
$H,L$ & high-/low-resolution role \\
\bottomrule
\end{tabular}
\end{table}

\subsection{The Atom-Role-Aware MLIP}
\label{sec:role_mlip}

The atom-role-aware model keeps the input, output and readout of
Section~\ref{sec:uniform_mlip} and partitions the atoms into two roles before the backbone
runs. The rest of this section turns that question into three pieces: a static role partition,
two capacity axes, and the block schedules that realize them. Let $\tau(i)\in\{H,L\}$ be the
role of atom $i$, where $H$ is high resolution and $L$
low resolution, and write $\mathcal H=\{i:\tau(i)=H\}$,
$\mathcal L=\{i:\tau(i)=L\}$ and $\mathcal A=\mathcal H\cup\mathcal L$. The role is a fixed
function of element identity and molecular topology. Standard assigns $\tau(i)=L$ iff
$m_{\mathrm H}(i)=1$, where $m_{\mathrm H}(i)=\mathbf 1[z_i=1]$. Flash assigns
$\tau(i)=L$ iff $m_{\mathrm H}(i)\vee m_{\mathrm W}(i)=1$, where $m_{\mathrm W}(i)=1$ when
$i$ is the oxygen of an intact water molecule, identified by its two covalent O--H bonds
in the supplied molecular topology. The remaining atoms have role $H$. Here \emph{resolution}
means capacity---width and update count---not spatial resolution; no atom is removed or
coarsened. Figure~\ref{fig:framework}(b--d) shows the three allocations this subsection
defines.

The checkpoint fixes the routing rule; each mask is an input that must be emitted on every
inference path and must match the checkpoint. This does not remove the obligation: Appendix~\ref{sec:app_arch}
documents a harness that emitted the water-oxygen mask for only one input format and would
have run Flash checkpoints as Standard. The masks depend only on
element identity and a local predicate, so they are static, interpretable and exactly
reproducible chemical priors \citep{marrink2007martini, kmiecik2016coarsegrained,
ryckaert1977numerical, hess1997lincs, hopkins2015longtimestep}. At inference in molecular
dynamics, the mask is computed on the first frame of each trajectory and held fixed for
that trajectory; it is not re-derived from instantaneous coordinates during integration.
Training uses the per-structure mask supplied by the data pipeline. Fixing the mask therefore makes
the conservative readout (Eq.~\eqref{eq:readout}, below) a fixed function of the coordinates
along a run, rather than a
piecewise function whose role assignment can switch between frames.

Capacity can be reduced along two independent axes. Width is per angular degree, so
reducing the channel count $D_s$ of role $s$ narrows the state and the tensor-product cost
without removing a degree. Update count is per atom, because a block can read a subset of
rows as queries and write only those rows. Neither operation removes an atom or a
coordinate, so the readout remains an all-atom sum. A learned router would have to preserve
each angular degree when it gates a row; a router fixed by element identity and a local
predicate has no such constraint and can be applied before the equivariant stack. This is
why the router is static.

For a role $s\in\{H,L\}$ the per-atom state space is
\begin{equation}
\mathcal X_s \;=\; \bigoplus_{\ell=0}^{l_{\max}}
\bigl(\mathbb R^{D_s}\otimes V_\ell\bigr),
\qquad \dim V_\ell=2\ell+1,
\label{eq:state_space}
\end{equation}
where $V_\ell$ is the irreducible representation of $\mathrm{SO}(3)$ of degree $\ell$,
$D_s$ is the number of channels per degree, and $\rho_s(R)$ denotes the induced rotation
action on $\mathcal X_s$. Standard uses $D_H=D_L=256$ and one shared state
$x\in\mathbb R^{N\times9\times256}$; Flash keeps two persistent states
$x^{H}\in\mathbb R^{N_H\times9\times256}$ and
$x^{L}\in\mathbb R^{N_L\times9\times32}$, where $N_H=|\mathcal H|$ and
$N_L=|\mathcal L|$. Both keep $l_{\max}=2$, so reducing $D_L$ narrows channels without
removing an angular degree. Standard can share one tensor because both roles have width 256;
Flash cannot, because 256 and 32 have different shapes. \emph{Persistent} means the tensor is
carried across the whole schedule and updated only by blocks that target that role.

The neighbour graph has an edge $(i,j)$ iff
$d_{ij}=\|\mathbf r_i-\mathbf r_j\|\le r_c$, with minimum-image distances when a periodic
cell is present. Edge contributions are weighted by the smooth cutoff
\begin{equation}
u(d)=\operatorname{clamp}\!\left(1-21\xi^5+35\xi^6-15\xi^7,\,0,\,1\right),
\qquad \xi=d/r_c,
\label{eq:cutoff}
\end{equation}
which vanishes with its first two derivatives at $d=r_c$.

Partition the adjacency matrix by target (row) and source (column),
$G=(G_{HH},G_{HL};G_{LH},G_{LL})$, and for block $b$ set
$G^{(b)}_{ts}=M^{(b)}_{ts}G_{ts}$, where $M^{(b)}$ is a binary $2\times2$ block mask. The
mask selects which of the four direction types $H\to H$, $H\to L$, $L\to H$ and $L\to L$
the block reads. Table~\ref{tab:block_masks} lists the four role-pair masks (A = all atoms; -f = Flash variant). A2H targets $H$
and reads both source roles; H2A targets both roles and reads only $H$ sources; A2L
targets $L$ and reads both; A2A, the uniform block, reads all four. The Flash variants
A2H-f and A2L-f use the A2H and A2L masks and differ in the state spaces and projections.

\begin{table}[t]
\centering
\caption{The four role-pair masks. A block reads only the source roles listed and writes only
the target roles listed.}
\label{tab:block_masks}
\small
\begin{tabular}{@{}lcc@{}}
\toprule
Block & Targets & Sources \\
\midrule
A2A & H, L & H, L \\
A2H & H & H, L \\
H2A & H, L & H \\
A2L & L & H, L \\
\bottomrule
\end{tabular}
\end{table}

Block $b$ has target set $T_b$, the roles with a nonzero target row in $M^{(b)}$. For a
target atom $i$, its masked neighbourhood is
$\mathcal N_i^{(b)}=\{j\in\mathcal N(i):M^{(b)}_{\tau(i),\tau(j)}=1\}$, and its state
update is
\begin{equation}
x_i^{(b+1)}=x_i^{(b)}+\mathcal B_b\Bigl(x_i^{(b)},
\bigl\{(\tilde x_{j\to i}^{(b)},\mathbf r_{ij}):
j\in\mathcal N_i^{(b)}\bigr\}\Bigr),
\qquad \tau(i)\in T_b,
\label{eq:block_update}
\end{equation}
where $\tilde x_{j\to i}^{(b)}=P_{b,\tau(j)\to\tau(i)}x_j^{(b)}$, $\mathcal B_b$ is the
E2Attention block of Section~\ref{sec:attention_block},
$\mathbf r_{ij}=\mathbf r_j-\mathbf r_i$, and $P_{b,s\to t}$ maps source-role channels into
target-role channels. Atoms outside $T_b$ pass through unchanged. In Eq.~\eqref{eq:block_update},
$\mathcal B_b$ denotes the full block increment; Eq.~\eqref{eq:aggregate} spells out its
residual, attention and FFN terms, so the residual is added once rather than once per term.

The projections are per-order linear maps,
\begin{equation}
P_{b,s\to t}
=\bigoplus_{\ell=0}^{l_{\max}}\left(W^{(\ell)}_{b,s\to t}\otimes I_{V_\ell}\right),
\qquad W^{(\ell)}_{b,s\to t}\in\mathbb R^{D_t\times D_s},
\label{eq:projection}
\end{equation}
and satisfy $P_{b,s\to t}\,\rho_s(R)=\rho_t(R)\,P_{b,s\to t}$ for every
$R\in\mathrm{SO}(3)$. A projection therefore changes channel width but not angular
structure \citep{geiger2022euclidean, passaro2023reducing}. Standard has $D_H=D_L$ and uses
the identity. Flash uses the identity within a
role and learned projections across roles: $P_{\uparrow}=P_{L\to H}$ maps 32 to 256 channels
per degree and $P_{\downarrow}=P_{H\to L}$ maps 256 to 32. These are the up and down
projections of the schedule.

A \emph{schedule} is the ordered list of attention blocks the backbone executes; its length is
the schedule depth. The \emph{key/value bank} is the set of source rows a block attends over.
The denominator in the update counts below is this depth. Standard executes
$[\mathrm{A2H},\mathrm{H2A}]\times2$. A2H updates only $H$, reading both
roles; H2A updates both roles, reading only $H$. High-resolution atoms therefore receive
$4/4$ updates and low-resolution atoms $2/4$, relative to the four-block schedule. Standard
allocates block participation only;
every atom keeps 256 channels. Flash executes
$[\mathrm{A2H}\text{-}\mathrm{f},\mathrm{A2H}\text{-}\mathrm{f},
\mathrm{A2L}\text{-}\mathrm{f}]\times2$. A2H-f takes $x^{H}$ as queries against the bank
$[x^{H};P_{\uparrow}(x^{L})]$ and updates $x^{H}$ only; A2L-f takes $x^{L}$ as queries
against $[P_{\downarrow}(x^{H});x^{L}]$ and updates $x^{L}$ only, where $[\,a;b\,]$
concatenates along the atom
axis. High-resolution atoms receive $4/6$ updates and low-resolution atoms $2/6$, relative to
the six-block schedule: repeating A2H-f keeps $H$ at four updates as the schedule deepens,
and A2L-f supplies the direct $L\to L$ path. Flash
allocates both persistent width and block participation. Table~\ref{tab:allocations}
summarises the two schedules and the uniform baseline.

\begin{table}[t]
\centering
\caption{Capacity allocations. Widths are channels per angular degree in the
high-/low-resolution stream; updates count how many blocks in a full schedule update an
atom of each role.}
\label{tab:allocations}
\small
\begin{tabular}{@{}lcccc@{}}
\toprule
Configuration & Widths (H/L) & Block schedule & H updates & L updates \\
\midrule
Baseline & 256/256 & 4 uniform blocks & 4/4 & 4/4 \\
Standard & 256/256 & [A2H, H2A]$\times2$ & 4/4 & 2/4 \\
Flash & 256/32 & [A2H-f, A2H-f, A2L-f]$\times2$ & 4/6 & 2/6 \\
\bottomrule
\end{tabular}
\end{table}

In pure water $\mathcal H$ is empty, so A2H-f has no target and A2L-f carries the $L\to L$
propagation; in mixed systems A2L-f also carries $H\to L$. When both roles are present,
both schedules preserve bidirectional cross-role exchange.

Both configurations apply role-specific atomic-energy heads to their routed rows:
Standard reads the two role partitions of its shared state, while Flash reads its two
persistent states. Each role has its own state and capacity, so each gets its own head;
merging before differentiation keeps the force the gradient of one energy rather than a sum
of per-head forces. The selected atomic energies are merged by the role mask before any
gradient is taken. The merged readout is
\begin{equation}
E \;=\; \sum_{i\in\mathcal H}\varepsilon^{H}_i
\;+\; \sum_{i\in\mathcal L}\varepsilon^{L}_i,
\qquad
\mathbf{F}_i \;=\; -\,\frac{\partial E}{\partial \mathbf{r}_i},
\label{eq:readout}
\end{equation}
and reduces to the uniform readout when no atom is low-resolution. Four invariants follow,
separating this from the reduced-fidelity schemes of Section~\ref{sec:related}: no atom
removed, no coordinate degree of freedom removed, cross-role interaction preserved both
ways, and an all-atom energy plus per-atom conservative force as the output signature.
Figure~\ref{fig:framework} sets both routed allocations beside the uniform one: the block
stream of each, the two capacity axes, and the low-resolution-atom update count.

\paragraph{Conservativity and smoothness.}
No atom leaves the model; the role changes how many updates and how many channels an atom
receives, not whether it appears in the energy. The role mask is fixed before the
backbone runs: at inference it is computed on the first frame of a trajectory and held
fixed for that trajectory, and in training it is the per-structure mask supplied by the
data pipeline. Along a run, therefore, $\tau$ is constant and Eq.~\eqref{eq:readout} is a
single differentiable function of the coordinates, not a piecewise function whose pieces
switch between frames. Interatomic contributions are multiplied by the cutoff of
Eq.~\eqref{eq:cutoff}, which goes to zero smoothly at $r_c$. The atomic energies of both
roles are summed into one scalar before any derivative is taken, and in the deployed model
every force is the gradient of that scalar; there is no separate force head. The forces are
therefore conservative by construction. Continuity is not uniform across the two attention
paths: Flash's unnormalized softplus contributions vanish with the cutoff, while Standard's
normalized softmax carries a $10^{-3}$ denominator floor; the weight still vanishes with the
smooth cutoff, but the floor controls the limit when the last neighbour is removed (Section~\ref{sec:attention_block}, Appendix~\ref{sec:app_ablation}). Away
from those events, and with the mask fixed, $E$ is differentiable. The mask and the Standard
regulariser are therefore two caveats: smoothness holds for a fixed mask and away from
cutoff events, and if the role were recomputed from instantaneous coordinates or changed
mid-trajectory, $E$ would be piecewise smooth and could jump at a reassignment.

\subsection{Inside the Block: E2Attention}
\label{sec:attention_block}

Each block $\mathcal B_b$ is an E2Attention block of E2Former-V2
\citep{huang2026formerv}, in equivariant form
\citep{liao2022equiformer, liao2023equiformerv}. We summarize its internals because the
special-block formulas below are written in these terms. The operator is not new here.
What changes between blocks is which rows are targets, which rows enter the key/value
bank, and at what width they are read. In the attention step, the query--key product is an
equivariant tensor product: it couples angular degrees so that each output transforms as
an irreducible representation of the same degree, and the edge features enter as
equivariant weights
\citep{vaswani2017attention, jaegle2021perceiver}. Appendix~\ref{sec:app_arch} gives the
architecture overview.

\paragraph{Initialization.}
At block $0$, each atom receives
\begin{equation}
\begin{split}
x_i^{(0)}=\frac{1}{\mathcal Z_i}\bigoplus_{\ell=0}^{l_{\max}}
\sum_{j\in\mathcal N(i)}&u(d_{ij})\,
\mathrm{MLP}_\ell^{\tau(i)}\!\left(\mathbf b(d_{ij}),e_{z_i},e_{z_j}\right)\\
&\times\mathbf Y_\ell(\hat{\mathbf r}_{ij}),
\end{split}
\label{eq:init}
\end{equation}
where $\mathbf b(d)$ is a radial-basis embedding of the distance, $e_z$ is an element
embedding, $\mathbf Y_\ell(\hat{\mathbf r})$ is the vector of degree-$\ell$ spherical
harmonics of the unit direction, and $\mathcal Z_i=\sum_{j\in\mathcal N(i)}u(d_{ij})$ is the per-atom
normalizer.

\paragraph{Queries, keys and values.}
Write $\bar x_i=\mathrm{RMSNorm}(x_i)$ and $\bar y_j=\mathrm{RMSNorm}(y_j)$, where $y_j$ is
the source state: $y_j=x_j$ in a same-space block and
$y_j=P_{b,\tau(j)\to\tau(i)}x_j$ in a cross-space one. The invariant features that form
queries and keys are
\begin{equation}
\begin{aligned}
c_{i,\ell}&=\sum_m\left(A_\ell\bar x_i^m\right)\odot
\left(B_\ell\bar x_i^m\right),\\
q_i&=\Phi_q\!\left([c_{i,0};\dots;c_{i,l_{\max}}]\right)\in\mathbb R^{h d_h},
\end{aligned}
\label{eq:qkv}
\end{equation}
where $A_\ell$ and $B_\ell$ are learned channel maps, $k_j$ is formed from $\bar y_j$ in
the same way, and $v_j=W_V\bar y_j$; $m$ indexes the $2\ell+1$ components of degree $\ell$,
and $h$ heads of dimension $d_h$ are used. The edge
features are
\begin{equation}
[g_{ij},\eta_{ij}]=\mathrm{MLP}_e\!\left(\mathbf b(d_{ij}),e_{z_i},e_{z_j}\right)
\in\mathbb R^h\times\mathbb R^h,
\label{eq:edge}
\end{equation}
where $g_{ij}$ is a per-head scale and $\eta_{ij}$ a per-head edge gate.

\paragraph{Attention weights.}
Standard uses a softmax over the masked neighbourhood. The per-edge geometric feature of
degree $\ell$ is
$s_{ij,\ell}=\sum_m(D^\ell W_{\mathrm{dot}}\bar x_i)_m
(\mathbf Y_\ell(\hat{\mathbf r}_{ij}))_m$, and with
$s_{ij}=[s_{ij,0};\dots;s_{ij,l_{\max}}]$ the logit and weight are
\begin{equation}
\begin{aligned}
\lambda_{ij}&=\left(W_{\mathrm{dir}}s_{ij}\right)\odot\eta_{ij}\odot
\frac{\langle q_i,k_j\rangle}{\sqrt{d_h}},\\
\alpha_{ij}&=\frac{u(d_{ij})\,e^{\lambda_{ij}}}
{\epsilon+\sum_{j'\in\mathcal N_i^{(b)}}u(d_{ij'})\,e^{\lambda_{ij'}}},
\end{aligned}
\label{eq:std_attn}
\end{equation}
where $\epsilon=10^{-3}$ is the denominator floor used by the implementation, $W_{\mathrm{dir}}\in\mathbb R^{h\times(l_{\max}+1)}$ projects the degree-wise
scalars to $h$ heads, $\eta_{ij}$ is the per-head gate, and
$\langle q_i,k_j\rangle=(q_i^1\!\cdot k_j^1,\dots,q_i^h\!\cdot k_j^h)\in\mathbb R^h$
collects the per-head dot products of the head slices $q_i^a,k_j^a\in\mathbb R^{d_h}$;
the exponential and division are per head. The floor keeps the denominator nonzero when the
last neighbour is removed; with the smooth cutoff the weight still vanishes, so the floor
matters only at that edge case, discussed in
Section~\ref{sec:role_mlip} and Appendix~\ref{sec:app_ablation}. Flash replaces the softmax
with a softplus gate,
\begin{equation}
\alpha_{ij}=\operatorname{softplus}\!\left(g_{ij}\odot
\frac{\langle q_i,k_j\rangle}{\sqrt{d_h}}\right)\odot\eta_{ij}\,u(d_{ij}).
\label{eq:flash_attn}
\end{equation}
Equation~\eqref{eq:flash_attn} is the OMol25 production form, which keeps the per-head dot
gate $g_{ij}$. The SPICE retrain used for the ablation and Table~\ref{tab:ablation} sets
$g_{ij}=1$ (no dot gate); the gate is priced separately in Section~\ref{sec:ablation}.

\paragraph{Aggregation and residual.}
Per head $a$, the block aggregates
$m_i^a=\sum_{j\in\mathcal N_i^{(b)}}\alpha_{ij}^a v_j^a$, concatenates the heads into
$m_i$, and applies the output projection and an equivariant feed-forward network,
\begin{equation}
\begin{aligned}
m_i^a&=\sum_{j\in\mathcal N_i^{(b)}}\alpha_{ij}^a v_j^a,\qquad
x_i'=x_i+W_O m_i,\\
x_i^{(b+1)}&=x_i'+\mathrm{FFN}_{S^2}(x_i'),
\end{aligned}
\label{eq:aggregate}
\end{equation}
where $m_i=[m_i^1;\dots;m_i^h]$. This is the update written in
Eq.~\eqref{eq:block_update}.

\paragraph{The four special blocks.}
With these definitions, the block types are
\begin{align}
\text{A2H:}\quad & x_i'=x_i+W_O\!\!\sum_{j\in\mathcal N(i)}\!\!
\alpha_{ij}W_V\bar x_j,
\qquad i\in\mathcal H,\label{eq:a2h}\\
\text{H2A:}\quad & x_i'=x_i+W_O\!\!\sum_{j\in\mathcal N(i)\cap\mathcal H}\!\!
\alpha_{ij}W_V\bar x_j,
\qquad i\in\mathcal A,\label{eq:h2a}\\
\text{A2H-f:}\quad & x_i^{H\prime}=x_i^{H}+W_O^H\!\!\sum_{j\in\mathcal N(i)\cap\mathcal H}\!\!
\alpha_{ij}^H W_V^H\bar x_j^H
\notag\\
& \qquad\qquad +W_O^H\!\!\sum_{j\in\mathcal N(i)\cap\mathcal L}\!\!
\alpha_{ij}^H W_V^H\overline{P_{\uparrow}x_j^L},
\qquad i\in\mathcal H,\label{eq:a2hf}\\
\text{A2L-f:}\quad & x_i^{L\prime}=x_i^{L}+W_O^L\!\!\sum_{j\in\mathcal N(i)\cap\mathcal H}\!\!
\alpha_{ij}^L W_V^L\overline{P_{\downarrow}x_j^H}
\notag\\
& \qquad\qquad +W_O^L\!\!\sum_{j\in\mathcal N(i)\cap\mathcal L}\!\!
\alpha_{ij}^L W_V^L\bar x_j^L,
\qquad i\in\mathcal L,\label{eq:a2lf}
\end{align}
where $P_{\uparrow}=P_{L\to H}$ and $P_{\downarrow}=P_{H\to L}$, and each update is followed by
$\mathrm{FFN}_{S^2}$. The uniform A2A block is A2H with target set $\mathcal A$.

\paragraph{Properties.}
Every operation above is $\mathrm{SO}(3)$-equivariant. The attention weights are invariant
under rotations; $W_V$ and $W_O$ act within the multiplicity channels of a fixed degree;
the projection of Eq.~\eqref{eq:projection} commutes with the rotation action; and the $S^2$
feed-forward network is equivariant, so a composition of these maps is equivariant.
Angular completeness holds
because both roles keep every degree up to $l_{\max}$; the cross-role projection preserves
angular information, and its per-degree rank is at most $D_L$; the full projection has rank
at most $(l_{\max}+1)D_L$, which is a capacity choice, not a dropped angular order. A block
costs $\mathcal O(N_Q(l_{\max}+1)hD^2+E_Q(l_{\max}+1)hD)$ for $N_Q$ target atoms and $E_Q$
target edges, with $D$ the width of the target stream; Flash adds
$(l_{\max}+1)(N_LD_LD_H+N_HD_LD_H)$ for the two cross-role projections. Connectivity follows from the masks: in Standard a low-resolution
atom reaches a low-resolution atom through $L\to H\to L$; in Flash the A2L bank contains
$L$, so one block suffices. Both schedules are strongly connected per cycle.

\paragraph{Derivation sketch.}
Equivariance follows block by block. The features $c_{i,\ell}$ contract the components of
each irreducible representation with themselves, and $s_{ij,\ell}$ contracts the target
state with spherical harmonics; both are invariant, hence so are $q_i$, $k_j$, the logits
and $\alpha_{ij}$. The value $v_j=W_V\bar y_j$ carries the rotation
action of the source degree, and $W_O$ maps it back to the target degree; the residual and
the feed-forward network preserve the action. The condition
$P_{b,s\to t}\rho_s=\rho_tP_{b,s\to t}$ is what lets a source in $\mathcal X_s$ enter a
target in $\mathcal X_t$ without mixing degrees. Smoothness holds for the Flash softplus
path: the cutoff is polynomial and vanishes with its first two derivatives at $r_c$, and
the softplus attention is unnormalized, so each edge contribution vanishes at the cutoff.
The Standard path uses a normalized softmax; its $10^{-3}$ denominator floor controls the
limit at cutoff events, while the weight still vanishes with the smooth cutoff
(Appendix~\ref{sec:app_ablation}).
We therefore do not claim a globally smooth Standard surface. With the mask fixed, the energy
is differentiable away from those cutoff events, and $\mathbf F_i=-\partial E/\partial\mathbf r_i$
is the exact gradient there. Fixing the mask removes trajectory-level role switching, which
would otherwise be a second discrete source; if the mask were re-evaluated from instantaneous
geometry and changed, $E$ would be only piecewise smooth.

\subsection{Reachable-Row Key/Value Construction}
\label{sec:locality}

In an A2H block the queries are the high-resolution atoms, $\mathcal H$, and the key/value
bank needs to contain only the rows those queries can reach,
\begin{equation}
\mathcal U \;=\; \bigcup_{i\in\mathcal H}\mathcal N(i),
\label{eq:reachable_rows}
\end{equation}
where $\mathcal N(i)$ includes periodic images. In a solvated system $\mathcal U$ is much
smaller than the full atom set; projecting only $\mathcal U$ instead of every atom is a measured
part of the saving. The attention reads the bank only through the neighbour index, with no pooling across atoms and
a per-row normalization that does not mix atoms. Removing rows
outside $\mathcal U$ therefore removes inputs no query can address and leaves every query's
attention input unchanged. We build the bank over $\mathcal U$ at inference; it adds no parameters and leaves the energy
definition unchanged, although enabling it during training changes numerics at the bit level. The construction is used in Flash, where the high-resolution query set is a small fraction of
atoms; the two pure-water boxes are empty-union endpoints. On the three non-degenerate
systems it reduces Flash latency by 23.1--25.5\% ($1.30$--$1.34\times$), with the empty-union
endpoints reaching 27.3\% ($1.375\times$) and carrying no headline claim (Section~\ref{sec:experiments},
Figure~\ref{fig:filtering}).

On the solvated peptide of Section~\ref{sec:experiments}, $\mathcal U$ contains 322 of the
5{,}992 primary atoms (5.37\%); the projection runs on 322 rows instead of 5,992, 18.6$\times$
fewer. The expanded attention bank has 10,546 image rows; that is not the projection
denominator. The construction is algebraically lossless but not bit-exact: the smaller row count
changes the GEMM kernel selected by the library, and the corresponding backward operation is
nondeterministic. A runtime gate disables the construction above a union fraction of 0.9;
Standard reaches that bound. The size of $\mathcal U$ is not the fraction of atoms assigned to
the high-resolution role.

\subsection{Implementation}
\label{sec:implementation}

The production Standard configuration uses four layers at a 6\,\AA{} cutoff. The production
Flash configuration uses six layers at a 5\,\AA{} cutoff and carries its low-resolution stream
at 32 channels per degree against 256 in the high-resolution stream. The choices are
deliberate on the allocation axes: Standard isolates the update-count axis on the allocation axes, because every atom
keeps the 256-channel width; its FFN and role-head differences are separate and are not
attributed to participation alone; Flash adds the width axis and the water-oxygen role. Water oxygen is the
next most abundant role and the most homogeneous in bulk water, while solute oxygen stays
high-resolution because it is chemically diverse. The 5\,\AA{} cutoff and six-layer depth are
empirical choices from the ablation, not a claim of optimality; Appendix~\ref{sec:app_ablation}
isolates the axes it can, and state layout remains unmeasured
(Appendix~\ref{sec:app_bounds}). The two configurations therefore differ in depth, cutoff,
routing, state layout, attention parameterisation and the feed-forward network (FFN) at once;
their comparison is not a
single-variable ablation.

The implementation is compiled PyTorch \citep{ansel2024pytorch} with custom tile-level
kernels \citep{tillet2019triton}, configured through Hydra \citep{yadan2019hydra} and trained
data-parallel across GPUs in fp32 with TF32 disabled. Production efficiency measurements use
\texttt{torch.compile} at inference; the single-variable ablation of
Appendix~\ref{sec:app_ablation} reports uncompiled batch-1 latency, and each table states its
mode. The high-resolution and low-resolution heads are trained by warm-starting on direct
forces and then fine-tuning conservatively \citep{fu2025learning}: the warm-start stage
predicts the force directly, and the fine-tuning stage differentiates the merged energy of
Eq.~\eqref{eq:readout}, which is the weight set all accuracy results use. The routing mask is
an input-format obligation and not only a model setting: it must be emitted on every inference
path, because a harness covering only one format benchmarks Flash as Standard
(Appendix~\ref{sec:app_arch}). Every accuracy table states its force mode---in the caption for
the main tables, in a column for the appendix tables. The controlled SPICE retrain and ablation
recipe is 951{,}005 training frames and the 50{,}195-frame SPICE test split (which the
ablation uses as its validation set), 200{,}000 steps,
micro-batch 64, one node of eight A100-40GB, learning rate $4\times10^{-4}$ with a 4{,}000-step
warmup, weight decay 0, gradient clip 100, energy:force weights 0.2:0.9, an atom-wise $L_2$ force loss,
seed 6666, and validation every 10k steps (Appendix~\ref{sec:app_ablation}). The production
OMol25 schedule is not recorded, so we do not claim the ablation recipe
is the production recipe. The production checkpoints carry 22.224M active parameters for
Standard and 22.876M for Flash (Appendix~\ref{sec:app_params}).

Source code is available at \url{https://github.com/IQuestLab/UBio-MolFM}.

The eight numerical gates that certify the implementation---rotation equivariance,
finite-difference agreement with the autograd force, size consistency, angular-order creation,
empty high-resolution stream, empty low-resolution stream, gradient flow and no-op
equivalence---are given in full in Appendix~\ref{sec:app_gates}.

\section{Experiments}
\label{sec:experiments}

Atom-role-aware allocation improves the force-accuracy--throughput frontier against a
uniform-capacity predecessor. The routed cell has 2.8\% higher force error than a uniform retrain
at 45.3\% lower latency and 41.3\% lower peak memory. Both production configurations remain
structurally stable and locally structured within the measured horizons; thermodynamic accuracy
is not claimed. Two findings qualify the design hypotheses: the role-stratified error imbalance
anticipated in Section~\ref{sec:intro}, and the short-range softening that qualifies H2.

\subsection{Experimental setup}
\label{sec:setup}

\textbf{Accuracy data.} OMol25 is evaluated on the full 2{,}762{,}021-sample composition
validation split \citep{levine2025open}; SPICE is evaluated on its 50{,}195-frame test set
\citep{eastman2023spice}. Both production configurations are retrained from scratch on SPICE
at their OMol25 architecture, except that the SPICE Flash recipe drops the scalar dot gate
(Section~\ref{sec:ablation}); the gate is priced separately. Many recent MLIP papers do not
report SPICE, so we cite published SPICE rows where available and use SPICE mainly for
controlled architectural comparisons rather than retraining external models under a different
recipe. They are evaluated at one common aggregation caliber (Appendix~\ref{sec:app_spice}). Accuracy is force MAE per Cartesian
component in meV/\AA{} and energy MAE in meV/atom. The OMol25 accuracy tables print the exact
component-weighted caliber; the SPICE table uses its own common aggregation (Appendix~\ref{sec:app_spice}). Figure~\ref{fig:pareto}(c--f) uses the training-time batch caliber
(8.14 and 11.26) because the Pareto axis was frozen earlier, and we flag that difference
rather than mixing the two. Table~\ref{tab:uniformreduce} is the exception: the controlled evaluator applies one
SPICE split, force aggregation and evaluation path to the V2 baseline and every U-* row; its V2
force MAE is 6.08, whereas 7.04 is the separately transcribed published value.

\textbf{Systems and campaigns.} Table~\ref{tab:systems} lists the systems. Each timing table
names its device and timing boundary, and every ratio uses measurements from the same
campaign, with the one backend exception stated below; no other ratio mixes campaigns.

\textbf{Molecular dynamics.} The 20\,ps matrix is Standard/Flash $\times$ water / 150\,mM
NaCl / DEKSAN $\times$ NVT/NPT/NVE $\times$ three seeds (104729, 130363, 155921) $=54$ legs,
each 40{,}000 steps at 0.5\,fs. Every leg completed; NPT used a Monte Carlo barostat with
acceptance in $[0.30,0.40]$, and NVE legs continued from their NVT endpoint. The 1\,ns
campaign is a separate NVT run on the 32{,}437-atom protein at 300\,K with a Langevin
thermostat, 2{,}000{,}000 steps at 0.5\,fs, one seed (104729), fixed cell. The 20\,ps
matrix's device is not recorded. The full protocol, including boxes,
cutoffs and the equilibration source, is in Appendix~\ref{sec:app_water}; quantities not
recorded there are marked as such rather than inferred.

\textbf{Metrics and statistics.} Every OMol25 trained result is single-seed; the SPICE seed
check in Section~\ref{sec:seedvar} adds two seeds at one recipe but does not extend to
OMol25. Accuracy differences are read as levels, not as statistical tests; the same-seed
rerun in Section~\ref{sec:ablation} bounds implementation variance, not seed variance. Out-of-memory
outcomes are retained as results and never interpolated. Pure water is the all-low-resolution
limit of Flash, in which the high-resolution stack is bypassed.

\begin{table}[t]
\centering
\small
\caption{Systems used in the paper. Sizes are the full simulated system. The 20\,ps matrix
uses one 216-molecule water box and a 49.96\,\AA{} NaCl box with 11 Na$^+$ and 11 Cl$^-$; the
6{,}000/18{,}000-atom water
boxes belong to the efficiency ladder, and the 64/512-molecule boxes to the separate 12-leg
water-only matrix (Appendix~\ref{sec:app_water}).}
\label{tab:systems}
\begin{tabular}{@{}lrrl@{}}
\toprule
System & Size & Type & Used for \\
\midrule
Water box (efficiency) & 6{,}000 / 18{,}000 atoms & solvent & efficiency ladder \\
Water box (MD) & 216 molecules & solvent & 20\,ps matrix \\
150\,mM NaCl (MD) & 49.96\,\AA{} box & electrolyte & 20\,ps matrix \\
DEKSAN peptide & 5{,}992 atoms & solute & MD, latency, ablation \\
Nucleic acid & 15{,}961 atoms & solute & efficiency, Figure~\ref{fig:framework} \\
Solvated protein & 32{,}437 atoms & solute & efficiency, 1\,ns MD \\
\bottomrule
\end{tabular}
\end{table}

\subsection{Accuracy}
\label{sec:accuracy}

On OMol25, Standard reaches 8.146\,meV/\AA{} force MAE against E2Former-V2's published 8.38,
a 2.8\% reduction, whereas Flash reaches 11.272, a 34.5\% increase that remains below the
1\,kcal\,mol$^{-1}$\,\AA$^{-1}$ chemical-accuracy threshold \citep{chmiela2017machine}
(Table~\ref{tab:accuracy}). On SPICE, after retraining and evaluation at one common caliber,
Standard reaches 5.89\,meV/\AA{} against E2Former-V2's 7.04, while Flash reaches 8.17
(Table~\ref{tab:spice}); metric and aggregation details are in Appendix~\ref{sec:app_spice}.
On the SPICE water subset, Standard reads 5.57\,meV/\AA{} against E2Former-V2's 5.26 (a
5.9\% increase) and Flash 16.28, a 3.1$\times$ regression; the All column is 95.6\% weighted on
PubChem and dimers and does not expose either. Energy is reported alongside force: on OMol25, Standard and Flash reach 2.190 and
3.941\,meV/atom against E2Former-V2's 1.16, and on SPICE they reach 0.30 and 0.39 against 0.31.
The advance is in efficiency and scale at competitive force accuracy; on OMol25 the energy
comparison is not in our favour and is not used as a headline.

Table~\ref{tab:accuracy} also separates the four OMol25 domains. Standard is worse than
E2Former-V2 on biomolecules (5.104 vs 4.96) and metals (25.555 vs 24.40), and better on
electrolytes (8.002 vs 10.11) and organics (17.072 vs 19.69). Flash is worse on all four
domains. The total row alone would hide both patterns; the domain table in
Appendix~\ref{sec:app_role} carries the full decomposition with its published-convention
caveat. For the explicit-solvent biomolecules that motivate the paper, role-aware allocation
is not a uniform accuracy win; it is a trade that pays on the electrolyte and organic domains
and on throughput, and we state that rather than claiming a uniform improvement.

\begin{table}[t]
\centering
\caption{Accuracy on OMol25 Val-Comp against uniform-capacity potentials. $E$ is energy MAE in
meV/atom; $F$ is per-component force MAE in meV/\AA{}. Best and second-best values are bold and
underlined, respectively; all E2Former-V3 results use conservative forces
\citep{bigi2024dark}. Baselines are as
published; E2Former-V2's force convention is unstated, so its row is read as a level.
E2Former-V3 rows are single-seed. The UMA-S row is
version-unpinned and is not the UMA-S-1.1 checkpoint of Table~\ref{tab:uma}.}
\label{tab:accuracy}
\setlength{\tabcolsep}{2.1pt}
\scriptsize
\resizebox{\linewidth}{!}{%
\begin{tabular}{llrrrrrrrrrr}
\toprule
Dataset & Model
& \multicolumn{2}{c}{Biomolecules}
& \multicolumn{2}{c}{Electrolytes}
& \multicolumn{2}{c}{Metals}
& \multicolumn{2}{c}{Organics}
& \multicolumn{2}{c}{Total} \\
\cmidrule(lr){3-4}\cmidrule(lr){5-6}\cmidrule(lr){7-8}
\cmidrule(lr){9-10}\cmidrule(lr){11-12}
& & $E$ & $F$ & $E$ & $F$ & $E$ & $F$ & $E$ & $F$ & $E$ & $F$ \\
\midrule
All & eSEN-sm-cons \citep{fu2025learning}
& \underline{0.59} & \textbf{4.61} & \textbf{1.01} & \underline{8.08} & 2.30 & 28.86 & \textbf{0.84} & \textbf{11.11} & \underline{1.27} & \underline{8.25} \\
All & UMA-S \citep{wood2025uma}
& \textbf{0.53} & 5.69 & 2.69 & 11.65 & 4.63 & 37.85 & \underline{1.00} & \underline{13.15} & 3.62 & 12.02 \\
All & MACE-OMOL \citep{levine2025open}
& 1.76 & 6.05 & 2.15 & 10.87 & \textbf{2.15} & \underline{24.51} & 2.31 & 23.70 & 1.84 & 10.10 \\
All & E2Former-V2-Cons. \citep{huang2026formerv}
& 1.01 & \underline{4.96} & \underline{1.52} & 10.11 & \underline{2.18} & \textbf{24.40} & 1.26 & 19.69 & \textbf{1.16} & 8.38 \\
\midrule
All & E2Former-V3-Standard
& 2.01 & 5.10 & 2.27 & \textbf{8.00} & 2.47 & 25.56 & 1.89 & 17.07 & 2.19 & \textbf{8.15} \\
All & E2Former-V3-Flash
& 5.00 & 7.68 & 3.92 & 11.45 & 2.86 & 28.40 & 2.45 & 20.07 & 3.94 & 11.27 \\
\bottomrule
\end{tabular}}
\end{table}

\begin{table}[t]
\centering
\scriptsize
\setlength{\tabcolsep}{2.2pt}
\caption{Accuracy on the SPICE test set. $E$ is energy MAE in meV/atom; $F$ is per-component
force MAE in meV/\AA{}. The \emph{All} column is structure-count-weighted for $E$ and
component-pooled for $F$ (Appendix~\ref{sec:app_spice}); the two bases differ from the published
source and are recomputed, not transcribed. Best and second-best values are bold and
underlined. E2Former-V3 rows are 200{,}000-step retrains from scratch on SPICE, single seed,
and use conservative forces. MACE rows are MACE-OFF23 \citep{kovacs2025maceoff}; the
E2Former-V1 row is \citep{li2025former}.}
\label{tab:spice}
\resizebox{\linewidth}{!}{%
\begin{tabular}{l *{8}{rr}}
\toprule
Model & \multicolumn{2}{c}{PubChem} & \multicolumn{2}{c}{Monomers} & \multicolumn{2}{c}{Dimers} & \multicolumn{2}{c}{Dipeptides} & \multicolumn{2}{c}{Solv.\ Amino} & \multicolumn{2}{c}{Water} & \multicolumn{2}{c}{Qmugs} & \multicolumn{2}{c}{All} \\
\cmidrule(lr){2-3}\cmidrule(lr){4-5}\cmidrule(lr){6-7}\cmidrule(lr){8-9}\cmidrule(lr){10-11}\cmidrule(lr){12-13}\cmidrule(lr){14-15}\cmidrule(lr){16-17}
& E & F & E & F & E & F & E & F & E & F & E & F & E & F & E & F \\
\midrule
MACE Small & 1.41 & 35.68 & 1.04 & 17.63 & 0.98 & 16.31 & 0.84 & 25.07 & 1.60 & 38.56 & 1.67 & 28.53 & 1.03 & 41.45 & 1.27 & 32.43 \\
MACE Medium & 0.91 & 20.57 & 0.63 & 9.36 & 0.58 & 9.02 & 0.52 & 14.27 & 1.21 & 23.26 & 0.76 & 15.27 & 0.69 & 23.58 & 0.80 & 18.63 \\
MACE Large & 0.88 & 14.75 & 0.59 & 6.58 & 0.54 & 6.62 & 0.42 & 10.19 & 0.98 & 19.43 & 0.83 & 13.57 & 0.45 & 16.93 & 0.77 & 13.39 \\
E2Former-V1 33M & 0.67 & 8.90 & 0.49 & 7.10 & 0.43 & 4.01 & 0.51 & 5.63 & 1.10 & 19.20 & 0.96 & 13.50 & 0.65 & 10.20 & 0.60 & 8.12 \\
eSEN-Direct \citep{fu2025learning} & \multicolumn{14}{c}{---} & 0.56 & 10.98 \\
eSEN-Cons.\ \citep{fu2025learning} & \multicolumn{14}{c}{---} & \textbf{0.23} & \underline{6.36} \\
E2Former-V2-Direct \citep{huang2026formerv} & 0.65 & 9.49 & 0.46 & 4.59 & 0.28 & 2.40 & 0.33 & 5.63 & 0.56 & 13.37 & 0.70 & 7.43 & 0.63 & 11.69 & 0.54 & 8.32 \\
E2Former-V2-Cons.\ \citep{huang2026formerv} & \underline{0.37} & \underline{8.00} & \underline{0.19} & \underline{3.25} & \underline{0.18} & \underline{2.36} & \underline{0.20} & \underline{4.45} & \underline{0.30} & \underline{9.51} & \underline{0.30} & \textbf{5.26} & \textbf{0.25} & \underline{9.43} & 0.31 & 7.04 \\
\midrule
E2Former-V3-Standard (4L, 6\,\AA{}) & \textbf{0.36} & \textbf{6.68} & \textbf{0.18} & \textbf{2.62} & \textbf{0.17} & \textbf{2.01} & \textbf{0.14} & \textbf{3.79} & \textbf{0.22} & \textbf{7.94} & \textbf{0.24} & \underline{5.57} & \textbf{0.25} & \textbf{7.98} & \underline{0.30} & \textbf{5.89} \\
E2Former-V3-Flash (6L, 32\,ch, 5\,\AA{}) & 0.43 & 9.10 & 0.41 & 3.42 & 0.31 & 3.20 & 0.23 & 6.05 & 1.01 & 22.98 & 0.90 & 16.28 & \underline{0.42} & 12.55 & 0.39 & 8.17 \\
\bottomrule
\end{tabular}}
\end{table}

\subsubsection{Seed variability on SPICE}
\label{sec:seedvar}

Across the two additional SPICE seeds, Standard spans 0.2717--0.2731\,meV/atom in energy and
5.32--5.37\,meV/\AA{} in force; Flash spans 0.3948--0.4419 and 7.36--8.11
(Table~\ref{tab:seedspice}). The headline run is a third seed at the same recipe but is not
included in those ranges; it is higher on force for both configurations and on energy for Standard, while Flash's
headline energy lies below its additional-seed range.
Table~\ref{tab:spice} remains the single-seed headline. Standard's force range
stays below E2Former-V2's 7.04\,meV/\AA{}; Flash's range remains above it, so seed variation
does not reverse the accuracy ordering. The check bounds seed variation on SPICE and is not a
multi-seed OMol25 evaluation; a multi-seed OMol25 estimate would require over 30{,}000
additional GPU-hours, so the OMol25 headline remains single-seed by design.

\begin{table}[t]
\centering
\small
\caption{Two additional same-recipe SPICE seeds against the single-seed headline. Ranges cover
the two additional seeds only; the headline run is a third seed but is not included in the
range. $E$ is energy MAE in meV/atom; $F$ is per-component force MAE in meV/\AA{}.}
\label{tab:seedspice}
\begin{tabular}{@{}lrrrr@{}}
\toprule
Model & Additional $E$ & Additional $F$ & Headline $E$ & Headline $F$ \\
\midrule
E2Former-V3-Standard & 0.2717--0.2731 & 5.32--5.37 & 0.30 & 5.89 \\
E2Former-V3-Flash & 0.3948--0.4419 & 7.36--8.11 & 0.39 & 8.17 \\
\bottomrule
\end{tabular}
\end{table}

\subsection{Efficiency and scaling}
\label{sec:efficiency}

On the frozen H200 Pareto campaign, Standard is $1.75$--$1.78\times$ and Flash
$4.00$--$4.88\times$ faster than E2Former-V2 across the three solvated systems. Across all
five systems the ranges are $1.75$--$1.84\times$ and $4.00$--$5.85\times$; the upper ends come
from the two pure-water boxes, and $5.85\times$ is the all-low-resolution limit. Both are non-dominated
on the force-accuracy--throughput frontier (Figure~\ref{fig:pareto},
Table~\ref{tab:h200pareto}). Table~\ref{tab:h200main} prints the campaign the ratios are
computed from: on the 32{,}437-atom protein, Standard is 533.866\,ms and Flash 194.719\,ms,
against E2Former-V2's 949.617\,ms.
Memory follows the same ordering: Standard uses less peak memory than E2Former-V2 on every
system, and Flash uses less still (Table~\ref{tab:h200pareto}).

\subsubsection{Comparison with UMA-S-1.1}
\label{sec:uma}

Table~\ref{tab:uma} uses the same H200 Pareto campaign and timing boundary, which includes
graph construction and the conservative force path. UMA-S-1.1 is the checkpoint whose latency
is measured here; its published force MAE is 9.54\,meV/\AA{} (Table~\ref{tab:external}).
The UMA-S row in Table~\ref{tab:accuracy} is a different, version-unpinned model at
12.02\,meV/\AA{}. UMA-S-1.1 takes 610.150\,ms on water 6k and 628.265\,ms on DEKSAN 6k; Standard
takes 122.839 and 124.404\,ms, and Flash 44.915 and 55.295\,ms. Standard is 4.97--5.05$\times$
faster and Flash 11.36--13.58$\times$ faster on these two systems. On water 18k, RNA 16k and
protein 32k, the tested UMA-S-1.1 configuration runs out of memory, whereas Standard and Flash
complete at 331.438/104.070, 262.964/96.197 and 533.866/194.719\,ms. Inference settings for
each model are in Table~\ref{tab:config} (Appendix~\ref{sec:app_protocol}).

\begin{table}[t]
\centering
\small
\caption{Latency against UMA-S-1.1 on the H200 Pareto campaign, in ms per energy-and-force
evaluation. All rows use the timing boundary that includes graph construction and the
conservative force path. OOM is the measured outcome, not a missing value. UMA-S-1.1 is the
measured checkpoint; the UMA-S row in Table~\ref{tab:accuracy} is a different, version-unpinned
model.}
\label{tab:uma}
\begin{tabular}{@{}lrrr@{}}
\toprule
System & UMA-S-1.1 & Standard & Flash \\
\midrule
Water 6{,}000 & 610.150 & 122.839 & 44.915 \\
DEKSAN 5{,}992 & 628.265 & 124.404 & 55.295 \\
Water 18{,}000 & OOM & 331.438 & 104.070 \\
Nucleic acid 15{,}961 & OOM & 262.964 & 96.197 \\
Protein 32{,}437 & OOM & 533.866 & 194.719 \\
\bottomrule
\end{tabular}
\end{table}

\begin{figure}[t]
\centering
\includegraphics[width=\linewidth]{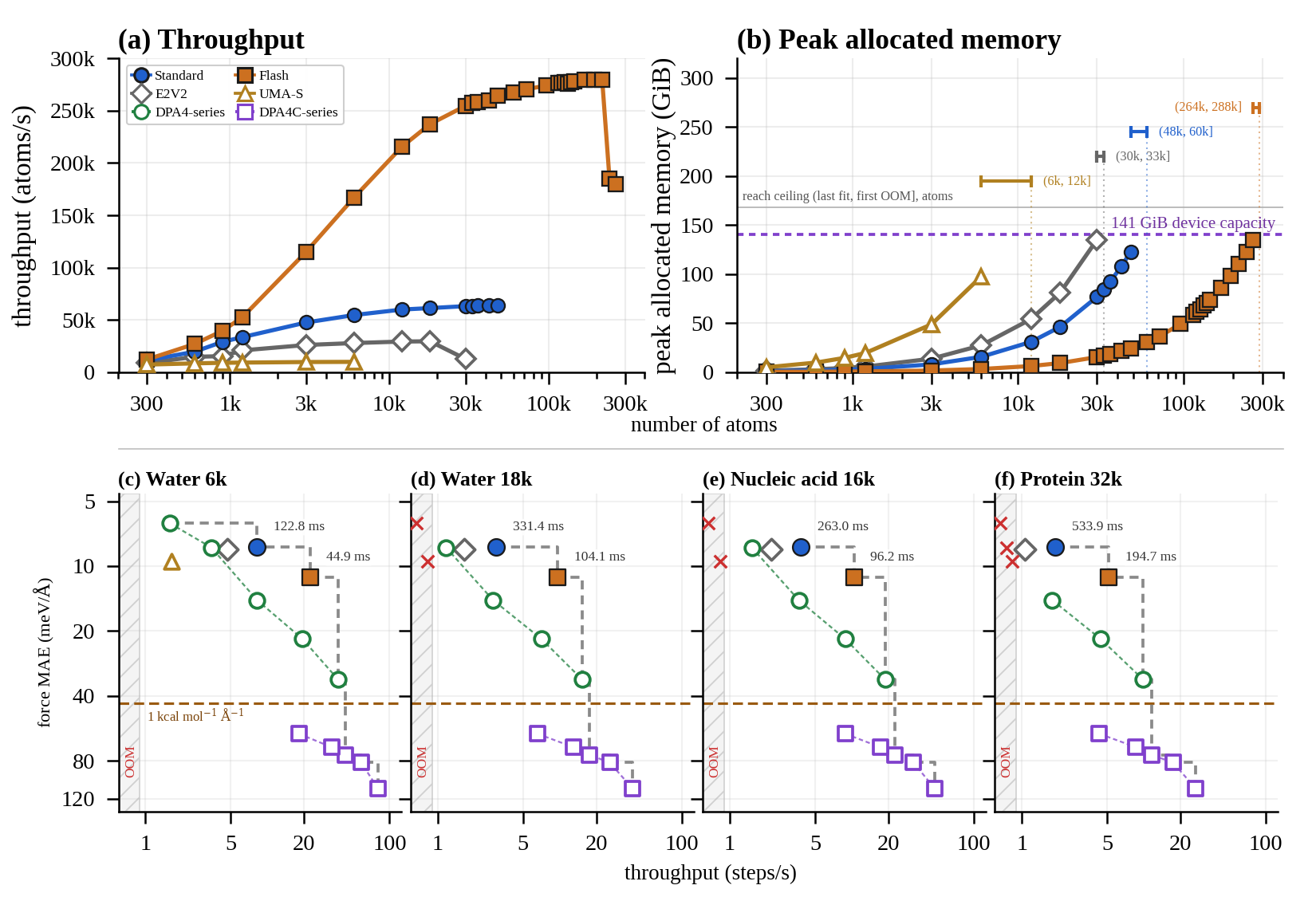}
\caption{Efficiency under device-specific protocols. \textbf{(a,b)} Single-step throughput and
peak allocated memory on the H200 water-box ladder; Flash and Standard use the opt-in exact
cell-list neighbor backend, and their exhaustive-search counterparts are tabulated in
Table~\ref{tab:ladder}. Latencies at or above 33{,}000 atoms use a pinned thread pool, so both curves are device
measurements and reach rests on the memory
column. \textbf{(c--f)} Force MAE on the OMol25 composition validation split
(Table~\ref{tab:accuracy}) against H200 throughput for benchmark systems (see
Table~\ref{tab:h200pareto} for details). The DPA4- and DPA4C-series each span
Nano/Mini/Neo/Air/Plus. Panels (a,b) and (c--f) are separate H200 campaigns with different
timing boundaries, not pooled. Red crosses mark OOM outcomes, and gray dashed steps trace the
empirical frontier. The accuracy axis of (c--f) uses the training-time batch caliber
(8.14/11.26), not the exact caliber printed in Table~\ref{tab:accuracy} (8.15/11.27).
UMA-S is plotted at the published 9.54\,meV/\AA{} of UMA-S-1.1, the checkpoint whose latency
is measured here (Table~\ref{tab:uma}); it is not the version-unpinned UMA-S row at
12.02\,meV/\AA{} in Table~\ref{tab:accuracy}.}
\label{fig:throughput}
\label{fig:pareto}
\end{figure}

On the H200 water-box ladder of Figure~\ref{fig:throughput}, with Flash and Standard measured
with the exact cell-list backend, Standard is $1.30$--$2.07\times$ and Flash
$1.39$--$8.01\times$ faster than E2Former-V2 over the shared sizes up to 18{,}000 atoms. The
lower bounds are at 300 atoms and are launch-bound; they are not device-speed statements. At
30{,}000 atoms, the last size all three complete, the ratios are $4.97\times$ and
$20.0\times$. In the fine reach scan, E2Former-V2 and Standard then run out of memory at 48{,}000 and 60{,}000 atoms,
where Flash reaches 264{,}000 at 135\,GiB; the exhaustive-search records are in
Appendix~\ref{sec:app_cost}. The E2Former-V2 denominator is its frozen exhaustive-search
ladder, not a cell-list measurement; the effect is small and against us on the shared points,
but it is stated rather than left implicit. For Flash at 1\,fs per step these latencies correspond to
about 1.9\,ns/day at 6k, 0.9 at 16k, 0.8 at 18k and 0.4 at 32k on one H200, before integration
and output overhead. The 264{,}000-atom result is a single-step energy-and-force evaluation,
not an MD trajectory; the largest MD system in this work is the 32{,}437-atom protein.

The default exhaustive-search Flash curve declines beyond $\sim$100{,}000 atoms because two shared
implementation costs grow: the exhaustive neighbor search (0.69\,s of the 2.28\,s step at
264{,}000 atoms, 40\% of GPU-busy time) and the per-call allocation and release of
neighbor-search scratch buffers (0.60\,s of host time). The cell-list backend removes both and
holds throughput flat to 216{,}000 atoms. The Flash cell-list curve drops at its last two
ladder points, 240{,}000 and 264{,}000 atoms, at the GPU-memory boundary: the step's peak
allocation is 135\,GiB, 96\% of the 141\,GiB device capacity, and the frequent reclamation and
re-allocation of transient buffers stall the step there.

\begin{table}[t]
\centering
\small
\caption{The five-system H200 Pareto campaign. The first three rows are latency in ms; the
last two rows are speedups ($\times$) computed from E2Former-V2's latency in the same column.
All rows share the timing boundary that includes graph construction and the conservative force
path. E2Former-V2 is a speed-only checkpoint whose weights differ from those behind its published
accuracy. The Flash cells in the two water columns are the all-low-resolution limit, in which no atom is
high-resolution; the three solvated systems span $4.00$--$4.88\times$. Peak memory is in
Table~\ref{tab:h200pareto}; the five-system campaign of Table~\ref{tab:h200} uses a different
boundary and is not interchangeable with this table.}
\label{tab:h200main}
\resizebox{\linewidth}{!}{%
\begin{tabular}{lrrrrr}
\toprule
 & Water 6{,}000 & Water 18{,}000 & Peptide 5{,}992 & Nucleic acid 15{,}961 & Protein 32{,}437 \\
\midrule
Standard (ms) & 122.839 & 331.438 & 124.404 & 262.964 & 533.866 \\
Flash (ms) & 44.915 & 104.070 & 55.295 & 96.197 & 194.719 \\
E2Former-V2 (ms) & 214.668 & 608.483 & 221.041 & 459.515 & 949.617 \\
\midrule
Standard / V2 & 1.75 & 1.84 & 1.78 & 1.75 & 1.78 \\
Flash / V2 & 4.78 & 5.85 & 4.00 & 4.78 & 4.88 \\
\bottomrule
\end{tabular}}
\end{table}

\subsection{Role-stratified error}
\label{sec:role}

We use \emph{high-resolution} (HR) for role $H$ and \emph{low-resolution} (LR) for role $L$;
Standard reduces only the update count, while Flash reduces width as well, so ``low-resolution''
is not a synonym for narrow width under Standard. HR atoms comprise
49.1--49.6\% of force components but carry 76--78\% of absolute error in both routed
configurations (Table~\ref{tab:rolecompact}). Standard reduces participation only, whereas Flash
reduces both persistent width and participation and routes more atoms low-resolution; their
HR/LR error ratios nevertheless remain close, 3.634 and 3.324. The ratio alone does not
identify capacity placement; the two configurations are not a causal ablation. The chemistry reading remains a conjecture
(Appendix~\ref{sec:app_bounds}).

\begin{table}[t]
\centering
\small
\caption{Role-stratified force error on the full OMol25 validation split, exact
component-weighted caliber. HR share is the fraction of force components carried by
high-resolution atoms; error share is the fraction of absolute error they carry. The full
per-domain table is in Appendix~\ref{sec:app_role}.}
\label{tab:rolecompact}
\begin{tabular}{@{}lrrr@{}}
\toprule
Model & HR share of components & HR share of error & HR/LR error ratio \\
\midrule
E2Former-V3-Standard & 49.57\% & 78.13\% & 3.634 \\
E2Former-V3-Flash & 49.12\% & 76.24\% & 3.324 \\
\bottomrule
\end{tabular}
\end{table}

\subsection{Uniform versus role-aware at the same recipe}
\label{sec:matched}

Appendix~\ref{sec:app_ablation} retrains a uniform architecture on SPICE at the same
200{,}000-step recipe and reaches 6.745\,meV/\AA{}. That is the same-recipe control, and it
beats E2Former-V2's published 7.04.
The controlled-evaluator V2 reads 6.08 but is a re-evaluation, not a same-recipe retrain, so it
is not the control for this comparison.
Against that control (Table~\ref{tab:matched}), the single-state routed model (\textbf{S+1})
reaches 6.934\,meV/\AA{}---2.8\%
worse---at 392.5\,ms and 19.29\,GB instead of 717.1\,ms and 32.84\,GB, a 45.3\% latency and
41.3\% memory reduction. Adding the s2 FFN (\textbf{S+2}) reaches 5.887\,meV/\AA{} at 411.5\,ms
and 22.43\,GB, better than the uniform control at lower cost, although that cell changes two
axes. The cleanest comparison is S+1: at roughly half the cost, the single-state mixed-resolution
bundle has 2.8\% higher force error than a uniform model. It changes the state layout and
allocation together, not routing alone; S+2 additionally changes the FFN.
These are 200{,}000-step, single-seed, SPICE-only cells; they are not the production
checkpoints, and the ordering need not carry to OMol25 (Appendix~\ref{sec:app_ablation}).

\begin{table}[t]
\centering
\small
\caption{Same-recipe uniform versus routed cells on SPICE (200{,}000 steps, single seed,
one node of eight A100-PCIE-40GB). \textbf{A} is a uniform retrain at this recipe, not E2Former-V2.
\textbf{S+1} changes the uniform state layout to single-state mixed resolution; the block
schedule and water routing move with that family. \textbf{S+2} additionally changes the FFN.
Latency and peak memory are on DEKSAN, one A100-PCIE-40GB, batch 1, fp32, without
\texttt{torch.compile}, median of ten; training used one node of eight A100-PCIE-40GB.}
\label{tab:matched}
\resizebox{\linewidth}{!}{%
\begin{tabular}{@{}lrrrr@{}}
\toprule
Cell & $F$ MAE (meV/\AA{}) & $E$ MAE (meV/atom) & DEKSAN latency (ms) & DEKSAN peak memory (GB) \\
\midrule
A (uniform) & 6.745 & 0.331 & 717.1 & 32.84 \\
S+1 (allocation only) & 6.934 & 0.313 & 392.5 & 19.29 \\
S+2 (allocation + FFN) & 5.887 & 0.297 & 411.5 & 22.43 \\
\bottomrule
\end{tabular}}
\end{table}

\subsection{Single-variable ablation}
\label{sec:ablation}

The production configurations differ on six axes (depth, cutoff, routing, persistent-state
layout, attention parameterisation and FFN), so comparing them does not attribute a difference
to one axis. We separate the axes with 16 reported training cells on SPICE, most differing from
their control on one axis at a fixed recipe of 200{,}000 steps, one seed and eight A100-PCIE-40GB GPUs;
the cumulative attention rung flips two switches and the family cells change the state layout
(Appendix~\ref{sec:app_ablation}). The two
cumulative endpoints are the same configurations at the same seed as the reference runs behind
Table~\ref{tab:spice}; retraining reproduces them to 0.25\% and 0.09\% in validation force.
These are same-seed reruns, so they bound implementation variance, not seed variance; we do not
interpret differences below that spread as real.

Table~\ref{tab:ablation} lists the production axes with a single-axis revert from production
Flash. Depth is the largest single-axis revert: four layers instead of six costs 34.4\% force
and 45.8\% energy, while using 3.9\% less peak memory and saving 7.0\% of latency on the
solvated box and 33.0\% on the 3{,}490-atom gas-phase protein. In the cumulative chain, adding the sixth layer costs 3.9\% latency on
the solvated box and 51.6\% on the protein, because the added blocks are all-to-high-resolution
(Appendix~\ref{sec:app_ablation}). The attention parameterisation splits into two switches.
Adding the dot gate improves validation force by 18.0\% and energy per atom by 14.8\%;
switching the activation to exp-softmax improves force by 14.1\% and energy by 9.3\%. Both cost
latency and memory (18.5\% and 17.8\% latency, 7.4\% and 1.4\% peak memory on the solvated
box), so the gate is both the larger accuracy axis and the more expensive in memory. The
cumulative rung that flips both switches gives 41.8\% in the same direction
(Appendix~\ref{sec:app_ablation}). Reverting the FFN to the s3 form costs 21.8\% force
for 2.0\% latency. Moving the cutoff from 5 to 6\,\AA{} changes force by 0.9\%, below the 1\% noise floor, and
energy by 11.1\%, at 45.6\% higher latency and 63.2\% higher memory; we do not interpret the
force change. The light-stream width is the most
expensive axis in the other direction: 256 channels instead of 32 improve force by 6.4\% at
62.0\% higher latency and 89.9\% higher peak memory.

The light-head revert and its cumulative parent are the
same trained cell, F+6; the production 16$\times$4 partition improves force by
3.0\% over 4$\times$16 at 1.2\% higher latency. The mixed-resolution family has only a cumulative
estimate: introducing the single-state form costs 2.8\% force while improving energy per atom
by 5.4\% and removing 45.3\% latency and 41.3\% memory on the solvated box. The block schedule
and water-oxygen routing are bound to the mixed-resolution family in both production
configurations, so neither has a separate estimate. The light-stream knobs exist only in the
two-state family, because the Standard family constructs its light modules without calling them
(Appendix~\ref{sec:app_ablation}).

\begin{table}[t]
\centering
\small
\caption{Single-axis reverts from production Flash on SPICE. Production here is the SPICE
retrain; the OMol25 production checkpoint keeps the dot gate. Each row changes one axis of the
SPICE configuration; the attention axis is split into its gate and activation switches,
and the light-head row reuses the cumulative cell F+6.
$\Delta$ is the reverted cell against production Flash, so positive $\Delta F$ MAE and
$\Delta E$ MAE are worse and positive latency and peak-memory changes are more expensive.
All cells are 200{,}000-step, single-seed models; accuracy is validation MAE at step 200k;
latency is the median of ten batch-1 inferences on one A100-PCIE-40GB in fp32 without
\texttt{torch.compile}. $^{\dagger}$The protein baseline for the reported relative latency
changes $+7.6\%$ and $+3.0\%$ is not available.
The gas-phase protein in the $\Delta$ latency 3{,}490-atom column is the 3{,}490-atom
structure of Appendix~\ref{sec:app_ablation}, not the 32{,}437-atom production protein; the
peak-memory column is measured on DEKSAN.}
\label{tab:ablation}
\setlength{\tabcolsep}{4pt}
\resizebox{\linewidth}{!}{%
\begin{tabular}{lllrrrrrr}
\toprule
Axis & Production & Reverted & $\Delta F$ MAE (\%) & $\Delta E$ MAE (\%) & $\Delta$ latency DEKSAN peptide (\%) & $\Delta$ latency 3{,}490-atom protein (\%) & $\Delta$ peak memory DEKSAN (\%) \\
\midrule
Depth & 6 layers & 4 layers & $+34.4$ & $+45.8$ & $-7.0$ & $-33.0$ & $-3.9$ \\
Attention (gate) & no dot gate & dot gate & $-18.0$ & $-14.8$ & $+18.5^{\dagger}$ & $+7.6^{\dagger}$ & $+7.4$ \\
Attention (activation) & softplus & exp-softmax & $-14.1$ & $-9.3$ & $+17.8^{\dagger}$ & $+3.0^{\dagger}$ & $+1.4$ \\
FFN & s2, grid 12 & s3 & $+21.8$ & $+15.7$ & $+2.0$ & $-6.3$ & $-2.3$ \\
Cutoff & 5\,\AA{} & 6\,\AA{} & $-0.9$ & $-11.1$ & $+45.6$ & $+27.1$ & $+63.2$ \\
Light width & 32 channels & 256 channels & $-6.4$ & $-6.0$ & $+62.0$ & $+16.0$ & $+89.9$ \\
Light heads & 16$\times$4 & 4$\times$16 & $+3.0$ & $+4.2$ & $-1.2$ & $+0.2$ & $-2.3$ \\
\bottomrule
\end{tabular}}
\end{table}

\subsubsection{Uniform reductions at near-matched cost}
\label{sec:matcheduniform}

Four conventional reductions of E2Former-V2 were trained on SPICE under the same recipe and
single seed as the production models: half depth (U-2L), half representation width (U-half),
half attention heads (U-h64), and width plus heads (U-half+h64). The U-* cells and the baseline
row come from one controlled SPICE evaluator; in that evaluator the untouched E2Former-V2
baseline reads 6.08\,meV/\AA{}, while the separately transcribed Table~\ref{tab:spice} row reads
7.04. The two baselines are not interchangeable, and the 7.04 row is not used to grade the U-*
cells. Against 6.08, all four have higher force error: U-half 7.01 (+15.3\%), U-h64 7.11
(+16.9\%), U-2L 8.47 (+39.3\%) and U-half+h64 9.01 (+48.2\%). Standard reaches 5.89\,meV/\AA{}
in force and 0.30\,meV/atom in energy, and Flash reaches 8.17 and 0.39
(Table~\ref{tab:uniformreduce}).

U-2L and Standard form the closest cost pair, at 119.8 versus 120.3\,ms at 6k atoms and 225.8
versus 219.6\,ms at 12k, with peak memory 18.93 versus 17.96\,GB at 6k and 37.32 versus
35.38\,GB at 12k; the match is near, not exact. These 6k/12k numbers are batch-1 H200
step-only measurements on a separate ladder, not the A100 ablation, the five-system campaign of
Table~\ref{tab:h200}, or the H200 Pareto campaign. Only Standard/U-2L has comparable latency;
the other reductions are slower at both sizes. Standard's lower error is therefore a comparison
at similar cost, not a causal result about chemical routing, and no random-routing or
matched-FLOPs control was trained. The broader claim that role allocation beats every uniform
reduction is not established.

\begin{table}[H]
\centering
\scriptsize
\caption{Uniform reductions and production points on SPICE (U-* cells: same recipe and seed;
baseline: controlled evaluator). The baseline is
the controlled-evaluator V2 (6.08\,meV/\AA{}); Table~\ref{tab:spice}'s transcribed 7.04 is not
interchangeable. U-2L halves depth, U-half width, U-h64 heads, U-half+h64 width and heads.
Latency/memory at 6k/12k, H200, batch 1, step-only ladder. Force is conservative
per-component MAE.}
\label{tab:uniformreduce}
\resizebox{\linewidth}{!}{%
\begin{tabular}{@{}lrrrrrr@{}}
\toprule
Model & $E$ (meV/atom) & $F$ (meV/\AA{}) & 6k latency (ms) & 6k peak (GB) & 12k latency (ms) & 12k peak (GB) \\
\midrule
E2Former-V2 baseline & 0.3029 & 6.08 & 202.7 & 30.12 & 388.1 & 59.51 \\
U-2L & 0.4465 & 8.47 & 119.8 & 18.93 & 225.8 & 37.32 \\
U-half & 0.3845 & 7.01 & 178.5 & 27.71 & 340.4 & 54.84 \\
U-h64 & 0.3689 & 7.11 & 157.7 & 20.76 & 291.1 & 40.96 \\
U-half+h64 & 0.5330 & 9.01 & 130.9 & 18.41 & 247.3 & 36.36 \\
\midrule
E2Former-V3-Standard & 0.30 & 5.89 & 120.3 & 17.96 & 219.6 & 35.38 \\
E2Former-V3-Flash & 0.39 & 8.17 & 37.4 & 3.57 & 59.4 & 6.84 \\
\bottomrule
\end{tabular}}
\end{table}

\subsection{Physical reliability}
\label{sec:gates}

\begin{figure}[H]
\centering
\includegraphics[width=\linewidth]{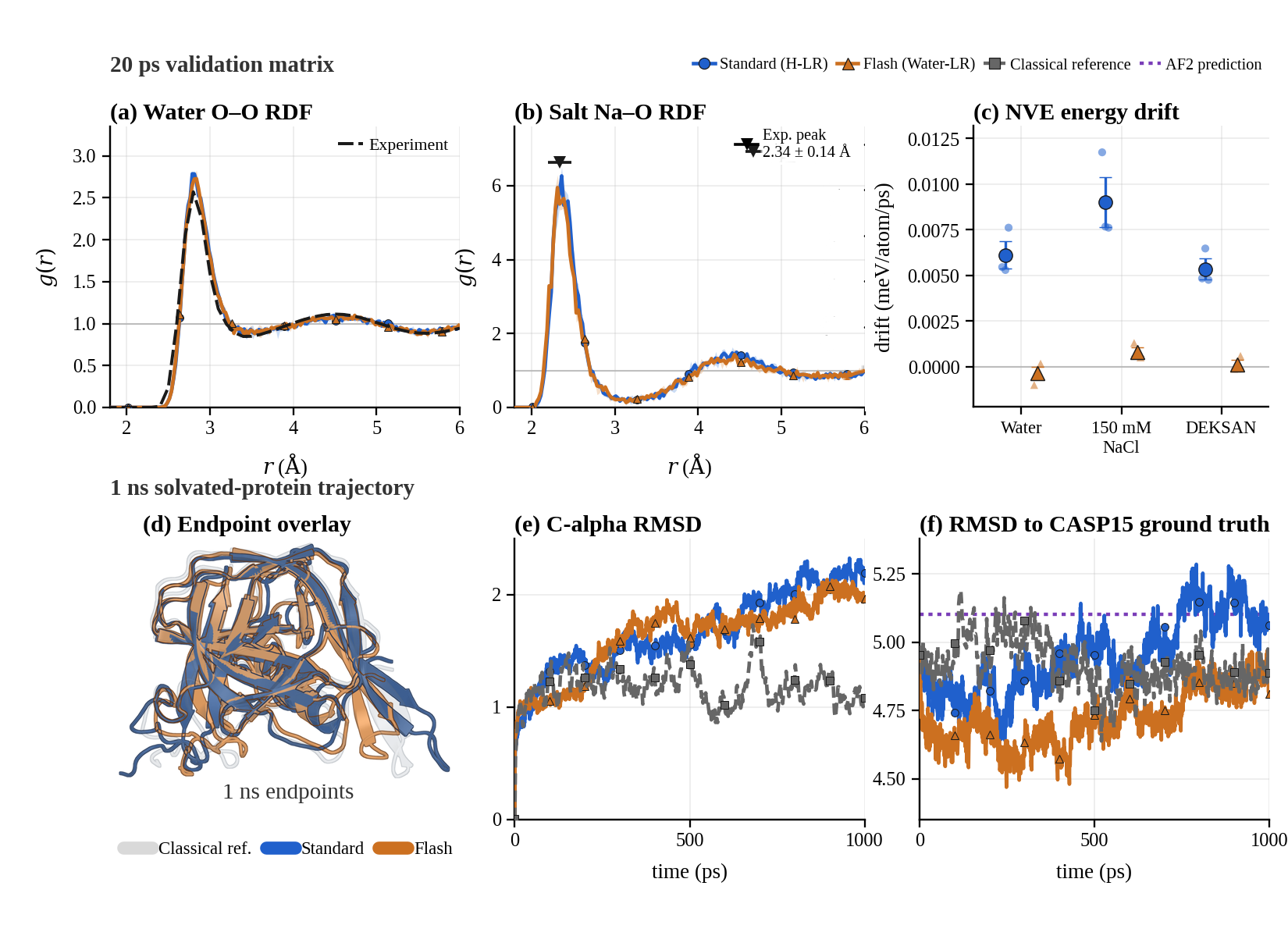}
\caption{Physical reliability across two campaigns. \textbf{(a,b)} Water O--O and salt Na--O
radial distribution functions from 20\,ps NVT simulation results, three-seed means over
10--20\,ps, with the seed range shaded. The water reference is the 295.1\,K X-ray curve of
\citet{skinner2013benchmark}; the salt marker is the $2.34\pm0.14$\,\AA{} Na--O peak of
\citet{mancinelli2007hydration}. \textbf{(c)} Signed total-energy drift fitted over 2--20\,ps
in NVE. The largest mean, 0.0090\,meV/atom/ps, corresponds to 0.162\,meV/atom over the 18\,ps
fit window. \textbf{(d--f)} Separate 1\,ns, fixed-cell NVT trajectories of one 32{,}437-atom
solvated protein with one seed: endpoint overlay, C$\alpha$ RMSD (\AA{}) from the equilibrated start,
and C$\alpha$ RMSD (\AA{}) to the CASP15 T1123-D1 experimental structure. Gray dashed curves in
\textbf{(e,f)} are the AMBER ff14SB/TIP3P GROMACS control propagated from the same equilibrated
start \citep{maier2015ff14sb,jorgensen1983comparison,abraham2015gromacs}. The purple dotted line
in \textbf{(f)} is an AlphaFold2-based CASP15 submission (group TS270), scored against the same
ground truth \citep{kryshtafovych2023critical}; it is an external reference, not a DeepMind
CASP entry. Here H-LR and Water-LR are the historical routing names for Standard
(hydrogen-low-resolution) and Flash (hydrogen-plus-water-oxygen-low-resolution); HR/LR in the
text denote the high-/low-resolution role.}
\label{fig:physical}
\end{figure}

Because averaged error does not rank usability \citep{fu2022forces, liu2023discrepancies,
raja2024stabilityaware}, we gate the reduced-capacity water path on model-independent aqueous
references (Appendix~\ref{sec:app_water}) rather than on force error
\citep{omranpour2024perspective}. Across water, a 150\,mM salt solution and the 5{,}992-atom
DEKSAN peptide, all 54 production legs complete without non-finite values, overlap or water
fusion, and the three-seed mean signed constant-energy drift remains below
0.01\,meV/atom/ps over the 2--20\,ps fit window. The recorded water and ion radial-distribution
peak positions remain within a few hundredths of an \AA{}ngstrom of experiment
\citep{skinner2013benchmark,mancinelli2007hydration} (Figure~\ref{fig:physical}; detailed
curves are in Appendix~\ref{sec:app_water}). Peak heights, coordination numbers and the second
shell are not reproduced as closely. In the 54-leg matrix, first-shell water O--O coordination
is 4.40--4.44 against the experimental 4.3--4.7 and peak heights are 6--8\% high; the separate
12-leg water-only matrix gives 4.78--4.97 coordination and a second shell displaced by
0.05--0.11\,\AA{}. The thermodynamic observables are worse: the 54-leg constant-pressure
densities are 1.0720--1.0762\,g/cm$^3$ against 0.997, and water
compressibilities are 2.66--3.42$\times10^{-5}$/bar against 4.52$\times10^{-5}$ (roughly
25--40\% too stiff). A correct RDF is not by itself evidence of a usable potential.

Separately, both production conservative checkpoints used for the 1\,ns campaign complete 1000\,ps on the
32{,}437-atom solvated protein without water fusion, nonbonded clash or solute collapse
(Table~\ref{tab:ns1main}, Appendix~\ref{sec:app_ns1}). The Standard leg uses the 4L/6.0\,\AA{}
configuration (22.224M active parameters) and the Flash leg the 6L/5.0\,\AA{} configuration
(22.876M; Appendix~\ref{sec:app_params}). This is one system, one ensemble (NVT), one seed, and a fixed
cell; it answers the duration question for that system and nothing else. It does not retire the failure onsets recorded elsewhere at 40, 58, 60, 511.9 and 680\,ps,
which are different systems and different configurations.

\subsubsection{High-temperature NVE stability}
\label{sec:nve100}

The 2--20\,ps NVE window is too short to stress conservation at elevated temperature. A
separate campaign ran 18 trajectories: Standard and Flash at 300, 500 and 1000\,K, three seeds
each, on water-512 (512 water molecules). Each trajectory is 10\,ps of NVT pre-equilibration
followed by 100\,ps of NVE at 0.5\,fs, and all 18 completed. We report the endpoint change
$\Delta E=(E_{\mathrm{tot}}(t)-E_{\mathrm{tot}}(0))/N$ and the linear drift slope over the full
100\,ps; uncertainties are standard deviations over the three seeds (Table~\ref{tab:nve100}).
The largest mean endpoint change is 0.92\,meV/atom (Standard, 1000\,K), the largest absolute
change among individual trajectories is 1.27\,meV/atom, and the measured mean temperatures at
the nominal 1000\,K condition are 988.3\,K (Standard) and 1008.3\,K (Flash). This is water-512
only and does not establish general stability for other systems or conditions.

\begin{table}[t]
\centering
\small
\caption{100\,ps NVE stability on water-512 after 10\,ps of NVT pre-equilibration at 0.5\,fs.
Each cell is mean $\pm$ standard deviation over three seeds, so each row is three trajectories.
$\Delta E$ is the endpoint total-energy change per atom; drift is the linear slope over the full
100\,ps. All 18 trajectories completed. ``$\sim 0$'' denotes a mean below the reported
precision.}
\label{tab:nve100}
\begin{tabular}{@{}lrrr@{}}
\toprule
Model & Nominal temperature (K) & Endpoint $\Delta E$ (meV/atom) & Drift (meV/atom/ps) \\
\midrule
Standard & 300 & $0.58 \pm 0.09$ & $+0.0054 \pm 0.0023$ \\
Standard & 500 & $0.84 \pm 0.24$ & $+0.0085 \pm 0.0020$ \\
Standard & 1000 & $0.92 \pm 0.29$ & $+0.0094 \pm 0.0026$ \\
\midrule
Flash & 300 & $0.01 \pm 0.01$ & $+0.0002 \pm 0.0002$ \\
Flash & 500 & $\sim 0 \pm 0.04$ & $\sim 0 \pm 0.0002$ \\
Flash & 1000 & $0.28 \pm 0.11$ & $+0.0034 \pm 0.0006$ \\
\bottomrule
\end{tabular}
\end{table}

At 1.80\,\AA{}, both production
high-resolution streams reproduce 99.2--100.1\% of the DFT water-dimer repulsion wall, whereas
Flash's reduced-capacity water path reproduces 49.7\%. Its minimum remains at the reference
position and depth, but neither averaged force error nor the radial distribution function
reveals this short-range softening (Appendix~\ref{sec:app_water}). The proton-transfer gate
passes for both production configurations, but Flash's surface is the softer of the two; the
full record is in Appendix~\ref{sec:app_water}.

The design envelope explains why the paper stops at the two production points. Three
low-resolution-capacity variants sat within 0.4\% of one another on validation force, 0.4\% on
held-out force, 1.2\% on validation energy and 1.8\% on held-out energy, yet their dynamics
ranged from a clean nanosecond to 968 of 1000 water pairs fused from 40\,ps at a minimum O--O
distance of 2.033\,\AA{}; the variant with the best held-out energy (0.399) still froze an ion
pair at 60\,ps, while a variant 0.5\% worse (0.401) fused at 40\,ps, so averaged error does not
rank dynamics (Appendix~\ref{sec:app_capacity}). A separate ensemble-and-box scan placed one variant's failure
onset between 16.2 and 189.0\,ps, an 11.7-fold range. The production configurations were
selected with physical gates rather than force MAE alone, and more aggressive cuts are not
supported by the evidence.

\begin{table}[t]
\centering
\small
\caption{The completed 1\,ns NVT campaign on the 32{,}437-atom solvated protein
(one system, one ensemble, one seed), run on eight A100-PCIE-40GB. Both legs reached
2{,}000{,}000 steps / 1000\,ps. The
structural pass exists because the runner's scalar log can read healthy while water dissociates;
all quantities use periodic boundary conditions (PBC), with intra-residue covalent contacts
excluded. The Standard leg uses
the 4L/6.0\,\AA{} production conservative configuration and the Flash leg the 6L/5.0\,\AA{} production conservative configuration.
One seed cannot give uncertainties.}
\label{tab:ns1main}
\begin{tabular}{@{}lrr@{}}
\toprule
 & Standard & Flash \\
\midrule
Status & complete, 1000\,ps & complete, 1000\,ps \\
Non-finite values & none & none \\
C$\alpha$ RMSD from start, final / max (\AA{}) & 2.192 / 2.326 & 1.964 / 2.136 \\
Radius of gyration, final / range (\AA{}) & 17.340 / 16.71--17.47 & 16.120 / 16.06--17.06 \\
Min.\ nonbonded distance over run (\AA{}) & 1.242 & 1.239 \\
Contacts below 1.2\,\AA{} & 0 & 0 \\
Water O--H, mean / max (\AA{}) & 0.972 / 1.143 & 0.973 / 1.142 \\
Endpoint $\Delta E$ / 1000\,ps (meV/atom/ps) & $-0.00938$ & $-0.00994$ \\
Final temperature (K) & 298.8 & 298.5 \\
\bottomrule
\end{tabular}
\end{table}

The 100\,ps NVE test in Section~\ref{sec:nve100} and the 1\,ns NVT run in
Table~\ref{tab:ns1main} are both reported in this chapter. The NVE campaign tests energy
conservation at elevated temperature on water-512; the 1\,ns run tests duration on one solvated
protein. Neither substitutes for the other.

\subsubsection{What survives at Flash speed}
\label{sec:observables}

Table~\ref{tab:observables} collects the physical observables under both production
configurations and states the comparison basis for each. RDF peak positions, first-shell
coordination and the proton-transfer gate survive in both configurations; Standard's
short-range wall survives, Flash's does not; density and compressibility are not reproduced.
The 1\,ns protein RMSD is one system and one seed. The appendix tables hold the full records;
this table is a summary.

\begin{table}[t]
\centering
\small
\caption{Physical observables under the two production configurations. Entries are Standard /
Flash unless the row states otherwise. The scope column marks the evidence basis: the 54-leg
matrix uses three seeds, the wall and proton-transfer entries are single-system,
single-profile, and the 1\,ns RMSD is one protein and one seed. Full records are in the
appendix tables or text named in the last column. The wall row compares Standard's
high-resolution stream (100.1) with Flash's low-resolution stream (49.7); Standard's LR stream
is 73.4 (a routing those weights were not trained under) and Flash's HR stream is 99.2.}
\label{tab:observables}
\resizebox{\linewidth}{!}{%
\begin{tabular}{@{}lllll@{}}
\toprule
Observable & Standard / Flash & Reference & Scope & Source \\
\midrule
Water O--O first RDF peak (\AA{}) & 2.788 / 2.838 & 2.80 & 54-leg, 3 seeds & Table~\ref{tab:rdf54} \\
First-shell water O--O coordination & 4.44 / 4.40 & 4.3--4.7 & 54-leg, 3 seeds & Table~\ref{tab:rdf54} \\
Water density (g/cm$^3$) & 1.0720 / 1.0762 & 0.997 & water NPT, 3 seeds & Appendix~\ref{sec:app_water} (text) \\
Water compressibility ($10^{-5}$/bar) & 3.42 / 2.66 & 4.52 & water NPT, 3 seeds & Section~\ref{sec:gates}; Appendix~\ref{sec:app_bounds} \\
Short-range water wall, Standard HR / Flash LR at 1.80\,\AA{} (\% DFT) & 100.1 / 49.7 & 100 & single system, single profile & Table~\ref{tab:wall} \\
Proton transfer & pass / pass (Flash softer) & pass & single system, single profile & Figure~\ref{fig:barrier} (foreground) \\
1\,ns protein C$\alpha$ RMSD to CASP15 (\AA{}) & 5.064 / 4.809 & 4.889 (classical) & one protein, one seed & Appendix~\ref{sec:app_ns1} \\
\bottomrule
\end{tabular}}
\end{table}

\subsection{Inference-only optimization}
\label{sec:filtering}

\begin{figure}[H]
\centering
\includegraphics[width=\linewidth]{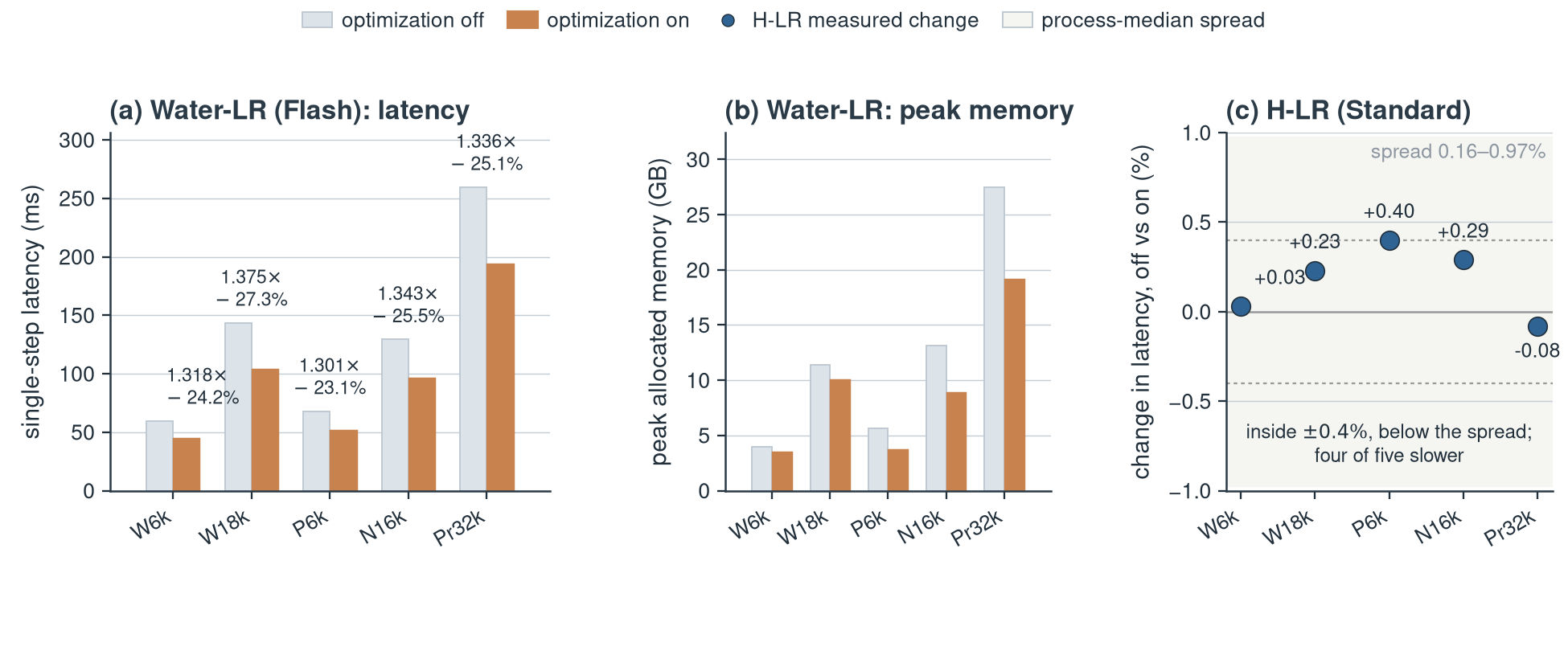}
\caption{Inference-time reachable-row construction, optimization off against on, over 66 timing
processes and two profiler runs. The Flash configuration gains $1.30$ to $1.375\times$ across
all five systems; the $1.375\times$ endpoint is the empty-union pure-water box and is not used
as headline evidence, so the non-degenerate range is $1.30$--$1.34\times$ with peak memory falling in every case---off$\rightarrow$on peak memory (GB) is
4.00$\rightarrow$3.55, 11.39$\rightarrow$10.12, 5.63$\rightarrow$3.77,
13.13$\rightarrow$8.93 and 27.49$\rightarrow$19.22 for the five systems; the Standard arm gains
nothing, its five deltas lying inside $\pm0.4\%$ --- four of the five marginally \emph{slower}
--- against a 0.16 to 0.97\% process-median spread, with peak memory unchanged to six
significant figures. Two points
retain high dispersion after being topped up to six processes (6.28\% and 10.75\%
process-median span) and that dispersion travels with them. Across all ten model--system pairs,
the maximum absolute energy difference is $2.2888\times10^{-5}$\,eV and the maximum
force-component difference is $6.3676\times10^{-4}$\,eV/\AA{}, both below the frozen
$1\times10^{-3}$ thresholds in their respective units. The kernel attribution behind the
saving is in Figure~\ref{fig:kernel_scaling}. H-LR and Water-LR are the routing names defined
in Figure~\ref{fig:physical}.}
\label{fig:filtering}
\end{figure}

Reachable-row construction adds no parameters and leaves the model definition and energy
function unchanged. It reduces Flash latency by 23.1--25.5\% ($1.30$--$1.34\times$) on the
three non-degenerate systems; the two pure-water boxes are empty-union endpoints (up to
27.3\%, $1.375\times$) and are not headline evidence. Standard's deltas remain within $\pm0.4\%$
(Figure~\ref{fig:filtering}). All ten model--system pairs pass
the frozen $10^{-3}$\,eV energy and $10^{-3}$\,eV/\AA{} force consistency thresholds. The
optimization is most effective when the high-resolution query set is small and spatially
concentrated, which is the Flash regime; Standard, whose high-resolution query set is the
whole non-hydrogen set, gains nothing from it.

\section{Discussion and Limitations}
\label{sec:discussion}

\subsection{What the results establish}

On the OMol25 validation split, Standard reaches 8.146\,meV/\AA{} force MAE, against
E2Former-V2's published 8.38, at 1.75--1.84$\times$ the throughput
across the five systems and lower peak memory on every system. Flash reaches 11.272 at
4.00--4.88$\times$ on the three solvated systems (4.00--5.85$\times$ including the pure-water
all-low-resolution limit), within the 1\,kcal\,mol$^{-1}$\,\AA$^{-1}$ chemical-accuracy
threshold \citep{chmiela2017machine} but less accurate than E2Former-V2. On SPICE water,
Standard reads 5.57 against 5.26 and Flash 16.28, a 3.1$\times$ regression that the All column
hides. Both are non-dominated on the force-accuracy--throughput frontier, and the same-recipe
SPICE control trades 2.8\% higher force error for 45.3\% lower latency and 41.3\% lower peak
memory. The accuracy comparison is a trade, not a uniform win: Standard loses on biomolecules
and metals, gains on electrolytes and organics, and energy MAE is 2.190 and 3.941\,meV/atom
against 1.16. High-resolution atoms are 49.1--49.6\% of force components but carry 76--78\% of
absolute error (HR/LR ratios 3.634 and 3.324), and all 54 20\,ps legs, the 100\,ps NVE check
and the 1\,ns protein run completed without numerical or structural failure.

\subsection{Limits of the evidence}

Every OMol25 trained result is single-seed, so no statistical claim attaches to an OMol25
accuracy difference; the SPICE check adds two seeds at one recipe but does not extend to
OMol25, and the same-seed rerun bounds implementation rather than seed variance. Published
baselines use unstated conventions while their latency is measured here, so cross-family
comparisons are levels rather than like-for-like rankings. Role-stratified MAE is absolute and
unnormalised by force magnitude, so a small low-resolution value does not by itself establish
that the narrow stream is adequate.

The configurations differ in depth, cutoff, routing, persistent-state layout, attention
parameterisation and FFN at once. The 16-cell SPICE ablation separates most axes, but the
mixed-resolution family has a cumulative estimate only, the block schedule and water-oxygen
routing move with it, and no state-layout variant was measured for accuracy. No
production-recipe-matched full-resolution control or matched-FLOPs control exists; SPICE cell A
matches only the 200{,}000-step ablation recipe and the production OMol25 schedule is
unrecorded. The numerical gates used random weights and were not re-run on the final
checkpoints. No trained random-routing control exists at the same low-resolution
fraction: per-sample, per-frame and fixed masks are distinct controls, an inference-only mask
is not one (Appendix~\ref{sec:app_bounds}), and the chemical prior's causal superiority over a
matched random routing remains unestablished.

\subsection{Dynamics horizon and thermodynamic bias}

The 20\,ps matrix is shorter than the water-box failure onsets recorded elsewhere (near 40 and
58\,ps, 60\,ps, 511.9\,ps, 680\,ps), the 1\,ns campaign is one system, one ensemble and one seed and does
not transfer to the 20\,ps matrix, and the 100\,ps NVE check is water-512 only. None establishes
general long-time stability or thermodynamic accuracy.

Variants within 0.4\% on force and 1.8\% on energy
ranged from a clean nanosecond to 968 of 1000 water pairs fused from 40\,ps at a minimum O--O
distance of 2.033\,\AA{}, and one variant's onset spanned 16.2--189.0\,ps. Densities are
1.0720--1.0762\,g/cm$^3$ against 0.997 and compressibilities 2.66--3.42$\times10^{-5}$/bar
against 4.52$\times10^{-5}$ (25--40\% too stiff), so a correct RDF is not by itself evidence of
a usable potential. At 1.80\,\AA{}, Flash's reduced-capacity water path reproduces 49.7\% of
the DFT water-dimer repulsion wall, against 99.2--100.1\% for the high-resolution streams.

\subsection{Choosing between Standard and Flash}

Standard keeps every atom at the full 256-channel width and changes only participation, so it
is the default when force or energy accuracy, short-range
repulsion or proton transfer matters, and for hydrogen-rich but not necessarily water-rich
systems.

Flash is the throughput, memory and reach point. Routing water oxygen with hydrogen makes its
advantage grow with the water and hydrogen fraction; within its limits it delivers 4.00--5.85$\times$ the throughput (the upper end is the pure-water all-low-resolution limit), less peak
memory, and a single-step 264{,}000-atom reach on one H200. Its water path reproduces only
49.7\% of the DFT wall and its proton-transfer surface is the softer of the two, so it should
not be used where short-range water repulsion or proton transfer is the target.

\subsection{Implications, open measurements, and claims not made}

Width and update count can be allocated independently, and a static
chemical prior suffices without a learned gate, a second potential or a boundary rule. The
tests that would strengthen the argument are a trained random-routing control, a
production-recipe-matched full-resolution control, an accuracy-valid state-layout ablation, the
200\,ps and 1\,ns matrix tiers, hydrogen-pair and ion-residence observables, and a cost model
predictive in magnitude.

We claim no reduced angular resolution, no learned or dynamic routing, no optimal partition, no
zero-shot solvent transfer and no demonstrated validity in reactive regimes.

\section{Conclusion}
\label{sec:conclusion}

E2Former-V3 allocates capacity by atom role while keeping one all-atom conservative potential.
A static role partition and kernel-level filtering move the force-accuracy--throughput frontier.
On OMol25, Standard reaches 8.146\,meV/\AA{} force MAE, against E2Former-V2's published 8.38, at 1.75--1.84$\times$ the throughput; Flash reaches 11.272 at
4.00--5.85$\times$, within the chemical-accuracy threshold \citep{chmiela2017machine} but less
accurate than E2Former-V2, and both are non-dominated on the aggregate frontier
(Figure~\ref{fig:pareto}). Standard loses on biomolecules and metals and gains on electrolytes
and organics, and energy MAE is 2.190 and 3.941\,meV/atom against 1.16.

Both operating points are usable in molecular dynamics within the measured horizons. Across
water, salt and peptide, all 54 20\,ps legs complete without numerical or structural failure;
thermostatted means hold 300\,K within 2.4\,K and RDF peaks match experiment within a few
hundredths of an \AA{}ngstrom. Both 1\,ns checkpoints complete on the 32{,}437-atom protein
without water fusion, clash or collapse. These results establish short-time stability and local
structure, not general long-time stability or thermodynamic accuracy: the 100\,ps NVE check is
water-512 only, the 1\,ns run is one protein, one ensemble and one seed, and water densities
remain about 7.5\% high.

High-resolution atoms are 49.1--49.6\% of force components but
carry 76--78\% of absolute error (HR/LR 3.634 under Standard, 3.324 under Flash). The ratio does
not identify capacity placement, and no random-routing control exists, so the chemical
partition's causal superiority remains unestablished. Flash's reduced-capacity water path
reproduces 49.7\% of the DFT repulsion wall at 1.80\,\AA{} against 100.1\% and 99.2\% for the
high-resolution paths, bounding its use where short-range repulsion or proton transfer matters.

Width and update count can be allocated independently, and a static
chemical prior suffices without a learned gate or boundary rule. Standard is the accuracy
point; Flash the throughput, memory and reach point. Within the present scope---explicitly
solvated, non-reactive systems, OMol25 single-seed evaluation, a 20\,ps matrix plus one 1\,ns
system---a static partition reduces cost at competitive accuracy. Next steps are mapping the
design space, training a random-routing control, extending the matrix to 200\,ps and more
solvents, and making the cost model predictive.

\clearpage
\bibliographystyle{plainnat}
\bibliography{refs}

\clearpage
\appendix
\noindent\textbf{Appendix roadmap.} The appendices follow the main text: architecture and
numerical certification; measurement protocol; checkpoint audit and role-stratified accuracy;
cost and the single-variable ablation; physical validation and capacity variants; limitations
and outstanding measurements.

\section{Architecture and Numerical Certification}
\label{sec:app_arch}

\begin{figure}[t]
\centering
\includegraphics{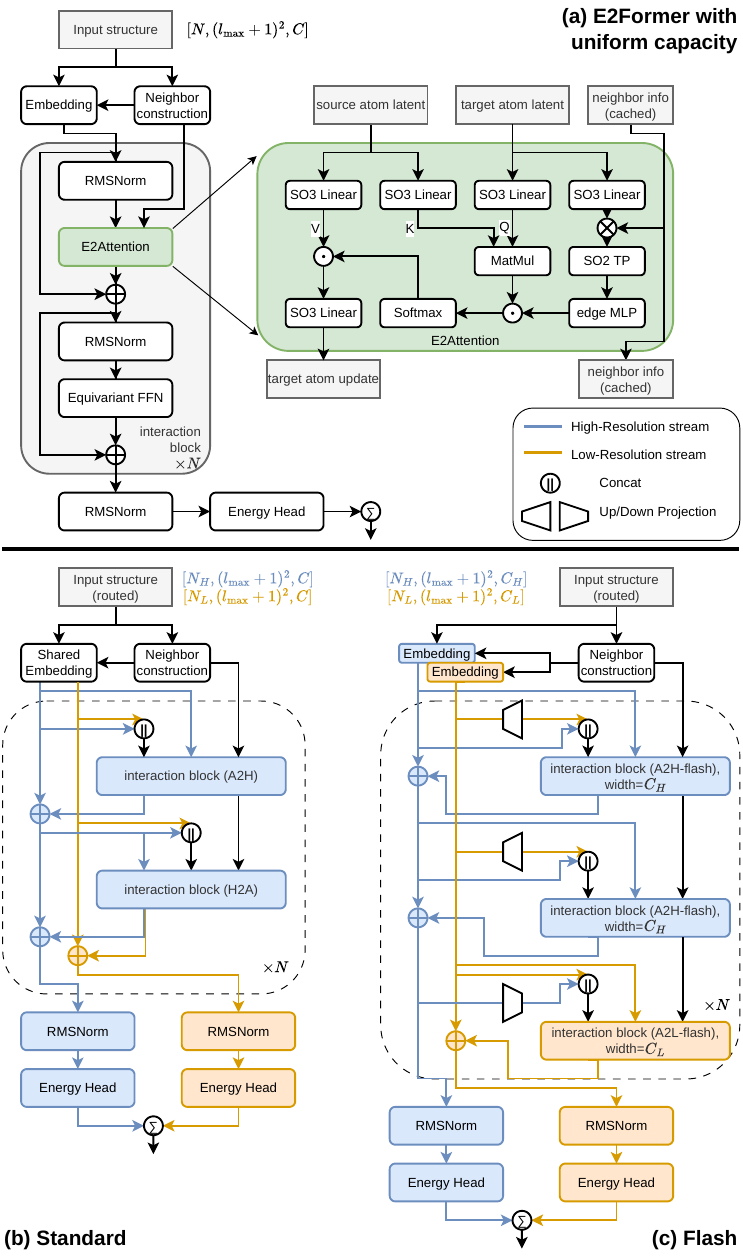}
\caption[The same chain under three capacity allocations.]{%
\textbf{The same chain under three capacity allocations.} \textbf{(a)} Uniform capacity:
E2Attention \citep{huang2026formerv} at the interaction block's core. \textbf{(b)} Interleaved,
used by Standard, allocates the update-count axis; \textbf{(c)} interleaved\_flash, used by Flash,
allocates both axes.}
\label{fig:arch}
\end{figure}

Figure~\ref{fig:arch} expands the E2Attention operator shared by the uniform-capacity
E2Former backbone and the two atom-role-aware allocations. From atomic numbers and coordinates,
neighbour construction supplies radial and directional edge features within the cutoff; the
embedding initialises equivariant atom states; repeated interaction blocks refine them; and the
energy readout produces atomic energies whose sum is differentiated with respect to the coordinates
to obtain conservative forces. In E2Attention \citep{huang2026formerv}, target atoms supply
queries, source atoms supply keys and values, and cached edge features modulate each target update.
Panels~(b) and~(c) show the interleaved and interleaved\_flash allocations used by Standard and
Flash; Section~\ref{sec:method} specifies their exact schedules and block semantics.

Both atom-role-aware allocations use separate high- and low-resolution energy heads: Standard applies
them to the two role partitions of its shared state, whereas Flash reads them from its two
persistent states; the selected atomic energies are summed before differentiation.

The routing mask is an input-format obligation and must be emitted on every inference path. The
benchmark harness originally emitted the water-oxygen mask for only one input format, which would
silently execute Flash checkpoints as Standard; after the fix, the control-system mask covered 1963
of 6085 atoms.

\clearpage

\subsection{Numerical certification}
\label{sec:app_gates}

Table~\ref{tab:gates} gives the eight gates that
Section~\ref{sec:implementation} names. All were run with random weights under eager execution
with TF32 disabled, and none was re-run on the final conservative checkpoints, so they
certify the implementation rather than the shipped models.

\begin{table}[h]
\centering
\small
\caption{Numerical correctness gates. Random weights, eager execution, TF32 disabled;
not re-run on the final conservative checkpoints.}
\label{tab:gates}
\resizebox{\linewidth}{!}{%
\begin{tabular}{ll}
\toprule
Gate & Result \\
\midrule
No-op equivalence against non-dual-stream & PASS. Identical parameter signature and counts; \\
configurations & energy and force deltas equal to those from rerunning the same code revision. \\
Rotation equivariance & PASS. Relative force deviation 0.002\% at each tested \\
 & low-resolution-stream FFN grid resolution. \\
Finite-difference forces & PASS on all six probe systems. Relative deviation of the \\
 & finite-difference force $-\mathrm{d}E/\mathrm{d}x$ from the autograd force was \\
 & $3.6\times10^{-4}$ to $1.5\times10^{-3}$ at a $10^{-3}$\,\AA{} displacement. \\
Angular-order creation & PASS. Perturbing the low-resolution portion of the key/value bank moved \\
 & the high-resolution attention output at all three degrees; a deliberately \\
 & broken variant gave exactly zero at the highest degree. \\
Empty high-resolution stream & PASS. Forces finite; perturbing low-resolution blocks moved them, \\
 & perturbing high-resolution blocks moved them by exactly zero, which is correct when no \\
 & high-resolution atoms exist. \\
Empty low-resolution stream & PASS. Finite energy and forces on an all-oxygen cluster. \\
Gradients & PASS. Non-zero gradients in high-resolution blocks, low-resolution blocks, all \\
 & up- and down-projections, both edge-degree embeddings and \\
 & both energy heads. \\
Size consistency & PASS. Relative energy error of two copies 60\,\AA{} apart \\
 & versus twice one copy: 3.7e-10 mixed, 7.3e-12 high-resolution-only, \\
 & 2.8e-8 low-resolution-only. \\
\bottomrule
\end{tabular}}
\end{table}

\FloatBarrier

\section[Measurement protocol]{Measurement Protocol and Reporting Conventions}
\label{sec:app_protocol}

\paragraph{Accuracy units and calibers.}
Accuracy is reported as energy MAE in meV/atom and force MAE in meV/\AA{}, the latter a
per-component MAE over the three Cartesian components of every valid atom, which must not
be conflated with the per-atom force-vector $L_2$ error used as the training objective.
Two independent accuracy-caliber distinctions are labeled per row and are not interchangeable:
the full 2,762,021-sample validation split versus its 1M subsample, and the training-time batch
mean (an unweighted mean over token-budget batches) versus the exact component-weighted MAE used
for the role-stratified results.

\paragraph{Efficiency and timing.}
Efficiency is reported as single-step latency in ms, throughput in atoms/s and peak
allocated device memory in the unit named in each table (decimal GB, $10^9$ bytes, or GiB,
$2^{30}$ bytes; 1 GiB = 1.0737 GB), with out-of-memory outcomes preserved rather than dropped.

\paragraph{Physical reliability observables.}
Physical reliability is reported through trajectory completion and structural-health checks,
temperature and constant-pressure acceptance, the coupled-coordinate proton-transfer barrier
in kcal/mol, the water-dimer oxygen-oxygen wall energy at 1.80\,\AA{} in kcal/mol,
radial-distribution peak positions and heights, first-shell coordination numbers, density in
g/cm$^3$, isothermal compressibility, signed constant-energy total-energy drift rate in
meV/atom/ps and, for the separate water-only analysis, final total-energy displacement in
meV/atom.

\paragraph{Timing qualifications.}
Unless a row states otherwise, all timings used fp32 with TF32 disabled, batch
size 1, frozen weights and compilation enabled, with the first compilation iteration
excluded and the median of the subsequent timed iterations reported. Each timing table
names its device and timing boundary, and numbers from different devices never appear in one
table; each value follows its own table's protocol. Where timing points were repeated, repetitions ran in separate processes.
Every OMol25 trained result is single-seed; the SPICE seed check in Section~\ref{sec:seedvar}
adds two seeds at one recipe. For Figure~\ref{fig:pareto}, every successful throughput
value was measured locally on one H200 using the implementation or code repository accompanying
the corresponding publication (this work's codebase for Standard and Flash), under the detailed
campaign in Table~\ref{tab:h200pareto}. Standard and Flash use the same production checkpoints
as their accuracy results; external-family points pair published accuracy with locally measured
throughput and retain the published metric-definition caveat, while E2Former-V2 additionally uses
an architecturally identical speed-only checkpoint whose weights differ from those behind its
published accuracy.
Two further protocol qualifications travel with the timings, both from the H200 water-box ladder
of Table~\ref{tab:ladder}. Latencies at or above 33000 atoms use a pinned intra-op thread pool,
and every run passes a scheduling-stability check; without the pin, scheduler quantization can
inflate wall time. The memory column is unaffected and the reach ceiling is a memory statement.
\texttt{compile=True} is correct for this ladder because it was measured on torch 2.8; it is not
the shipping configuration on torch 2.11, where E2Former's compiled forces come out identically
zero and eager execution is required, so the flag must not be carried into a torch-2.11 recipe.

Table~\ref{tab:config} collects the inference configurations behind the cross-model
comparisons: these are the settings used for the comparisons in this work, and UMA-S-1.1 is not
an unoptimized baseline, since its accelerated Triton backend is enabled.

\begin{table}[h]
\centering
\small
\setlength{\tabcolsep}{4pt}
\caption{Inference configuration for each model in the cross-model comparisons. All entries
are the settings used for the reported accuracy and timing comparisons.}
\label{tab:config}
\resizebox{\linewidth}{!}{%
\begin{tabular}{@{}l l l l c p{5.4cm}@{}}
\toprule
Model & Checkpoint / version & Backend / compile & Precision & TF32 & Notes \\
\midrule
E2Former-V3-Standard & production conservative checkpoint & PyTorch compile/Inductor $+$ FAISS-GPU & FP32 & off & --- \\
E2Former-V3-Flash & production conservative checkpoint & PyTorch compile/Inductor $+$ FAISS-GPU & FP32 & off & --- \\
UMA-S-1.1 & \texttt{uma-s-1p1} & PyTorch compile, \texttt{merge\_mole}, \texttt{umas\_fast\_gpu} Triton & FP32 & off & no activation checkpointing (disables Triton) \\
MACE-OMOL & \texttt{MACE-OMOL-extra-large} & fused cuEquivariance (\texttt{enable\_cueq=True}) & FP32 & --- & compilation off \\
DPA & \texttt{DPA-3.2-5M} / \texttt{DPA4-Beta} & fused SO(2) (\texttt{DP\_TRITON\_INFER=2}) & FP32 & off & compilation on \\
\bottomrule
\end{tabular}}
\end{table}

\subsection{SPICE Accuracy Protocol}
\label{sec:app_spice}

Table~\ref{tab:spice} evaluates the SPICE test set \citep{eastman2023spice} in the layout of
E2Former-V2 Table~1 \citep{huang2026formerv}, with every row expressed at one aggregation
caliber. Energy is in meV/atom; its \emph{All} value is the structure-count-weighted mean of
the seven subset MAEs, which reproduces the published MACE Large energy aggregate
digit-for-digit. Force is the mean absolute error per Cartesian component in meV/\AA{}; its
\emph{All} value is pooled over all force components. The source table does not use one basis:
the MACE and E2Former-V1 rows are structure-weighted on both metrics, whereas E2Former-V2 is
structure-weighted on energy but component-pooled on force. Table~\ref{tab:spice} therefore
retains the published subset cells but recomputes every derivable \emph{All} value on the stated
common basis rather than transcribing a mixed aggregate column.

The force convention was verified, not assumed, from the test file's
MACE force column. Its energy errors identify it as a MACE Large-quality model,
which fixes its force reading to 1.08--1.19 per component rather than 2.15--2.39 as per-atom
vector $L_2$. The two models from this work were retrained on SPICE from scratch and evaluated
by pooling the full 50195-frame test file. Relative to the published MACE-OFF23 partition, that
file contains 209 additional PubChem structures and 12 additional dimers, or 221 structures in
total; the resulting aggregate difference is approximately 0.1\%.

The published rows are transcribed rather than re-measured, and their training budgets are not
matched. For E2Former-V2-Cons., component pooling gives 7.043, consistent with its printed 7.05,
whereas structure weighting gives 6.28; its energy aggregate of 0.31 is structure-weighted and
is retained. The corresponding recomputed E2Former-V2-Direct aggregate is 0.54/8.32. The eSEN
rows publish only \emph{All}, not the seven subset values; their pooling rule is unstated and
they are 3.2M-parameter ablations, so they are context rather than commensurable rankings.

The transcription was checked cell by cell against E2Former-V2 Table~1. Its E2Former-V1 Water
force cell is 13.50, while the original E2Former-V1 Table~3 prints 13.52; Table~\ref{tab:spice}
follows E2Former-V2 Table~1 and therefore uses 13.50. Either value gives the same displayed
component-pooled aggregate of 8.12. The remaining published subset cells agree with the source.
The \emph{All} column is 95.6\% weighted on PubChem and dimers and is consequently nearly blind
to the regimes targeted here: solvated amino acids and water carry only 0.10\% and 0.17\% of its
energy weighting. On those two subsets Flash reads 22.98 and 16.28 in force MAE, against 7.94
and 5.57 for Standard.
\FloatBarrier

\section{Checkpoint Audit and External Families}
\label{sec:app_layout}

\subsection{Active Checkpoint Parameter Audit}
\label{sec:app_params}

Table~\ref{tab:activeparams} reports optimizer-state active-parameter counts for the
audited checkpoints used in the OMol25 and physical-validation analyses. A parameter is active iff its Adam/AdamW optimizer state
contains an \texttt{exp\_avg} tensor, which requires that it received a training gradient;
the count sums those tensor elements and excludes buffers and serialized inactive parameters.

\begin{table}[h]
\centering
\scriptsize
\caption{Training-active parameters for the audited mixed-resolution checkpoints used in the
OMol25 and physical-validation analyses. Counts are from checkpoint optimizer state, not
model-definition or serialized-tensor totals.}
\label{tab:activeparams}
\begin{tabular}{lr}
\toprule
Checkpoint & Active params (M) \\
\midrule
\multicolumn{2}{l}{\emph{Production, accuracy and warm-start checkpoints}} \\
Standard production, 4L, 6\,\AA{} & 22.223790 \\
Standard non-production, 6L, 5.5\,\AA{} & 33.619460 \\
Standard earlier, 6L, 6\,\AA{} & 33.147457 \\
Flash production, 6L, 32\,ch, 5\,\AA{} & 22.875905 \\
Flash non-production, 6L, 32\,ch, 5.5\,\AA{} & 22.875905 \\
Standard direct warm-start, 4L / 6L & 22.568880 / 34.997766 \\
Flash direct warm-start, 6L, 32\,ch & 23.226501 \\
Standard earlier conservative run, 4L & 22.540779 \\
\midrule
\multicolumn{2}{l}{\emph{Proton-transfer PES sweep}} \\
Standard 4L / Standard 6L warm-start & 22.568880 / 34.997766 \\
Flash 4L, 32 / 64\,ch & 12.346065 / 12.576657 \\
Flash 4L, 64\,ch, attention 16$\times$4 / 4$\times$8 & 12.582849 / 12.536337 \\
Flash 4L, 64\,ch, attention 8$\times$8 & 12.578721 \\
Flash 6L, 64\,ch, attention 16$\times$4 & 23.155649 \\
\midrule
\multicolumn{2}{l}{\emph{Water-wall and low-resolution-capacity checkpoints}} \\
Standard replicate & 22.540779 \\
Flash, 64 scalar $+$ 64 vector & 22.301834 \\
Flash, 128 scalar $+$ 128 vector & 22.560906 \\
Flash, 64 scalar $+$ 64 vector, two tails & 22.513618 \\
\bottomrule
\end{tabular}
\end{table}

\subsection{External families}
\label{sec:app_external}

Table~\ref{tab:external} lists the more recent external families whose published accuracies use
the same OMol25 validation split; their metric conventions and provenance are stated in the
caption.

\begin{table}[h]
\centering
\footnotesize
\caption{More recent external families. These are on the \emph{same} validation split as
Table~\ref{tab:accuracy}: their sources report the OMol25 composition validation split, which
carries the same four domains and total as ours, and the two baselines these sources share with
\citet{huang2026formerv} agree digit-for-digit after unit conversion (eSEN-sm-d 9.92\,meV/\AA{}
$\leftrightarrow$ 0.23\,kcal/mol/\AA{}; eSEN-sm-cons 8.25 $\leftrightarrow$ 0.19). Force MAE is
transcribed exactly at 43.3641\,meV/\AA{} per kcal/mol/\AA{}. Energy is not transcribed because
no source states an energy-referencing convention, and none states whether its force metric is
per-component or per-vector, so the metric caveat still applies. Locally measured H200 throughput
for these models is plotted in Figure~\ref{fig:pareto}; pairing it with these published accuracies
still combines two measurement environments, which the figure marks as a separate class.}
\label{tab:external}
\begin{tabular}{lrrrrr}
\toprule
Family & Nano & Mini & Neo & Air & Plus \\
 & \multicolumn{5}{c}{force MAE (meV/\AA{})} \\
\midrule
DPA4 \citep{li2026dpa} & 33.65 & 21.77 & 14.40 & 8.20 & 6.33 \\
DPA4C \citep{li2026universal} & 107.75 & 81.58 & 75.06 & 69.25 & 59.97 \\
\midrule
\multicolumn{6}{l}{UMA-S-1.1, as re-transcribed in \citet{li2026dpa}: 9.54\,meV/\AA{}} \\
\bottomrule
\end{tabular}
\end{table}

\FloatBarrier

\section{Role-Stratified Force Error}
\label{sec:app_role}

Table~\ref{tab:role} gives the two-configuration decomposition at the exact component-weighted
caliber, and Table~\ref{tab:domain} resolves it by chemical domain. Each configuration covers 2762021
structures, 283298012 atoms and 849894036 force components. Because Standard and Flash route
internally, their reported splits had to be \emph{proven} to match their own partitions rather
than assumed: three per-batch guards ran on every batch of the entire split, checking that the
batch carried a genuine water mask, that it matched the evaluator's independently read mask,
and that it agreed with the canonical routing returned by the model; routing was derived from
each checkpoint's own configuration and never passed in. Two limits on what this decomposition
can argue are stated here. These are absolute per-component MAEs, unnormalised by force
magnitude, so a small
low-resolution-role value is not on its own evidence that the narrow stream is adequate, and the
normalised version was not computed. Standard and Flash also differ at once in architecture,
training stage and routing, so no difference between them may be attributed to routing alone.

\begin{table}[h]
\centering
\footnotesize
\caption{Role-stratified per-component force MAE (meV/\AA{}) on the full 2762021-sample split
at the \emph{exact component-weighted} caliber. Each row uses the router bound to that
configuration; HR and LR denote its high- and low-resolution roles. HR/LR is the HR-role
column divided by the LR-role column. Single-seed.}
\label{tab:role}
\resizebox{\linewidth}{!}{%
\begin{tabular}{lrrrrr}
\toprule
Model & HR role & LR role & Overall & HR / LR & HR share of components \\
\midrule
\multicolumn{6}{l}{\emph{Exact component-weighted caliber}} \\
E2Former-V3-Standard, 4L, 6\,\AA{} & 12.839 & 3.533 & 8.146 & 3.634 & 49.57\% \\
E2Former-V3-Flash, 6L, 32\,ch, 5\,\AA{} & 17.495 & 5.264 & 11.272 & 3.324 & 49.12\% \\
\bottomrule
\end{tabular}}
\end{table}

The two configurations apply different strengths of HR/LR capacity reduction: Standard changes
participation only, whereas Flash changes both participation and persistent width and assigns a
larger fraction of the system to the low-resolution role. Their HR/LR error ratios nevertheless
remain close, 3.634 against 3.324. The ratio alone does not identify capacity placement;
the chemistry reading remains a conjecture. This is not a causal ablation: the configurations differ
simultaneously in architecture, training stage and routing.

\begin{table}[h]
\centering
\footnotesize
\caption{The same evaluation resolved by chemical domain, force-component MAE in meV/\AA{},
formatted \emph{high-resolution / low-resolution / all}, with each configuration evaluated
under its bound router. Domain atom counts are 91063984 (32.1\%), 171462133 (60.5\%),
13709053 (4.8\%) and 7062842 (2.5\%). Exact caliber, single-seed. The last column provides
the published all-atom comparison: E2Former-V2's four domain values and its total, transcribed
from Table~2 of
\citet{huang2026formerv}, which reports the same four domains and total on the same OMol25
composition validation split, in the source's own printed \emph{Total} box (no per-subset
high-resolution/low-resolution split is published, so this column has no such cell). It is \emph{not} on our
caliber: the source's force metric convention is unstated, whereas our three columns are
component-weighted, so it is read as a level and not as a per-domain tie-break, exactly as
Table~\ref{tab:accuracy} reads its total. Transcribed, not re-measured: E2Former-V2 has no
checkpoint that is accuracy-valid for the role-stratified evaluator, so it could not be
re-scored here even with unmatched weights.}
\label{tab:domain}
\resizebox{\linewidth}{!}{%
\begin{tabular}{llll}
\toprule
Domain & Standard & Flash & E2Former-V2 (published all) \\
\midrule
Biomolecules & 7.478 / 2.716 / 5.104 & 11.075 / 4.353 / 7.685 & --- / --- / 4.96 \\
Electrolytes & 12.595 / 3.635 / 8.002 & 17.831 / 5.476 / 11.446 & --- / --- / 10.11 \\
Metals & 41.764 / 6.825 / 25.555 & 46.121 / 7.950 / 28.396 & --- / --- / 24.40 \\
Organics & 26.485 / 5.759 / 17.072 & 30.971 / 7.099 / 20.065 & --- / --- / 19.69 \\
\midrule
Total & 12.839 / 3.533 / 8.146 & 17.495 / 5.264 / 11.272 & --- / --- / 8.38 \\
\bottomrule
\end{tabular}}
\end{table}

\begin{figure}[h]
\centering
\includegraphics[width=0.86\linewidth]{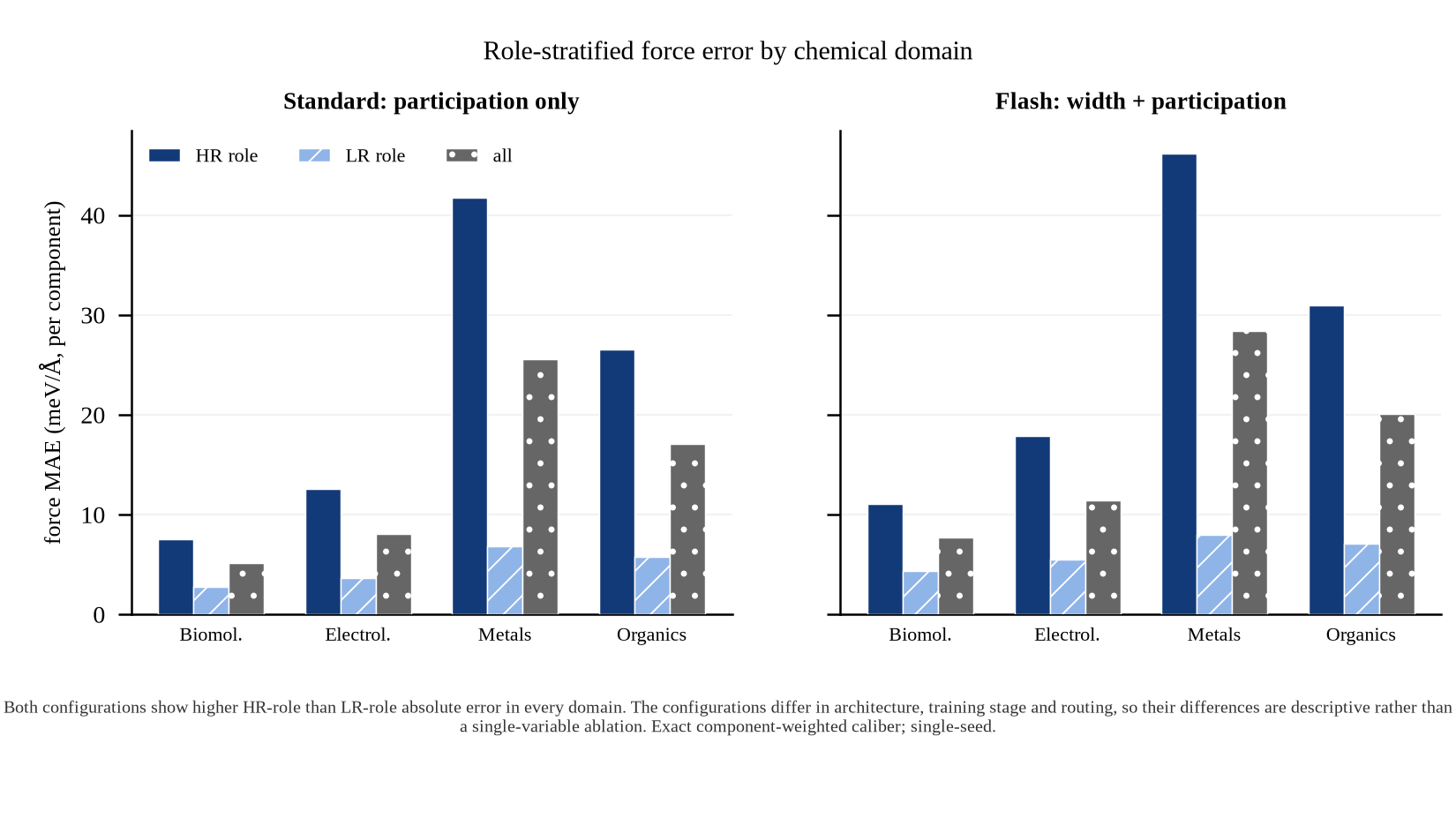}
\caption{Role-stratified force error by chemical domain. Each panel shows HR-role, LR-role and
all-atom per-component MAE for one production configuration under its bound router. The HR/LR
imbalance appears in all four domains under both the participation-only Standard allocation and
the stronger width-plus-participation Flash allocation. Exact component-weighted caliber, full
split; the two configurations differ at once in architecture, training stage and routing, so
their differences are descriptive rather than a single-variable ablation. Single-seed.}
\label{fig:domain}
\end{figure}

\begin{figure}[h]
\centering
\includegraphics[width=0.66\linewidth]{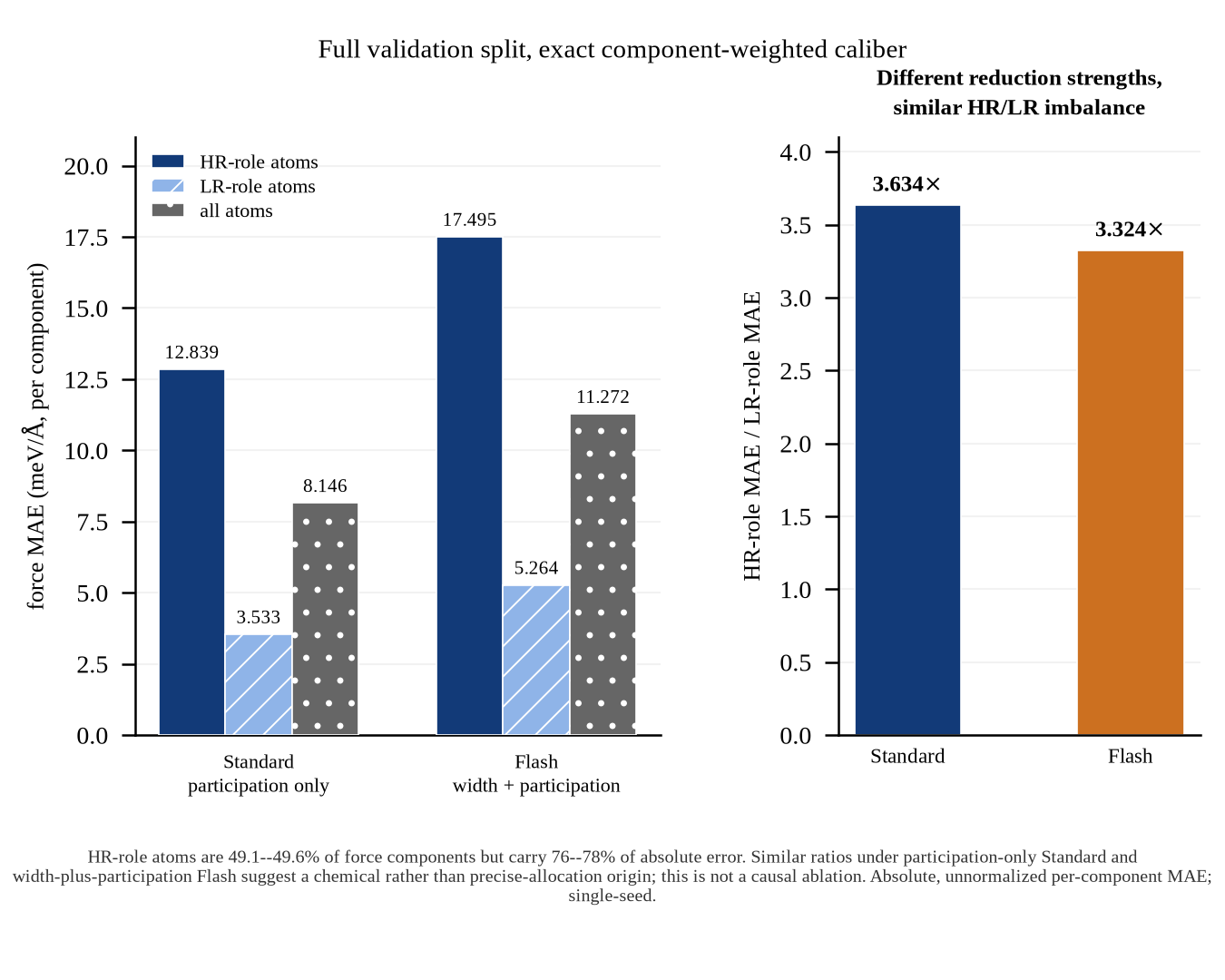}
\caption{Role-stratified per-component force MAE, full 2762021-sample split, exact
component-weighted caliber. Standard changes participation only; Flash changes persistent width
and participation and routes more atoms low-resolution. Even under these different reduction
strengths, their HR/LR ratios remain close, 3.634 and 3.324. HR-role atoms are
49.1 to 49.6\% of components but carry 76 to 78\% of the absolute error. This consistency
does not identify capacity placement, and the two configurations are not a causal ablation. Bounds in the text above and
Appendix~\ref{sec:app_bounds}. Single-seed.}
\label{fig:role}
\end{figure}

\FloatBarrier

\section{Cost Measurements}
\label{sec:app_cost}

\subsection{Reading the cost tables}

\begin{figure}[h]
\centering
\includegraphics[width=\linewidth]{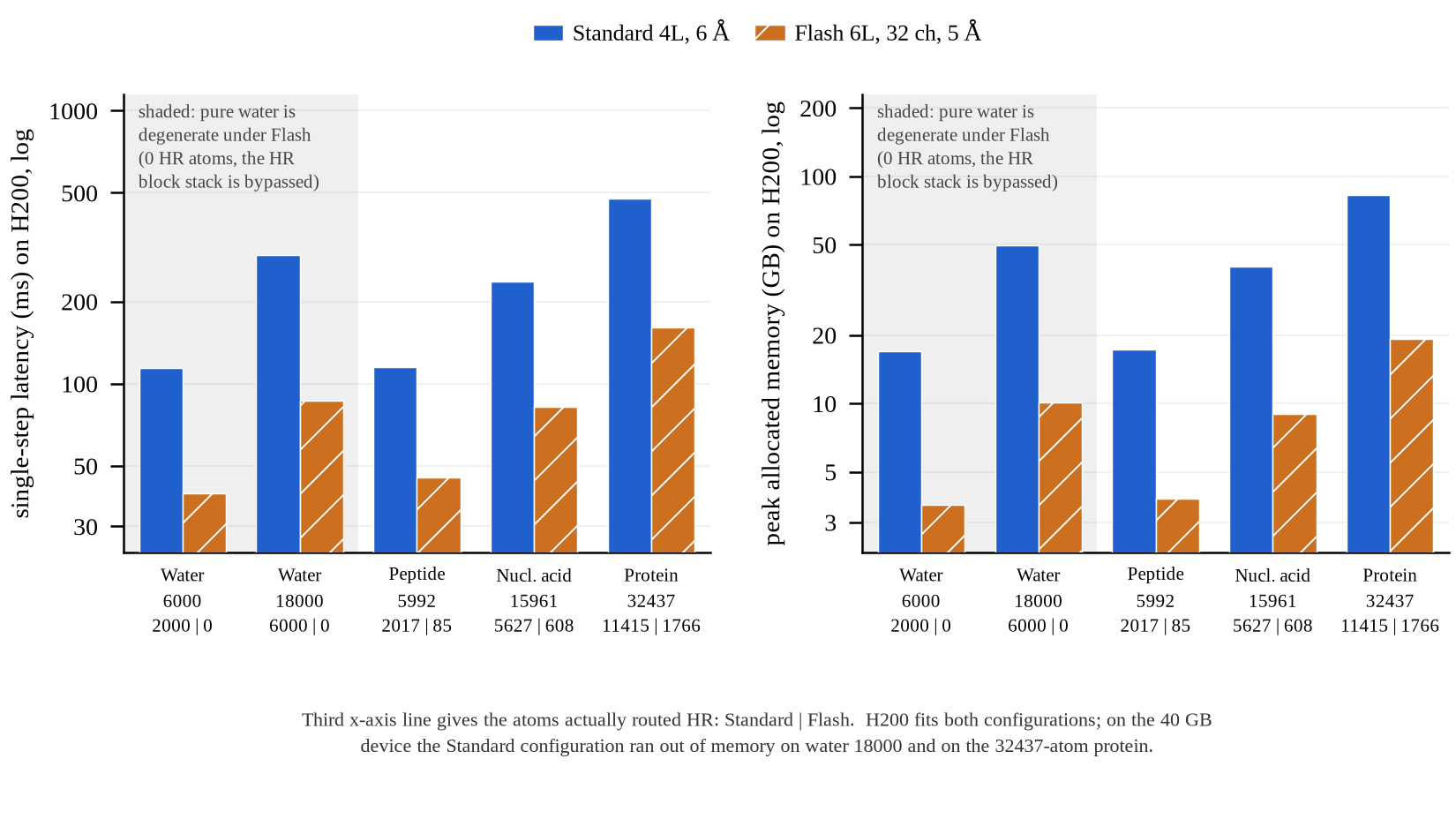}
\caption{Single-step latency and peak allocated memory across five benchmark systems,
H200 only, on log-scaled axes, for the two production configurations. On the 32437-atom
solvated protein Flash runs at 161.18\,ms and 17.929\,GiB against Standard at 477.38\,ms and
77.206\,GiB. The third x-axis line gives the atoms actually assigned to the high-resolution role
under each routing. The two shaded pure-water groups are degenerate under Flash: zero
atoms route high-resolution, so the high-resolution block stack is bypassed and only the narrow stream
executes; those groups carry no headline efficiency claim. The systems that exercise both
streams are the peptide with 85 high-resolution atoms, the nucleic acid with 608 and the protein
with 1766. Every configuration shown fits the H200; on the 40\,GiB device the Standard
configuration ran out of memory on water 18000 and on the 32437-atom protein, and those
outcomes are stated rather than interpolated.}
\label{fig:costscaling}
\end{figure}

Table~\ref{tab:h200} gives the five-system H200 step-only campaign, and
Figure~\ref{fig:costscaling} plots it; these values are separate from the Pareto campaign
behind the $1.75$--$1.84\times$ ratios. Section~\ref{sec:efficiency} quotes
Table~\ref{tab:h200main} for those latencies and ratios and Table~\ref{tab:h200pareto} for
peak memory; Figure~\ref{fig:pareto} is in the main text. Table~\ref{tab:a100} is the A100-40GB
companion table on the same five systems but a different device and timing protocol; it adds the high-resolution-routed
fraction and the two out-of-memory outcomes, which are preserved rather than interpolated.
Table~\ref{tab:ladder} is the H200 exhaustive-search counterpart of the cell-list water-box
ladder in Figure~\ref{fig:throughput}(a,b).
Its reach ordering is the memory column: at 48000 atoms the flash configuration allocates
24.718\,GiB on the A100 and 24.758\,GiB on the H200 ladder, within
0.2\%, so the devices are comparing the same boxes rather than two different
systems.
The separate 17-slot accuracy--throughput campaign is frozen under its own
protocol;
Table~\ref{tab:h200pareto} reports the two production rows together with the speed-only
E2Former-V2 row measured in the same campaign.
E2Former-V2 is the comparison object of Tables~\ref{tab:h200main} and~\ref{tab:h200pareto}
and of the accuracy claim of Section~\ref{sec:accuracy}; it has \emph{no} measurement under
the protocol of Table~\ref{tab:h200} and Figure~\ref{fig:costscaling}, so those two report the
production configurations only. The ratio claims use Table~\ref{tab:h200main}; the memory
claims use Table~\ref{tab:h200pareto}.

\subsection{A capacity cost model}

The per-atom cost of an E2Attention block has two capacity terms. The tensor-product and GEMM
work scales approximately with the number of updates $u_i$ and the square of the width $C_i$;
the node, edge, readout and autograd work scales with atoms and edges and is nearly independent
of width. A uniform model has $u=4$ and $C=256$ for every atom. Standard gives high-resolution
atoms $(u,C)=(4,256)$ and low-resolution atoms $(2,256)$; Flash gives $(4,256)$ and $(2,32)$.
If the block cost were purely $u C^2$, the capacity-only ratios would be
\begin{equation}
\frac{T_{\mathrm{Standard}}}{T_{\mathrm{uniform}}} = 1-\frac12 f_{\mathrm{LR}},
\qquad
\frac{T_{\mathrm{Flash}}}{T_{\mathrm{uniform}}} = f_{\mathrm{HR}} + \frac{2}{4}\left(\frac{32}{256}\right)^2 f_{\mathrm{LR}},
\label{eq:costmodel}
\end{equation}
where $f_{\mathrm{LR}}$ is the low-resolution atom fraction. On the nucleic acid of
Figure~\ref{fig:framework} ($f_{\mathrm{LR}}=0.647$ for Standard, $0.962$ for Flash) these
ratios are 0.68 and 0.046 for the 32-channel production Flash, i.e. 1.48$\times$ and
22.0$\times$. Neighbour search, memory traffic, the autograd graph and
the readout do not shrink with $C^2$. The model therefore explains the direction and the
ordering of the two configurations, not the absolute Flash ratio. It also predicts that
Flash's advantage grows with the water and hydrogen fraction, which is what the system ladder
shows.

\subsection{H200 and A100 step-only campaigns}

\begin{table}[h]
\centering
\small
\caption{Single-step latency and peak allocated memory on \textbf{H200} only. The water
columns are marked $^{\ast}$ because pure water is \emph{degenerate} under the Flash
routing: zero atoms route high-resolution and the high-resolution stack is bypassed, so they
carry no headline efficiency claim. Frozen weights, compiled, fp32, TF32 off; the Pareto
campaign of Table~\ref{tab:h200pareto} is separately frozen under a different timing boundary,
so the two may not share rows.}
\label{tab:h200}
\resizebox{\linewidth}{!}{%
\begin{tabular}{lrrrrr}
\toprule
& Water$^{\ast}$ & Water$^{\ast}$ & Peptide & Nucleic acid & Protein \\
Model & 6000 & 18000 & 5992 & 15961 & 32437 \\
\midrule
\multicolumn{6}{l}{\emph{Single-step latency} (ms)} \\
Standard, 4L, 6\,\AA{} & 114.45 & 294.95 & 114.86 & 236.84 & 477.38 \\
Flash, 6L, 32\,ch, 5\,\AA{} & 39.77 & 86.34 & 45.41 & 82.39 & \textbf{161.18} \\
\midrule
\multicolumn{6}{l}{\emph{Peak allocated memory} (GiB)} \\
Standard, 4L, 6\,\AA{} & 15.766 & 46.432 & 16.035 & 37.301 & 77.206 \\
Flash, 6L, 32\,ch, 5\,\AA{} & 3.324 & 9.440 & 3.533 & 8.348 & \textbf{17.929} \\
\bottomrule
\end{tabular}}
\end{table}

\begin{table}[h]
\centering
\small
\caption{Latency and peak memory on \textbf{A100-PCIE-40GB}; no H200 number
appears here. $N_{\mathrm{HR}}$ is the number of atoms actually assigned to the high-resolution role. Out-of-memory outcomes
are reported as measured. The nucleic-acid Standard row peaked at 37.258\,GiB on a 40\,GiB
device. Its H200 counterpart is Table~\ref{tab:h200}, and the two rows it shares with the H200
water-box ladder are in Table~\ref{tab:ladder} under a different protocol. Both water rows have
$N_{\mathrm{HR}}=0$ under Flash and are degenerate: the entire
high-resolution block stack is bypassed. Frozen weights, compiled, fp32, TF32 off. The Standard
and Flash rows use architecturally identical direct-force warm-start checkpoints; with frozen
weights, latency and memory are weight-independent.}
\label{tab:a100}
\begin{tabular}{lrrrrrrrr}
\toprule
& & \multicolumn{3}{c}{Standard, 4L, 6\,\AA{}} & & \multicolumn{3}{c}{Flash, 6L, 32\,ch, 5\,\AA{}} \\
\cmidrule(lr){3-5}\cmidrule(lr){7-9}
System & Atoms & $N_{\mathrm{HR}}$ & ms & GiB & & $N_{\mathrm{HR}}$ & ms & GiB \\
\midrule
Water & 6000 & 2000 & 270.75 & 15.717 & & 0 & 84.03 & 3.275 \\
Water & 18000 & 6000 & OOM & OOM & & 0 & 200.08 & 9.387 \\
Peptide & 5992 & 2017 & 278.57 & 15.983 & & 85 & 98.81 & 3.482 \\
Nucleic acid & 15961 & 5627 & 620.42 & 37.258 & & 608 & 188.54 & 8.297 \\
Protein & 32437 & 11415 & OOM & OOM & & 1766 & 394.76 & 17.873 \\
\bottomrule
\end{tabular}
\end{table}

\subsection{Exhaustive-search ladder}

\begin{table}[h]
\centering
\small
\caption{The exhaustive-search counterpart to the cell-list ladder in
Figure~\ref{fig:throughput}(a,b), measured on the \textbf{H200}. Selected anchors (300 to
264000 atoms), with no interpolated point: every filled
cell is a direct measurement, and the sizes between anchors were simply not run rather than
filled in. Boxes are identical across models, \texttt{compile=True}, fp32, TF32 off, frozen
weights. The 264000-atom row comes from one fresh process per size; the rows to 60000 atoms
from one process per model sweeping the size list.
Rows at or above 33000 atoms are marked $^{\ddagger}$ and are the re-pinned
measurement; Appendix~\ref{sec:app_protocol} gives the thread-pool pin and the scheduling
artifact it removes. With the pool pinned, per-atom cost is flat across the transition
($4.43$--$4.46\,\mu$s/atom for Flash and $15.75$--$15.86\,\mu$s/atom for Standard), so these
latencies are device measurements. Peak
memory was never affected: it is linear at $0.5116$\,GiB per 1000 atoms from 60000 to 264000
atoms with no break at the transition, which is why the reach claim rests on that column. The ceiling is bracketed at
$(264000, 288000]$ atoms by an actual out-of-memory outcome, not by the grid. The 264000-atom
Flash row reports memory only; the 2.28\,s step quoted in the text is a separate profiled wall
time, not this table's latency column. The E2Former-V2
row at 30000 atoms sits at 135.071\,GiB, essentially the whole device, and its latency should be
read with that saturation in mind.}
\label{tab:ladder}
\resizebox{\linewidth}{!}{%
\begin{tabular}{lrrrrrr}
\toprule
& \multicolumn{3}{c}{H200 latency, ms} & \multicolumn{3}{c}{H200 peak memory, GiB} \\
\cmidrule(lr){2-4}\cmidrule(lr){5-7}
Atoms & E2Former-V2 & Standard & Flash & E2Former-V2 & Standard & Flash \\
\midrule
300 & 34.48 & 26.75 & 23.36 & 1.669 & 1.136 & 0.403 \\
1200 & 57.39 & 38.64 & 28.79 & 5.776 & 3.424 & 0.853 \\
6000 & 215.08 & 111.20 & 38.58 & 27.426 & 15.745 & 3.309 \\
12000 & 408.81 & 202.05 & 60.31 & 54.255 & 31.022 & 6.367 \\
30000 & 2359.15 & 475.85 & 133.23 & 135.071 & 76.929 & 15.563 \\
48000 & OOM & 755.91$^{\ddagger}$ & 212.82$^{\ddagger}$ & OOM & 122.881 & 24.758 \\
60000 & --- & OOM & 272.69$^{\ddagger}$ & --- & OOM & 30.874 \\
264000 & --- & --- & --- & --- & --- & 135.227 \\
\bottomrule
\end{tabular}}
\end{table}

The cell-list ladder behind Figure~\ref{fig:throughput}(a,b), which is a different backend
from the exhaustive-search table above, gives the ratios quoted in Section~\ref{sec:efficiency}:
at 300, 18{,}000 and 30{,}000 atoms, Standard is 1.298, 2.072 and 4.968 times E2Former-V2,
and Flash is 1.389, 8.014 and 20.015 times. The 30{,}000-atom values are the 4.97$\times$ and
20.0$\times$ quoted there; the exhaustive-search table's 30{,}000-atom Flash ratio is
17.7$\times$. The two backends are not interchangeable.

\subsection{Pareto campaign}

\begin{table}[h]
\centering
\footnotesize
\caption{Selected locally measured rows from the frozen H200 accuracy--throughput campaign
used in Figure~\ref{fig:pareto}; each cell is latency in ms / peak allocated memory in GiB.
All rows share one timing boundary that includes graph construction and the conservative force path, and the aggregate is the median of process medians. The full campaign has
$17\times5=85$ model--system points at three scheduled processes per point, and
five high-spread points received three additional processes, giving 270 records in total:
225 successful, 30 out of memory and 15 blocked by checkpoint access. Standard is 1.75 to
1.84$\times$ faster than E2Former-V2 while allocating less peak memory than it on every system.
$^{\dagger}$E2Former-V2 is speed-only: its latency and peak memory are measured locally under
this protocol on an architecturally identical checkpoint, whereas its accuracy in
Figure~\ref{fig:pareto} is taken from the publication; the two quantities therefore come from
different weights.}
\label{tab:h200pareto}
\resizebox{\linewidth}{!}{%
\begin{tabular}{lrrrrr}
\toprule
& Water 6000 & Water 18000 & Peptide 5992 & Nucleic acid 15961 & Protein 32437 \\
\midrule
Standard & 122.839 / 15.746 & 331.438 / 46.410 & 124.404 / 16.015 & 262.964 / 37.275 & 533.866 / 77.165 \\
Flash & 44.915 / 3.309 & 104.070 / 9.423 & 55.295 / 3.512 & 96.197 / 8.314 & 194.719 / 17.895 \\
E2Former-V2$^{\dagger}$ & 214.668 / 27.43 & 608.483 / 81.38 & 221.041 / 28.03 & 459.515 / 60.46 & 949.617 / 126.90 \\
\bottomrule
\end{tabular}}
\end{table}

\subsection{Reachable-row mechanism and kernel attribution}

\begin{figure}[t]
\centering
\includegraphics[width=\linewidth]{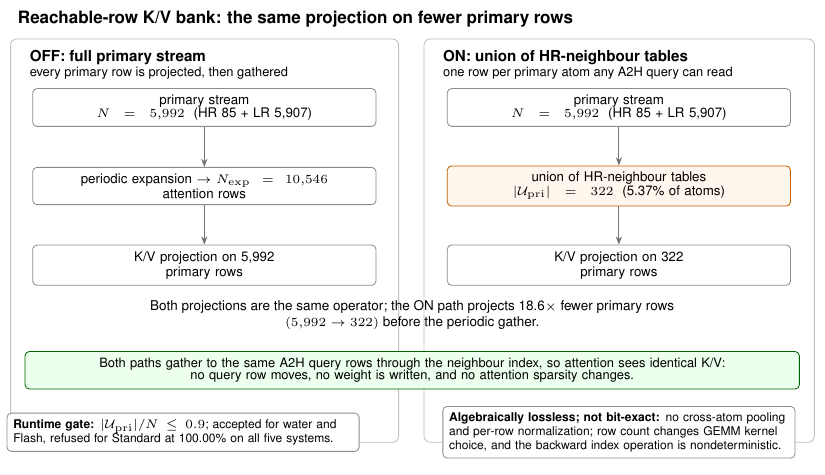}
\caption{The reachable-row key/value bank, drawn on the solvated peptide, the smallest system that exercises
both streams. Both paths are the same model on the same system, so the only thing that moves
between the panes is a row count. OFF projects every primary node ($5{,}992$); periodic images expand the attention bank to
$10{,}546$ rows. ON projects only the $322$ primary rows ($5.37\%$ of atoms) that some A2H query
can reach through the neighbour index, running the identical projection operator on
$18.6\times$ fewer projection rows ($5{,}992\rightarrow322$). Both then gather to the same A2H query rows, so attention sees
identical keys and values: no query row moves, no weight is written and no attention sparsity
changes. That equivalence is what makes the rewrite algebraically lossless --- the all-to-all
side is read only through a gather, with no pooling across atoms and a per-row normalization that does not mix atoms --- and it is also all it is: the path is not bit-exact, because the row count
changes the GEMM kernel choice and the scatter used in the backward pass is nondeterministic. These counts apply to the linear up-projection used in production; the tensor-product variant lifts all light
rows before slicing and does not save this projection, and losslessness also assumes no all-side self term.
The reachable-row count is also not the fraction of atoms assigned to the high-resolution role, and the runtime
gate refuses it above $0.9$ of atoms, which is the case for Standard at $100.00\%$ on all
five systems; the percentages in the accepted row are union sizes, not speedups. All three
qualifications that travel with the measurements this mechanism underlies are unchanged: the
attention-block figure is an inclusive compiled-block category rather than pure attention
kernel time, all ten model--system pairs meet the numerical-consistency thresholds, and the two high-dispersion Flash timing
points retain their $6.28\%$ and $10.75\%$ process-median spans.}
\label{fig:kv_union}
\end{figure}

\begin{figure}[t]
\centering
\includegraphics[width=\linewidth]{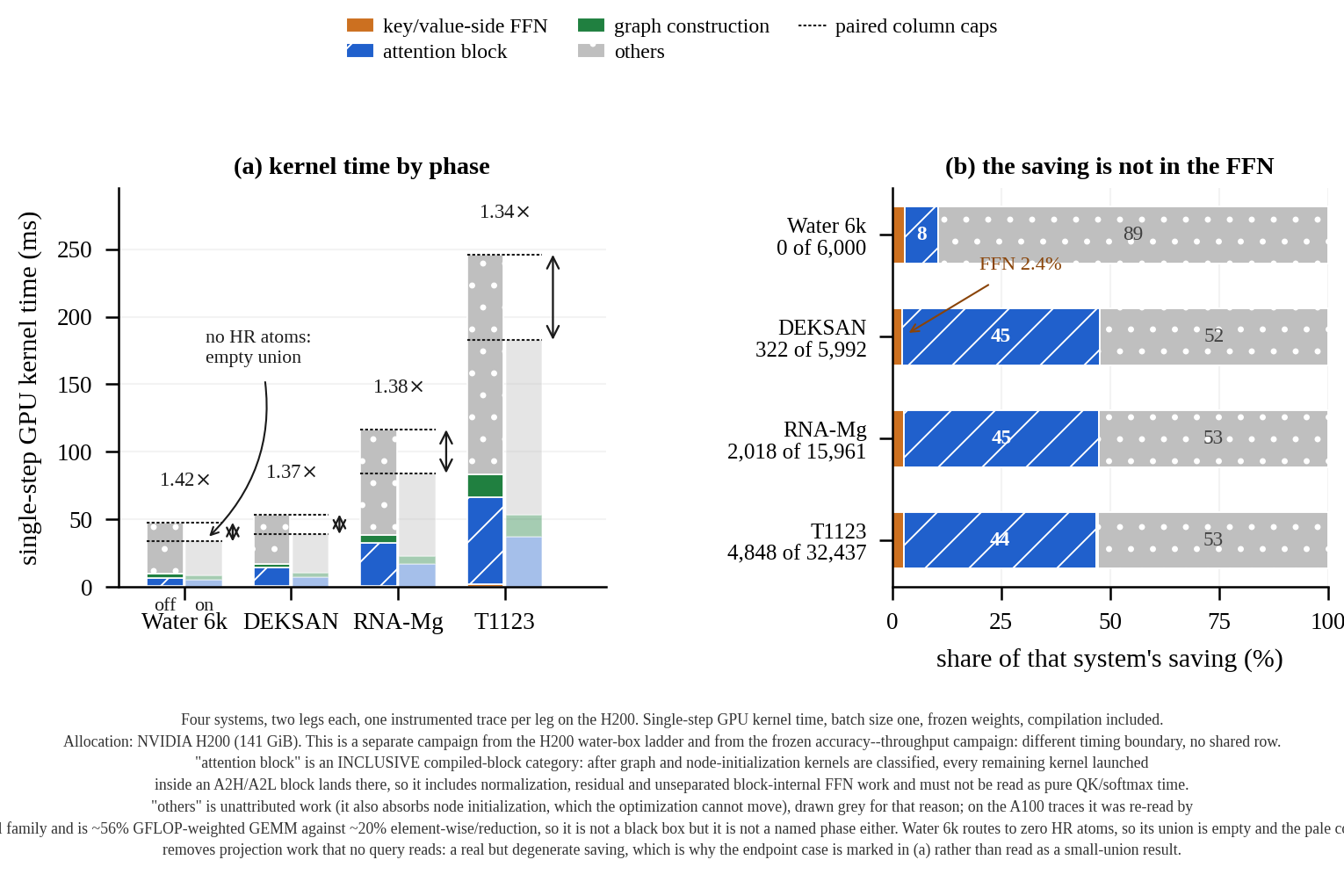}
\caption{Where the reachable-row saving actually lands. Panel (a) stacks single-step GPU
kernel time into four non-overlapping categories, optimization off against on, for four
systems spanning $6.0$k to $32.4$k atoms; the dashed caps mark each column's total and the
arrow the reduction between them. Panel (b) decomposes each system's own reduction, in the same
colours. The component the optimization is named for, the key/value-side FFN, carries $2.35$ to
$2.85\%$ of the saving on every system: a figure plotting it alone would attribute a
$14.50$\,ms effect on the peptide system to a $0.34$\,ms component. The inclusive attention block carries
$44.1$ to $45.3\%$ on the three non-degenerate systems, remarkably stable across a $5.4\times$
span in atom count, and $52.4$ to $52.8\%$ lands in \emph{others}, the unattributed category,
drawn grey for that reason. That bucket is decomposable --- on the A100 traces it was
re-read by kernel family and found to be roughly $56\%$ GFLOP-weighted GEMM against $20\%$
element-wise and reduction --- so it is not a black box, but it is not a named phase either
and no claim rests on it. The
leftmost system is an endpoint case, marked in (a): Flash routes pure water to \emph{zero}
high-resolution atoms, so its union is empty and the pale ON column deletes projection work no query can
read --- a real but degenerate saving, which is why it is not pooled with the others. The
qualifications that travel with the reachable-row measurements hold here unchanged: the
attention-block figure is an inclusive compiled-block category, not pure attention kernel time;
the rewrite is numerically consistent to the frozen thresholds across all ten model--system
pairs (Figure~\ref{fig:filtering}); and the two highest-dispersion Flash legs in this scan
retain $18.5\%$ and $16.4\%$ spreads across their three profiled wall samples. Allocation: NVIDIA H200 (141\,GiB), one device for
every row; the four systems and the timing boundary are not those of the separately frozen
campaign in Table~\ref{tab:h200}, so no number here is comparable with it. The same scan run
on the A100-PCIE-40GB reproduces every
off-to-on ratio to within $2.1\%$ and repeats the Standard null result, which is stated in
the text; it is not the source of any value drawn here.}
\label{fig:kernel_scaling}
\end{figure}

Figure~\ref{fig:kv_union} draws the mechanism those numbers come from and
Figure~\ref{fig:kernel_scaling} attributes its saving. That four-system kernel scan ran on the
\textbf{H200} (141\,GiB), one device for every row, and it is a separate campaign from every
table above, sharing no number with any of them. The same four systems were measured first on
the A100-PCIE-40GB, and the four off-to-on kernel-time ratios reproduce within $2.1\%$
(water6k $1.435\rightarrow1.420$, deksan $1.386\rightarrow1.371$, rna\_mg
$1.413\rightarrow1.384$, t1123 $1.368\rightarrow1.345$), with Standard again gaining nothing on
any system, its four H200 deltas lying inside $\pm0.4\%$ (the largest is $-0.30\%$). That cross-device agreement is the
evidence that the attribution is a property of the mechanism rather than of the device. The
mechanism itself is specified in Section~\ref{sec:locality}.

\FloatBarrier

\section{Single-Variable Architecture Ablation on SPICE}
\label{sec:app_ablation}

\newcommand{\dw}[1]{\textcolor{red!75!black}{$+#1$}}
\newcommand{\db}[1]{\textcolor{blue!70!black}{$-#1$}}

\subsection{Protocol and noise floor}

\paragraph{Protocol.}
The two production configurations differ on six axes at once, so their comparison does not
attribute a difference to any one of them (Appendix~\ref{sec:app_bounds}). The 16 cells below
vary one axis at a time on a fixed recipe: SPICE \citep{eastman2023spice}, 200{,}000 steps,
micro-batch 64, eight A100-40GB GPUs, learning rate $4\times10^{-4}$, no weight decay,
gradient clip 100, energy:force weights 0.2:0.9, an atom-wise $L_2$ force loss, conservative forces from
the energy, seed 6666. Two designs share the same cells: a cumulative chain starts from a
unit-capacity baseline, adds one axis per rung and ends at the production configuration, while
a revert design starts from production Flash and undoes one axis. Two single-switch cells
(F-4, F-6) isolate the row-normalisation and dot-product-gate switches on the Flash recipe; a
cell that flips both was trained but is not reported.

\paragraph{Noise floor.}
The two cumulative endpoints are the same configurations at the same seed as the reference runs
behind Table~\ref{tab:spice}. Retraining reproduces them to 0.25\% in validation force
(Standard, 5.887 against 5.901\,meV/\AA{}), 0.09\% for Flash (8.169 against 8.177), 0.5\% and
2.1\% in median latency, and 0.02\% in peak memory. Differences below 1\% in force, 5\% in
latency and about 2\% in energy per atom are not read as effects. Peak memory is effectively
deterministic.

\subsection{The 16 cells}

Table~\ref{tab:ablation_overview} gives, for every cell, what it changes relative to its
comparison cell (column ``vs'') and the resulting change in validation force, validation energy
per atom, latency and peak memory. Table~\ref{tab:ablation_config} gives the absolute
configuration of every reported cell: in the F+ chain the bold entries form a staircase, and
each revert row differs from F+7 in one entry. F-4 and F-6 were trained from scratch on the
same recipe and differ from F+7 only in the attention type, verified against the recorded
training configuration. Parameter counts are omitted: the single-state
family constructs light-stream modules that it never calls, so the totals are not comparable
across cells. Latency is the median of ten timed batch-1 iterations after one warmup, in fp32
without \texttt{torch.compile}, on one A100-40GB; peak is allocated memory. Timings use the
5{,}992-atom periodic DEKSAN system and a 3{,}490-atom gas-phase protein structure
with no solvent and no cell, so water-oxygen routing has nothing to act on there and both
configurations keep 1{,}766 of 3{,}490 atoms at high resolution. These numbers compare cells to
one another; they are not the H200 production throughputs of Appendix~\ref{sec:app_cost}.
Rows marked $^{\mathrm{b}}$ report within-session relative latency changes;
the protein baseline for the reported $+7.6\%$ and $+3.0\%$ is not available.

\begin{table}[!ht]
\centering
\scriptsize
\setlength{\tabcolsep}{4pt}
\hyphenpenalty=10000 \exhyphenpenalty=10000
\caption{Overview of the 16 reported cells. ``vs'' is the comparison cell; the change column
is before$\rightarrow$after. $\Delta$ columns are relative changes at step 200k and at
batch-1 inference on one A100-40GB; peak memory is allocated memory. Red: worse or more
expensive; blue: better or cheaper. Single seed.}
\label{tab:ablation_overview}
\begin{tabular}{@{}l c p{5.0cm} c c c c c@{}}
\toprule
Cell & vs & \raggedright Change (before $\rightarrow$ after) & \multicolumn{5}{c}{Relative change (\%)} \\
\cmidrule(lr){4-8}
 & & & $\Delta F$ MAE & $\Delta E$ MAE & \begin{tabular}{@{}c@{}}$\Delta$ latency\\DEKSAN\end{tabular} & \begin{tabular}{@{}c@{}}$\Delta$ latency\\protein\end{tabular} & \begin{tabular}{@{}c@{}}$\Delta$ peak memory\\DEKSAN\end{tabular} \\
\midrule
A      & --   & \raggedright uniform architecture (reference) & \multicolumn{5}{c}{--} \\
S+1    & A    & \raggedright uniform $\rightarrow$ single-state MR & \dw{2.8} & \db{5.4} & \db{45.3} & \db{32.4} & \db{41.3} \\
S+2    & S+1  & \raggedright s3 FFN $\rightarrow$ s2 FFN & \db{15.1} & \db{5.1} & \dw{4.8} & \dw{7.0} & \dw{16.3} \\
\addlinespace
F+1    & A    & \raggedright uniform $\rightarrow$ two-state MR & \dw{32.5} & \dw{28.5} & \db{60.8} & \db{44.9} & \db{60.9} \\
F+2    & F+1  & \raggedright light width 256 $\rightarrow$ 32 & \dw{5.4} & \dw{1.2} & \db{28.6} & \db{13.7} & \db{29.0} \\
F+3    & F+2  & \raggedright depth 4 $\rightarrow$ 6 & \db{25.6} & \db{28.4} & \dw{3.9} & \dw{51.6} & \dw{3.4} \\
F+4    & F+3  & \raggedright s3 FFN $\rightarrow$ s2 FFN & \db{15.2} & \db{12.9} & \dw{1.0} & \dw{3.6} & \dw{1.4} \\
F+5    & F+4  & \raggedright exp-softmax, dot gate $\rightarrow$ softplus, no dot gate & \dw{41.8} & \dw{38.0} & \db{8.7} & \db{18.3} & \db{3.6} \\
F+6    & F+5  & \raggedright cutoff 6\,\AA{} $\rightarrow$ 5\,\AA{} & \db{0.1} & \dw{8.9} & \db{32.6} & \db{18.8} & \db{38.7} \\
F+7    & F+6  & \raggedright light heads 4$\times$16 $\rightarrow$ 16$\times$4 & \db{3.0} & \db{4.1} & \dw{1.2} & \db{0.2} & \dw{2.3} \\
\addlinespace
F-1    & F+7  & \raggedright light width 32 $\rightarrow$ 256 & \db{6.4} & \db{6.0} & \dw{62.0} & \dw{16.0} & \dw{89.9} \\
F-2    & F+7  & \raggedright depth 6 $\rightarrow$ 4 & \dw{34.4} & \dw{45.8} & \db{7.0} & \db{33.0} & \db{3.9} \\
F-3    & F+7  & \raggedright s2 FFN $\rightarrow$ s3 FFN & \dw{21.8} & \dw{15.7} & \dw{2.0} & \db{6.3} & \db{2.3} \\
F-4$^{\mathrm{b}}$ & F+7 & \raggedright no dot gate $\rightarrow$ dot gate & \db{18.0} & \db{14.8} & \dw{18.5} & \dw{7.6} & \dw{7.4} \\
F-5    & F+7  & \raggedright cutoff 5\,\AA{} $\rightarrow$ 6\,\AA{} & \db{0.9} & \db{11.1} & \dw{45.6} & \dw{27.1} & \dw{63.2} \\
F-6$^{\mathrm{b}}$ & F+7 & \raggedright softplus $\rightarrow$ softmax & \db{14.1} & \db{9.3} & \dw{17.8} & \dw{3.0} & \dw{1.4} \\
\bottomrule
\end{tabular}
\end{table}

\begin{table}[!ht]
\centering
\scriptsize
\caption{Absolute configuration of every reported cell, from the recorded training configuration.
``vs'' is the comparison cell; the bold entry is the single setting that changes. Light
width/heads describe the light stream (256/32; heads$\times$scalars) and are undefined for A
and the single-state family, which has no separate light stream. Grid resolution is
omitted: it exists only for the s2 FFN.}
\label{tab:ablation_config}
\setlength{\tabcolsep}{4pt}

\begin{tabular}{l c l c c l l c l l}
\toprule
Cell & vs & MR stream & light width & $L$ & FFN & attention & $r_c$ (\AA{}) & light heads & H$_2$O--O light \\
\midrule
A      & --   & off       & --   & 4 & s3 & softmax$+$dot & 6 & --        & -- \\
S+1    & A    & \textbf{single}    & --   & 4 & s3 & softmax$+$dot & 6 & --        & no \\
S+2    & S+1  & single    & --   & 4 & \textbf{s2} & softmax$+$dot & 6 & --        & no \\
\addlinespace
F+1    & A    & \textbf{two-state} & 256  & 4 & s3 & softmax$+$dot & 6 & 4$\times$16 & yes \\
F+2    & F+1  & two-state & \textbf{32} & 4 & s3 & softmax$+$dot & 6 & 4$\times$16 & yes \\
F+3    & F+2  & two-state & 32   & \textbf{6} & s3 & softmax$+$dot & 6 & 4$\times$16 & yes \\
F+4    & F+3  & two-state & 32   & 6 & \textbf{s2} & softmax$+$dot & 6 & 4$\times$16 & yes \\
F+5    & F+4  & two-state & 32   & 6 & s2 & \textbf{softplus} & 6 & 4$\times$16 & yes \\
F+6    & F+5  & two-state & 32   & 6 & s2 & softplus & \textbf{5} & 4$\times$16 & yes \\
F+7    & F+6  & two-state & 32   & 6 & s2 & softplus & 5 & \textbf{16$\times$4} & yes \\
\addlinespace
F-1    & F+7  & two-state & \textbf{256} & 6 & s2 & softplus & 5 & 16$\times$4 & yes \\
F-2    & F+7  & two-state & 32   & \textbf{4} & s2 & softplus & 5 & 16$\times$4 & yes \\
F-3    & F+7  & two-state & 32   & 6 & \textbf{s3} & softplus & 5 & 16$\times$4 & yes \\
F-4    & F+7  & two-state & 32   & 6 & s2 & \textbf{softplus$+$dot} & 5 & 16$\times$4 & yes \\
F-5    & F+7  & two-state & 32   & 6 & s2 & softplus & \textbf{6} & 16$\times$4 & yes \\
F-6    & F+7  & two-state & 32   & 6 & s2 & \textbf{softmax} & 5 & 16$\times$4 & yes \\
\bottomrule
\end{tabular}
\end{table}

\paragraph{Reading the two designs.}
Table~\ref{tab:ablation_overview} holds both designs in one place: the F+ rows read each
cumulative rung against its parent, and the F$-$ rows read each revert against production
Flash. The single-axis reverts are also collected in Table~\ref{tab:ablation} (main text),
where the attention axis appears as its two single-switch rows. The two designs agree on four
axes; the cutoff axis is below the 1\% force floor and changes sign between the two designs,
so we do not read it. The revert estimate is the larger one for depth,
FFN and light width, by a factor of 1.2--1.4; for attention the cumulative rung flips both
switches and is 2.3--3.0$\times$ larger than either single-switch revert. The cutoff axis is
below the force floor in both designs, although its energy and cost changes are not. The
light-head revert reuses F+6, the parent in the cumulative step F+6$\rightarrow$F+7, so the
two designs share one comparison.

\subsection{Attention axis}
The SPICE-retrained Flash ablation uses softplus attention without the dot gate, while Standard uses
softmax attention with the dot gate with an exp-softmax normalisation; the difference bundles the row
normalisation, the activation and the dot-product scalar gate. Adding the gate while keeping
softplus (F-4) improves validation force by 18.0\% and energy per atom by 14.8\%; switching the
activation to exp-softmax while keeping the gate absent (F-6) improves force by 14.1\% and
energy by 9.3\%. The gate is the larger of the two, and both cost: the gate adds 18.5\% latency
and 7.4\% peak memory, the activation switch 17.8\% and 1.4\% (within-session pairs).
The OMol25 production checkpoint keeps the dot gate, so it captures the
larger single-switch gain; the SPICE-retrained Flash form drops it (not the OMol25 production
checkpoint), for a physical reason: the
mixed-resolution exp-softmax branch normalises each query over its neighbours with a $10^{-3}$
denominator floor, which controls the limit at cutoff events in this implementation, whereas
the softplus path is a per-edge activation with no row normalisation. That is a property of the
normalisation alone; the two single-switch cells price the gate and the activation separately
without requiring the combined cell.

\paragraph{Where the routing cost falls.}
The SPICE test results in Table~\ref{tab:spice} show larger force errors for production
Flash than Standard on solvated amino acids and water. The single-variable cells above
separate architecture effects on aggregate validation error; they do not establish
subset-specific routing effects.

\subsection{Axes without a separate estimate}
Two axes of the production comparison have no single-variable estimate. The block
schedule and water-oxygen routing are bound to the mixed-resolution family in both production
configurations: Standard uses a single-state schedule with water oxygen at high resolution, and
Flash uses a two-state schedule with water oxygen routed low, so neither can be changed without
the other in this codebase. In the single-state family the light-stream modules are
constructed but never called, so the light-stream embedding,
the attention head count and the scalar head are inert there; a probe
found no gradient on 228 of 228 light-family parameters and identical live weights across the
cells that vary them. Those axes are measured in the two-state family only. The Standard
track's only single-axis reverts are light-stream settings, which are inert there, and its FFN
revert is the same design cell as S+2. The unit-capacity
baseline is a retrain at this recipe, not the published E2Former-V2 checkpoint: it reaches
6.745\,meV/\AA{} against the 7.04 of Table~\ref{tab:spice}, and all deltas in this appendix are
internal to the campaign.

\subsection{Scope}
Every cell is a 200k-step model on one seed, and none is converged: validation error is still
falling by 7--10\% per 50k steps at the end of each run. Accuracy is
read from the runs' own validation logs. The ranking and the relative deltas are
stable over the last 50k steps, and the absolute values are not final. The campaign is SPICE
only, so the axis ordering need not carry to OMol25 or to the production checkpoints of
Table~\ref{tab:accuracy}. Latency is measured without compilation or activation
recomputation; the reported quantity is the deltas between cells.

\FloatBarrier

\section{Physical Validation}
\label{sec:app_water}

\subsection{Simulation protocols}
\label{sec:app_md_protocol}

\textbf{20\,ps matrix.} The 54 legs are Standard/Flash $\times$ water / 150\,mM NaCl / DEKSAN
$\times$ NVT/NPT/NVE $\times$ three seeds (104729, 130363, 155921), each 40{,}000 steps at
0.5\,fs. The matrix uses a 216-molecule water box and a 49.96\,\AA{} NaCl box holding
11 Na$^+$ and 11 Cl$^-$; the 64- and 512-molecule boxes belong to the
separate 12-leg water-only matrix (Tables~\ref{tab:waterstruct} and~\ref{tab:waternpt}), not
to these 54 legs. DEKSAN is the 5{,}992-atom solvated peptide. NVT legs use the production
Langevin thermostat at 300\,K; NPT
legs use a Monte Carlo barostat with acceptance in $[0.30,0.40]$; NVE legs continue from the
completed NVT endpoint without re-thermalization. Initial states are the equilibrated NPT
states. The Standard cutoff is
6\,\AA{} and the Flash cutoff is 5\,\AA{}; both are short-range models and use no PME.

\textbf{1\,ns campaign.} The 32{,}437-atom solvated protein is initialized from the same
equilibrated NPT state; the identity was verified against the initial state. The Standard
leg is the 4L/6.0\,\AA{} configuration (22.224M active parameters) and the Flash leg the
6L/5.0\,\AA{} configuration (22.876M). The run is
300\,K Langevin NVT, 0.5\,fs, 2{,}000{,}000 steps = 1000\,ps, seed 104729, fixed cell,
center-of-mass removal, one frame every 1000 steps (0.5\,ps), on eight A100-PCIE-40GB,
one process per configuration. The run was cut over to \texttt{torch.compile} partway through;
the Langevin RNG is not persisted across a restart, so no leg is bitwise reproducible across
the cutover and the campaign is gated on energy continuity rather than coordinate agreement.

\textbf{100\,ps high-temperature NVE campaign.} A separate campaign extends the constant-energy
stress test past the 20\,ps matrix on the 512-molecule water box: Standard and Flash $\times$
300, 500 and 1000\,K $\times$ three independent seeds, 18 trajectories in total, each 10\,ps of
NVT pre-equilibration followed by 100\,ps of NVE at a 0.5\,fs timestep. All 18 trajectories
completed. Endpoint energy changes and linear drift slopes, with seed-to-seed standard
deviations, are reported in Section~\ref{sec:nve100} and Table~\ref{tab:nve100} of the main
text and are not repeated here.

\textbf{Classical control.} The same start state was propagated for 1000\,ps in NVT with
GROMACS 2025.3 and the AMBER ff14SB/TIP3P force field: 300\,K \texttt{v-rescale} thermostat on
Protein and Non-Protein groups, 2\,fs timestep, 500{,}000 steps, Verlet cutoff 1.2\,nm, PME
electrostatics, h-bond constraints with LINCS, and one frame every 2\,ps
\citep{maier2015ff14sb,jorgensen1983comparison,abraham2015gromacs}.

\subsection{What this appendix measures}

\begin{figure}[h]
\centering
\includegraphics[width=0.72\linewidth]{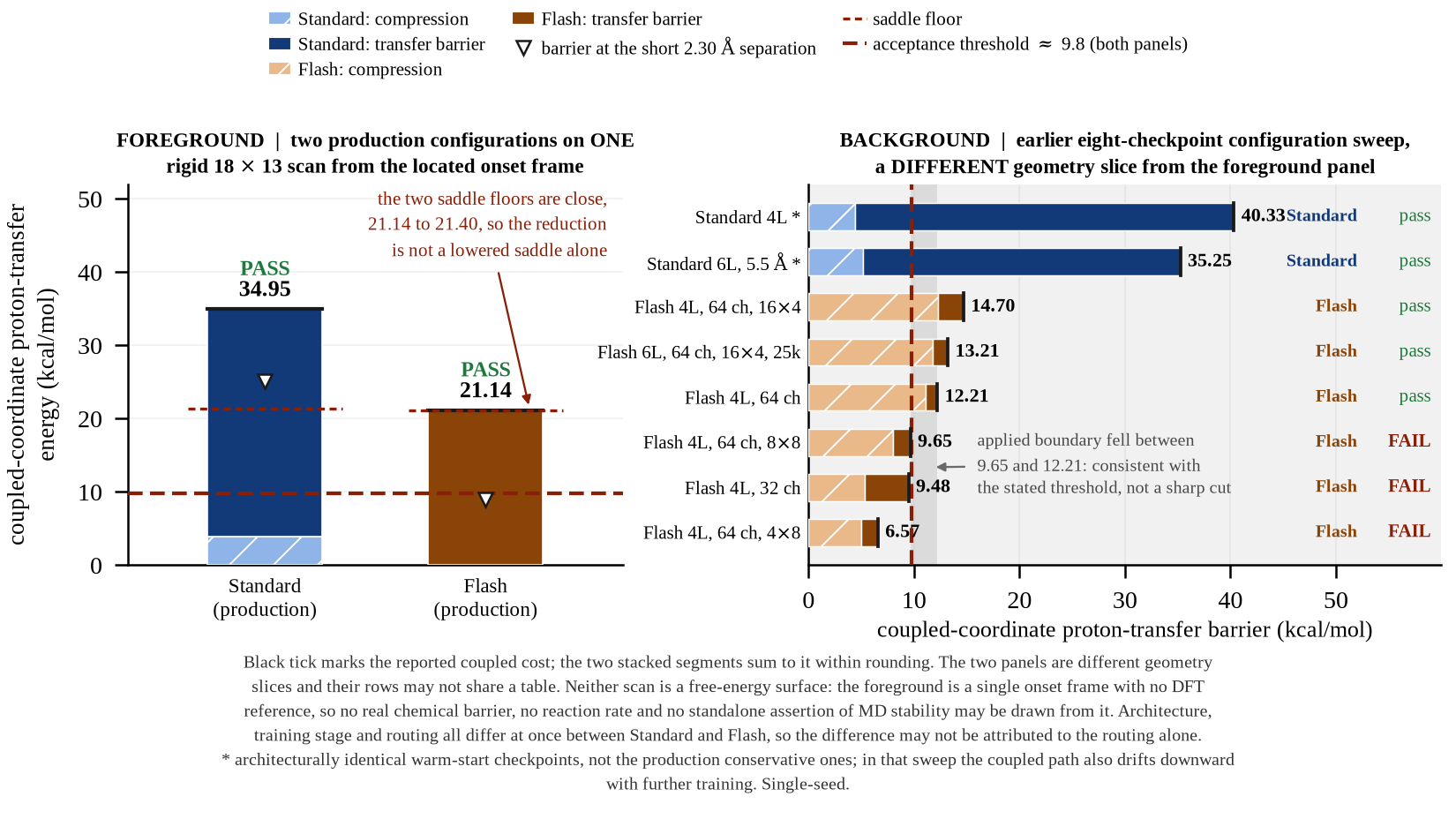}
\caption{Coupled-coordinate proton-transfer energetics, two panels on \emph{different
geometry slices} whose bars may not be read against each other. \textbf{Foreground}: the two
production configurations on one identical rigid $18\times13$ scan from the located
511.855\,ps failure-onset frame. Both pass the approximately 9.8\,kcal/mol threshold, at
coupled costs of 34.95 (Standard) and 21.14 (Flash); the marked saddle floors are close,
21.14 to 21.40, so Flash's reduction lives as much in compression and the product basin as in
the barrier. \textbf{Background}: the earlier eight-checkpoint configuration sweep on a
rigid 234-geometry slice, routing class marked on every bar so a Flash number is never read
against a Standard one. There the two Standard configurations clear the threshold threefold to
fourfold, at 40.33 and 35.25, while three of six Flash configurations fail, worst at 6.57.
Its qualifications stand: the applied boundary fell between 9.65 and 12.21 rather than sharply,
the two routings describe different surfaces, the coupled path drifts downward with further
training, and both Standard rows are warm-start rather than production conservative checkpoints.
All rows single-seed.}
\label{fig:barrier}
\end{figure}

\begin{figure}[h]
\centering
\includegraphics[width=0.72\linewidth]{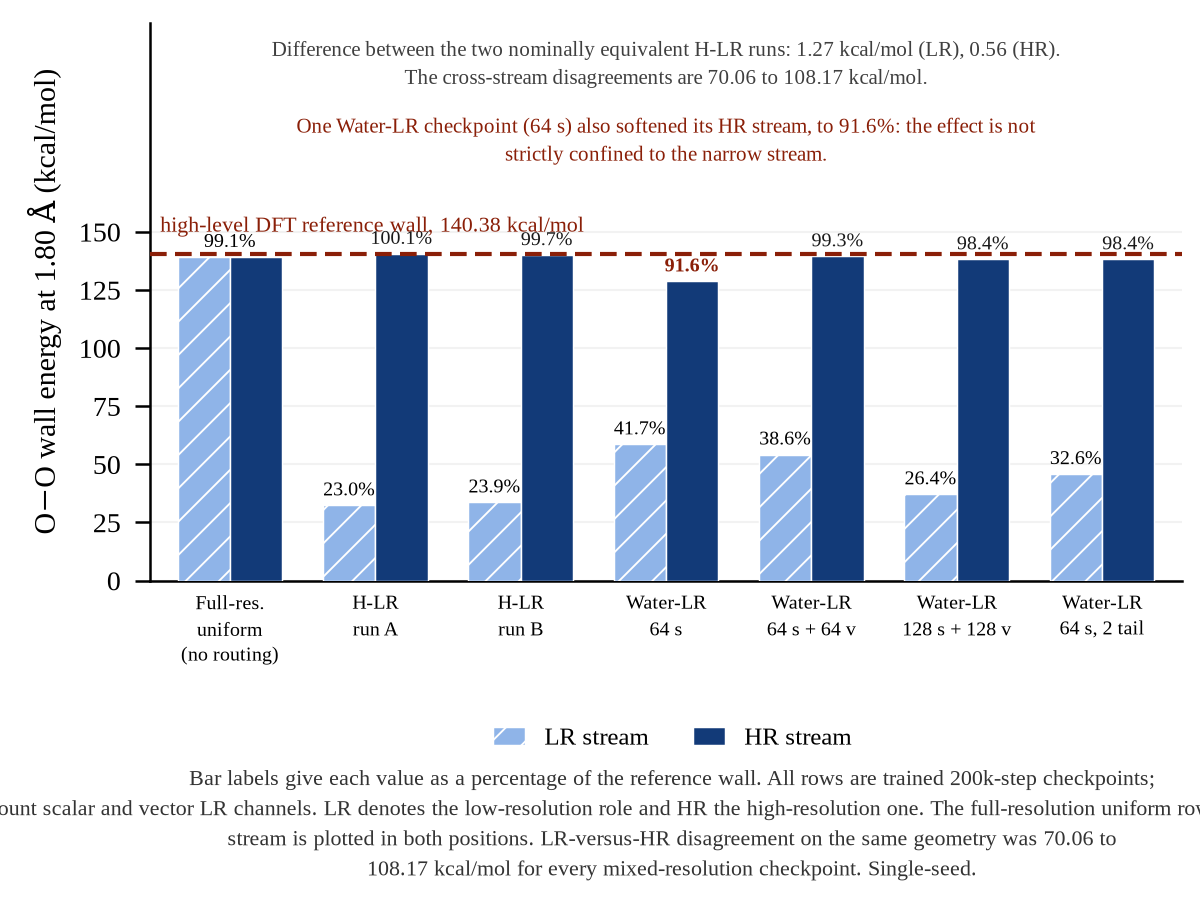}
\caption{Water-dimer oxygen-oxygen repulsive wall at 1.80\,\AA{} separation, LR stream
against HR stream, per checkpoint, with each bar labelled as a percentage of the
high-level DFT reference wall of 140.38\,kcal/mol ($\omega$B97M-V/def2-TZVPD;
\citealp{mardirossian2016omega,rappoport2010propertyoptimized}). LR and HR denote the low-resolution and
high-resolution roles. The reduced-capacity stream
systematically under-repels at short range, reproducing only 23.0 to 41.7\% of the
reference, while the HR streams of the remaining checkpoints sit at 98.4 to 100.1\%; LR-versus-HR
disagreement on the same geometry was 70.06 to 108.17\,kcal/mol for every
mixed-resolution checkpoint. The two nominally equivalent Standard runs differ by
1.27\,kcal/mol on the LR stream and 0.56\,kcal/mol on the HR stream. One Flash
checkpoint also softened its HR stream, to 91.6\%, so the effect is not strictly
confined to the narrow stream. The full-resolution uniform row applies no routing, so its
single stream is plotted in both positions. All sweep rows are 200k-step checkpoints; the
production rows are the production checkpoints. All rows are single-seed.}
\label{fig:wall}
\end{figure}

\begin{figure}[h]
\centering
\includegraphics[width=\linewidth]{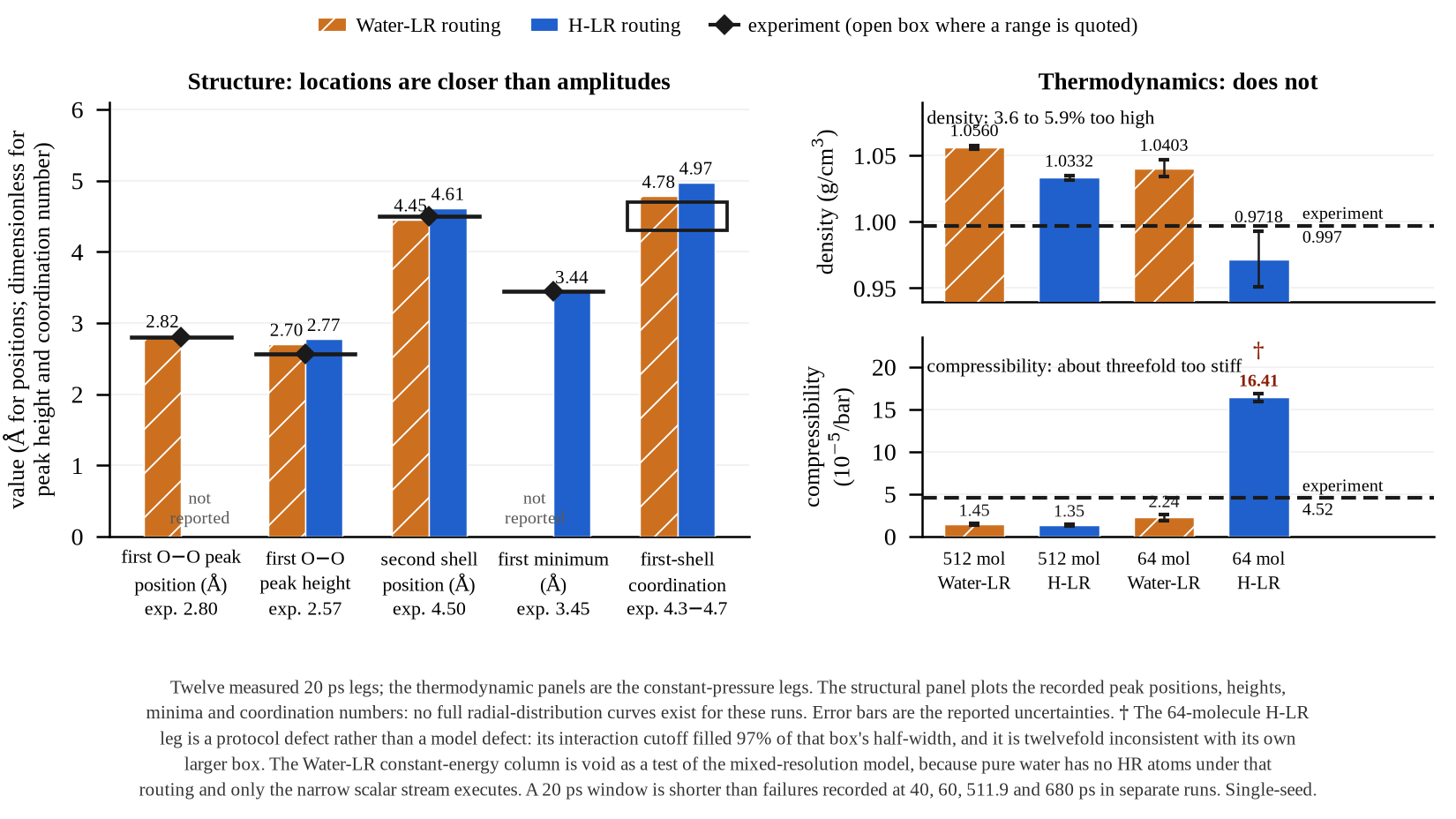}
\caption{Liquid water: some structural locations are close, but amplitudes and
thermodynamics are not uniformly reproduced. On the twelve water-only legs, the first
oxygen-oxygen peak sits at 2.82\,\AA{} against 2.80 experimental, and a first minimum at
3.44\,\AA{} against 3.45 (the two are drawn from different routings, as the table's
\emph{not reported} cells record); peak heights are 2.70 and 2.77 against 2.57, and
first-shell coordination is 4.78 and 4.97 against an experimental 4.3 to 4.7. The
constant-pressure legs give densities 3.6 to 5.9\% too high and compressibilities roughly
threefold too stiff. A correct radial distribution function is therefore not evidence of a
usable potential, and a 20\,ps window is shorter than the tabulated failures near 40, 58 and
60\,ps and the separate failures at 511.9 and 680\,ps. This panel plots scalar summaries
against their experimental values, so its y-axis mixes units, \AA{}ngstroms for positions
and dimensionless values for peak height and coordination number; for the full curves and
for the three-seed values on the 54-leg matrix, see Figure~\ref{fig:rdf} and
Table~\ref{tab:rdf54}. Error bars are the reported uncertainties. Two leg classes must be
read with care. The 64-molecule Standard leg, marked with a dagger, is a protocol defect rather
than a model defect: its interaction cutoff filled 97\% of that box's half-width and the leg
is twelvefold inconsistent with its own larger box, which invalidates its thermodynamic
readings but not its structure or energy conservation. And the Flash pure-water column is
the routing's \emph{degenerate limit} rather than a failed test: pure water has no high-resolution
atoms under that routing, so only the narrow scalar stream executes, and it cannot evidence
cross-role exchange or carry an efficiency headline. All results are single-seed.}
\label{fig:waterstruct}
\end{figure}

\begin{figure}[h]
\centering
\includegraphics[width=\linewidth]{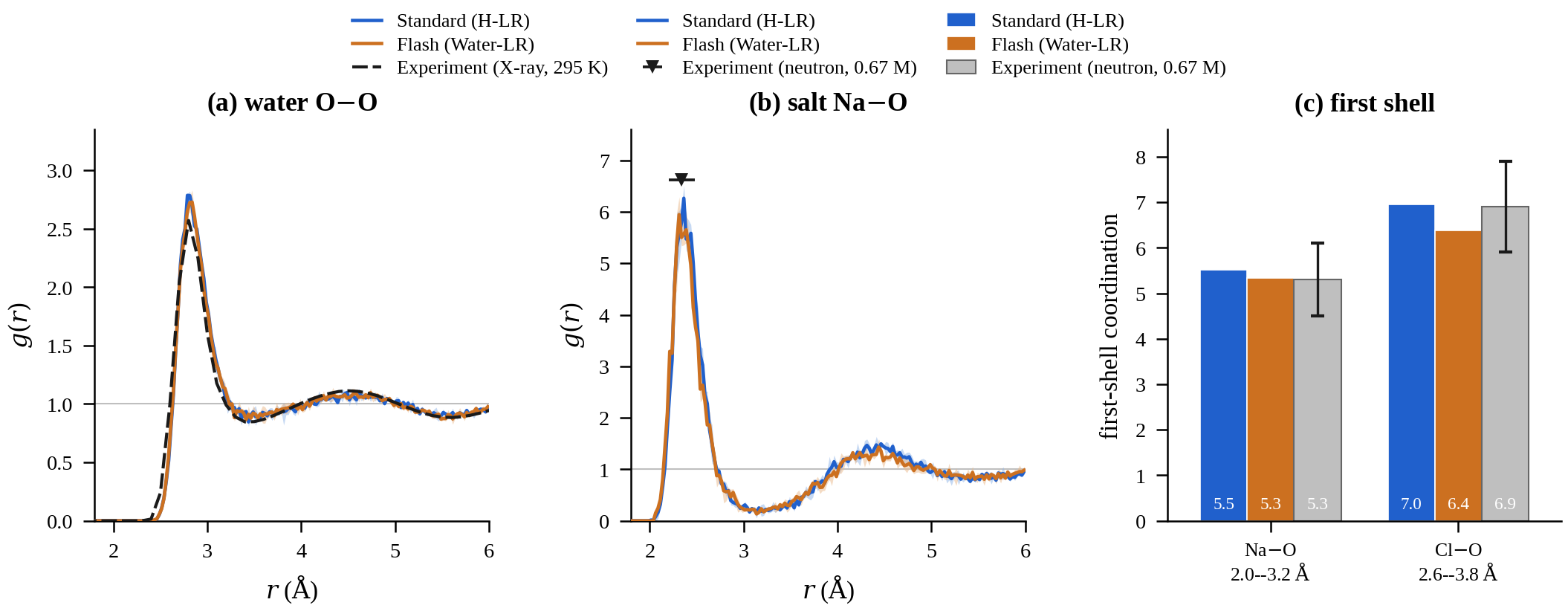}
\caption{The 54-leg production matrix's radial distribution functions, on the NVT legs only
(the legs with recorded curves). Curves are three-seed means with the seed-to-seed span
shaded; every curve is 50 frames over $t=10$--20\,ps. \textbf{(a)} Water O--O against the
X-ray measurement of \citet{skinner2013benchmark}: both routings nearly coincide with
experiment, with first peaks within 0.04\,\AA{} and the same first minimum and second shell.
The reference is a digitised rendering of that figure, not the authors' table
\citep{instadeep2025rdf}, and reproduces the published scalars to the last digit. \textbf{(b)}
Na--O, where Standard and Flash coincide to 0.05\,\AA{}; the marker is the measured first-peak
position of \citet{mancinelli2007hydration}, and no reference curve is drawn because that
source publishes figures without a table. We found no tabulated experimental Na--O or Cl--O
first-peak height: \citet{galib2017revisiting} report simulated heights and
\citet{powell1993chloride} a Cl--O position and coordination, but neither supplies the height
used here. \textbf{(c)} First-shell coordination integrated over each pair's published range
(Na--O 2.0--3.2\,\AA{}, Cl--O 2.6--3.8\,\AA{}), the like-for-like counterparts of the published
5.3(0.8) and 6.9(1.0); integrating to each curve's own first minimum gives smaller,
non-comparable values (Table~\ref{tab:rdf54}), so the conventions are not mixed. Four bounds
travel with the figure: the experimental ion values are at 0.67\,M against this box's
150\,mM, so concentration is not matched; \citet{mancinelli2007hydration} reports the Na--O
position as concentration-independent, which is what makes the dilute limit usable; the salt
panel is a position rather than a curve and is weaker evidence than the water panel; and
Na--Cl is measured but not converged (11 Na$^+$ and 11 Cl$^-$ over 50 frames, per-seed peak
heights 0.0 to 18.2), so it is not drawn. A correct RDF is not evidence of a usable potential
(Figure~\ref{fig:waterstruct}).}
\label{fig:rdf}
\end{figure}

\begin{figure}[t]
\centering
\includegraphics[width=\linewidth]{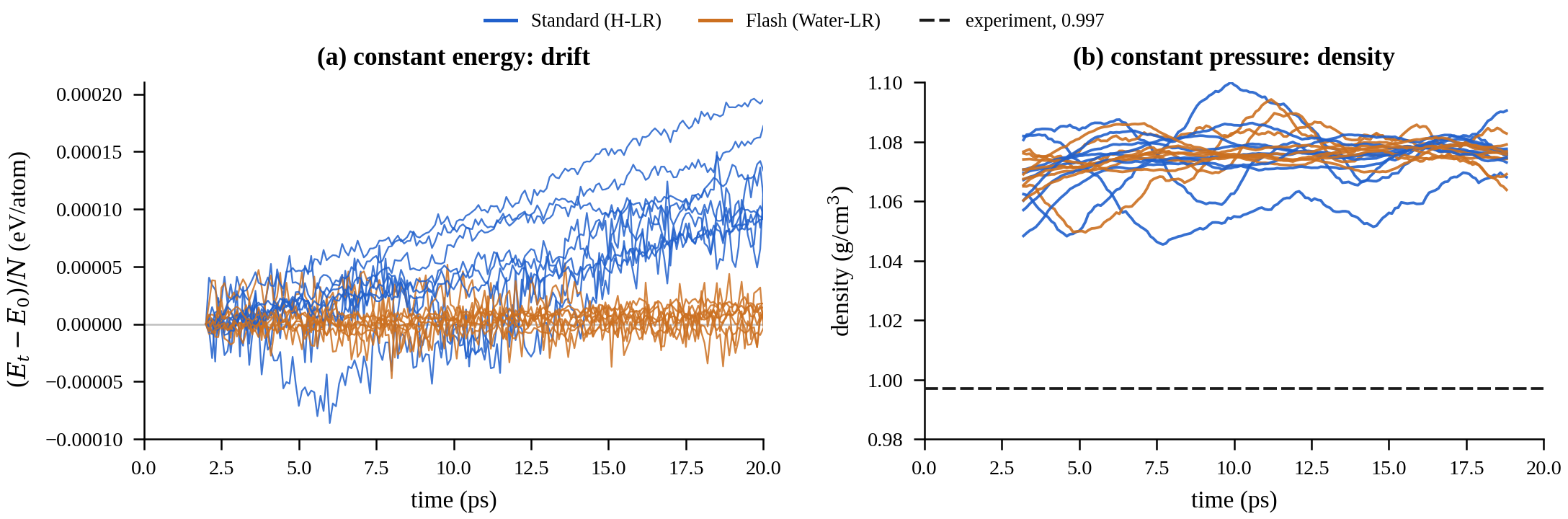}
\caption{The 20\,ps production dynamics matrix, drawn from per-frame traces: two models
$\times$ three systems $\times$ three ensembles $\times$ three seeds, 54 legs of 40000 steps.
All 54 complete with no failure, non-finite value, overlap or oxygen fusion, and every
constant-pressure leg's Monte-Carlo acceptance is inside $[0.30,0.40]$. One thin line per
seed. \textbf{(a)} Constant-energy drift, $(E_t-E_0)/N$ in eV/atom: the Flash traces are flat
and centred on zero on all three systems, while three of the Standard traces show the upward
slope tabulated in Table~\ref{tab:nvedrift} in meV/atom/ps. The fitted change over the
2--20\,ps window is the tabulated slope times $18/1000$ in eV/atom; it is a fitted trend
change, not the plotted displacement from $t=0$. \textbf{(b)} Constant-pressure density,
rolled over 25 frames because per-frame values are dominated by the box's instantaneous
fluctuation. Both models sit at 1.0720--1.0762\,g/cm$^3$ against an experimental 0.997; the
overshoot is drawn because it indicates a probable pressure and equation-of-state bias.
Stable, unfused, energy-conserving trajectories are not quantitatively accurate
thermodynamics. The duration bound is load-bearing: 20\,ps is shorter than the tabulated
failures near 40, 58 and 60\,ps and the separate failures at 511.9 and 680\,ps, so these legs
support short-time stability and local structure only. The 1\,ns tier of this matrix was not
run; the separate 1\,ns NVT campaign is reported in Table~\ref{tab:ns1main}. Hydrogen-pair
observables, ion residence and exchange rates, and the 200\,ps screening tier remain
unmeasured.}
\label{fig:dynamics}
\end{figure}

Table~\ref{tab:barrier} and Figure~\ref{fig:barrier} give the full measured record of the
coupled-coordinate proton-transfer gate. Section~\ref{sec:gates} retains only the short-range
repulsion result on the production checkpoints and points here for the complete physical-gate
record. Table~\ref{tab:wall} and Figure~\ref{fig:wall} report the oxygen--oxygen repulsion
scans. Tables~\ref{tab:waterstruct}, \ref{tab:rdf54}, \ref{tab:waternpt}
and~\ref{tab:waternve}, together with Figures~\ref{fig:waterstruct}, \ref{fig:rdf}
and~\ref{fig:dynamics}, give the liquid-structure and dynamics evidence.
Tables~\ref{tab:temperature}, \ref{tab:covalent} and~\ref{tab:nvedrift} give the temperature,
chemical-integrity and energy-conservation evidence for the 54-leg matrix.
Table~\ref{tab:ns1main} gives the separate production 1\,ns NVT campaign
that extends the duration question past this matrix's 20\,ps on one system.
Two statements are repeated here because they bound what these tables can be used for, and
they are \emph{not} the same statement. The 64-molecule Standard leg is a protocol defect
rather than a model defect, its interaction cutoff filling 97\% of that box's half-width:
that makes the leg uninformative on the thermodynamic axes, not invalid, and its structure,
integrity and energy-conservation evidence still stands. The Flash pure-water column is
a \emph{degenerate limit} of the routing rather than a failed measurement --- pure water has
no high-resolution atoms under it, so only the narrow stream executes --- which is why it cannot
evidence cross-role exchange or carry an efficiency headline, and not a reason to discard
what it does show. The body states the degenerate limit in its experimental setup
(Section~\ref{sec:experiments}); the 64-molecule protocol defect is stated here only, with its
evidence beside it.

\subsection{Proton-transfer and short-range gates}

\begin{table}[h]
\centering
\small
\caption{Coupled-coordinate proton-transfer barrier gate \citep{agmon1995grotthuss},
kcal/mol on a rigid
234-geometry slice zeroed at the uncompressed reactant. The stated acceptance threshold is
approximately 9.8\,kcal/mol; the applied boundary fell between 9.65 and 12.21 and is
therefore not a sharp cut, and the two routings describe different surfaces, so only the
ordering within a routing is meaningful. The last row stays \emph{unfilled for this scan}: the
production Flash configuration was never run on this 234-geometry slice and no other checkpoint
is substituted for it. It is not unmeasured --- it was scanned on the separate onset-frame
surface of the foreground panel of Figure~\ref{fig:barrier}, where it passes --- but the two
slices are different surfaces and may not share a row. The two Standard rows used architecturally identical warm-start
checkpoints rather than the production conservative ones. Single-seed.}
\label{tab:barrier}
\resizebox{\linewidth}{!}{%
\begin{tabular}{llrrrcl}
\toprule
Checkpoint & Routing & Layers & Step & Coupled barrier & Compression $+$ barrier & Gate \\
\midrule
Standard, 6\,\AA{} & Standard & 4 & 740\,000 & \textbf{40.33} & 4.4 $+$ 36.0 & pass \\
Standard, 5.5\,\AA{} & Standard & 6 & 550\,000 & \textbf{35.25} & 5.2 $+$ 30.1 & pass \\
Flash, 64\,ch, 16$\times$4 attention & Flash & 4 & 200\,000 & 14.70 & 12.3 $+$ 2.4 & pass \\
Flash, 64\,ch, 16$\times$4 attention & Flash & 6 & 25\,000 & 13.21 & 11.8 $+$ 1.4 & pass \\
Flash, 64\,ch & Flash & 4 & 200\,000 & 12.21 & 11.1 $+$ 1.2 & pass \\
Flash, 64\,ch, 8$\times$8 attention & Flash & 4 & 200\,000 & 9.65 & 8.0 $+$ 1.6 & \textbf{FAIL} \\
Flash, 32\,ch & Flash & 4 & 200\,000 & 9.48 & 5.3 $+$ 4.1 & \textbf{FAIL} \\
Flash, 64\,ch, 4$\times$8 attention & Flash & 4 & 200\,000 & 6.57 & 5.0 $+$ 1.6 & \textbf{FAIL} \\
\midrule
Flash, 32\,ch (production) & Flash & 6 & --- & --- & --- & not run on this slice \\
\bottomrule
\end{tabular}}
\end{table}

\begin{table}[h]
\centering
\small
\caption{Water-dimer oxygen-oxygen repulsive wall at 1.80\,\AA{}, against a high-level
DFT reference of 140.38\,kcal/mol ($\omega$B97M-V/def2-TZVPD;
\citealp{mardirossian2016omega,rappoport2010propertyoptimized}). The production block is the paper's two reported
checkpoints, scanned on the same 44-point grid and energy convention as the sweep
below it, so the two blocks are directly comparable; both checkpoints were re-verified and the scan is a deterministic single-point evaluation of
frozen weights. In that
block the routing column names the oxygen routing actually applied, and only the routing a
model was trained under is that model's own surface --- the scan's oxygen-LR flag is not
gated on the checkpoint's training config, so Standard's LR row is a routing those
weights never saw. All rows of the \emph{sweep} block are trained 200k-step checkpoints of a
4M-sample subset, \emph{not} production models; there the two nominally equivalent Standard runs
differ by 1.27\,kcal/mol on the LR stream and 0.56\,kcal/mol on the HR stream. The
sweep's 64-scalar Flash row is the only checkpoint measured that softened its
\emph{high-resolution} stream.}
\label{tab:wall}
\resizebox{\linewidth}{!}{%
\begin{tabular}{llrrrr}
\toprule
Checkpoint & Oxygen routing & LR-stream $E$ & \% of reference & HR-stream $E$ & \% of reference \\
 & & (kcal/mol) & & (kcal/mol) & \\
\midrule
\multicolumn{6}{l}{\emph{Production checkpoints}} \\
E2Former-V3-Standard, 4L, 6\,\AA{} (Standard) & LR &103.02 & 73.4 & 140.58 & 100.1 \\
E2Former-V3-Flash, 6L, 32\,ch, 5\,\AA{} (Flash) & LR &69.81 & \textbf{49.7} & 139.22 & 99.2 \\
\midrule
\multicolumn{6}{l}{\emph{200k-step configuration sweep (4M-sample subset), not production}} \\
Full-resolution uniform, no routing & --- & 139.16 & 99.1 & 139.16 & 99.1 \\
Standard, run A & LR &32.30 & 23.0 & 140.47 & 100.1 \\
Standard, run B & LR &33.57 & 23.9 & 139.91 & 99.7 \\
Flash, 64 scalar & LR &58.57 & 41.7 & \textbf{128.62} & \textbf{91.6} \\
Flash, 64 scalar $+$ 64 vector & LR &54.16 & 38.6 & 139.41 & 99.3 \\
Flash, 128 scalar $+$ 128 vector & LR &37.06 & 26.4 & 138.12 & 98.4 \\
Flash, 64 scalar, two tail blocks & LR &45.76 & 32.6 & 138.12 & 98.4 \\
\bottomrule
\end{tabular}}
\end{table}

\subsection{Liquid structure and thermodynamics}

\begin{table}[h]
\centering
\small
\caption{Liquid-water structure over the twelve measured 20\,ps \emph{water-only} legs (two
box sizes $\times$ two routings $\times$ three ensembles; the precedence for the paper's
liquid-structure evidence is Table~\ref{tab:rdf54}), against
the X-ray benchmark. This is not the 54-leg matrix: it predates it and covers water alone.
Cells marked ``not reported'' are an outstanding analysis gap: they are not zero and no value
is substituted for them. First-shell peak positions agree to within 0.05\,\AA{}; first O--O peak heights sit
5--8\% above experiment, the second shell is 0.11\,\AA{} long, and first-shell coordination
is 4.78--4.97 against 4.3--4.7.}
\label{tab:waterstruct}
\begin{tabular}{lrrr}
\toprule
Structural quantity & Flash & Standard & Experiment \\
\midrule
First O-O peak position & 2.82\,\AA{} & not reported & 2.80\,\AA{} \\
First O-O peak height & 2.70 & 2.77 & 2.57 \\
Second shell position & 4.45\,\AA{} & 4.61\,\AA{} & 4.50\,\AA{} \\
First minimum & not reported & 3.44\,\AA{} & 3.45\,\AA{} \\
First-shell coordination number & 4.78 & 4.97 & 4.3 to 4.7 \\
\bottomrule
\end{tabular}
\end{table}

\begin{table}[h]
\centering
\small
\caption{First-shell structure on the \textbf{54-leg production matrix}, from the stored
curves rather than from tabulated scalars, and only on the NVT legs, which are the ones for
which radial distribution functions were recorded. This is the table the paper's
liquid-structure claim rests on; Table~\ref{tab:waterstruct} is the water-only precedence.
Peak values are read off the three-seed mean curve and first-shell coordination is the
committed cumulative count integrated over the range the comparable experiment states, so it
need not reproduce the per-leg-peak convention of an earlier summary. Salt O--O is included
because the salt box carries water; Na--O and Cl--O are the ion-hydration shells. All curves
are three seeds $\times$ 50 frames, $t = 10$ to 20\,ps, single trajectory per seed.}
\label{tab:rdf54}
\emph{Coord.} is the committed cumulative count integrated over \textbf{the range the
comparable experiment states}, so it can be read against a published integral; on the
narrower integration to each curve's own first minimum the same curves give 5.07/4.97
(water O--O), 5.21/4.78 (salt O--O), 5.30/5.30 (Na--O) and 6.66/7.03 (Cl--O), which are
not comparable to a published number and are not used for comparison here.
\begin{tabular}{llrrrl}
\toprule
Curve & Model & Peak $r$ (\AA{}) & Peak $g$ & First min.\ $r$ (\AA{}) & Coord. \\
\midrule
Water O--O & Standard & 2.788 & 2.786 & 3.438 & 4.44 \\
Water O--O & Flash & 2.838 & 2.727 & 3.413 & 4.40 \\
\addlinespace
Salt O--O & Standard & 2.812 & 2.749 & 3.513 & 4.32 \\
Salt O--O & Flash & 2.812 & 2.753 & 3.413 & 4.28 \\
\addlinespace
Salt Na--O & Standard & 2.362 & 6.263 & 3.138 & 5.52 \\
Salt Na--O & Flash & 2.312 & 5.945 & 3.138 & 5.34 \\
\addlinespace
Salt Cl--O & Standard & 3.163 & 3.197 & 3.888 & 6.96 \\
Salt Cl--O & Flash & 3.213 & 2.838 & 3.987 & 6.38 \\
\bottomrule
\end{tabular}
\end{table}

\begin{table}[h]
\centering
\small
\caption{Constant-pressure legs of the separate 12-leg water-only matrix (64 and 512
molecules; not the 54-leg production matrix). Densities are 3.6 to 5.9\% too high and
compressibilities roughly threefold too stiff. The 64-molecule Standard leg is a protocol
defect rather than a model defect: its interaction cutoff filled 97\% of that box's
half-width, and its compressibility is twelvefold inconsistent with its own larger box. The
acceptance column is this matrix's own protocol, not the 54-leg $[0.30,0.40]$ range. The
experimental row is the IAPWS-95 reference at 298\,K and 1\,bar
\citep{wagner2002iapws, pi2009anomalies}.}
\label{tab:waternpt}
\begin{tabular}{lrrrr}
\toprule
Leg & Density (g/cm$^3$) & Deviation & Compressibility (1e-5 /bar) & Acceptance \\
\midrule
512 molecules, Flash & 1.0560 $\pm$ 0.0016 & $+$5.9\% & 1.45 $\pm$ 0.11 & 0.83 \\
512 molecules, Standard & 1.0332 $\pm$ 0.0015 & $+$3.6\% & 1.35 $\pm$ 0.13 & 0.83 \\
64 molecules, Flash & 1.0403 $\pm$ 0.0063 & --- & 2.24 $\pm$ 0.35 & 0.86 \\
64 molecules, Standard & 0.9718 $\pm$ 0.0209 & --- & \textbf{16.41 $\pm$ 0.45} & 0.86 \\
Experiment & 0.997 & --- & 4.52 & --- \\
\bottomrule
\end{tabular}
\end{table}

The 54-leg production matrix's NPT legs give 1.0720 $\pm$ 0.0046 (water Standard),
1.0762 $\pm$ 0.0025 (water Flash) and 1.0738--1.0758\,g/cm$^3$ (NaCl and DEKSAN),
against the reference 0.997\,g/cm$^3$, with Monte-Carlo acceptance 0.31--0.38; these are the values quoted in the
main text. The table above is the separate water-only matrix and must not be substituted for
them.

\begin{table}[h]
\centering
\small
\caption{Constant-energy (NVE) legs of the 12-leg water-only matrix (4 legs). The Flash rows are the
routing's \emph{degenerate limit}: pure water has no high-resolution atoms under it, so only the narrow
scalar stream executed. They are a valid measurement of that stream and are not evidence
about cross-role exchange. The 64-molecule Standard leg is the protocol defect described above,
which affects its thermodynamic readings rather than its conservation.}
\label{tab:waternve}
\begin{tabular}{lrrrr}
\toprule
Leg & Final drift (meV/atom) & RMS & Level shift, first to last 4\,ps & Shift / RMS \\
\midrule
512 molecules, Flash & $-$0.013 & 0.008 & $+$0.001 & 0.12 \\
64 molecules, Flash & $+$0.043 & 0.023 & $-$0.002 & $-$0.10 \\
512 molecules, Standard & $+$0.041 & 0.021 & \textbf{$+$0.041} & \textbf{1.95} \\
64 molecules, Standard & $+$0.130 & 0.093 & \textbf{$+$0.189} & \textbf{2.03} \\
\bottomrule
\end{tabular}
\end{table}

\subsection{Dynamics traces}

\subsection[Temperature, integrity, energy]{Temperature, Chemical Integrity and Energy Conservation}
\label{sec:app_integrity}

The main text claims that the production trajectories preserve chemical structure, hold
temperature stable and match experimental liquid structure. This subsection gives all three in full, on the same 54 legs, and states where the evidence stops.

\textbf{Temperature.} Table~\ref{tab:temperature} reports the measured kinetic temperature
over the 5 to 20\,ps window, target 300\,K, timestep 0.5\,fs. The first 5\,ps are excluded
because they carry the start-up transient from the initial velocity assignment, which dominates
the block means over that stretch. Over the remaining window the 54 leg
means average 299.1\,K; every one of the 18 system-model-ensemble cells sits within 3.4\,K of
the target, and 17 of 18 within 2.4\,K. There is no secular drift: per leg the second half
minus the first half stays within 7.1\,K, and the largest fitted slope magnitude of any leg is
0.78\,K/ps; both are dominated by the shortest, most per-atom-fluctuating legs.

The six constant-energy cells --- eighteen of the 54 legs --- are the exception, and the table
must be read with the protocol in mind. Because an NVE leg is continued from the completed
matching NVT leg rather than re-thermalized, it inherits that endpoint's instantaneous kinetic
energy as its conserved total energy, so an NVE leg's mean temperature stays wherever its
parent NVT leg happened to end. That is why the two water NVE cells
span 290.83 to 310.33\,K, while every NVE cell on the larger
NaCl and DEKSAN boxes lies within 3.2\,K of target. In none of them does the
temperature walk: the offset is set at the start and held. The honest statement is therefore
that the thermostatted ensembles hold 300\,K and the constant-energy legs conserve energy at
whatever temperature they started from, not that all 54 legs sit at 300\,K.

\begin{table}[h]
\centering
\small
\caption{Kinetic temperature over the 5 to 20\,ps window, target 300\,K, over the 54-leg
production matrix. ``Mean'' is the mean of the three leg means, ``SEM'' its three-replica
standard error, ``Leg range'' the span of the three leg means, and ``Frame range'' the
extremes reached by any saved frame. The first 5\,ps are excluded as start-up transient. The
NVE rows carry a protocol caveat rather than a modelling one: each NVE leg is continued from
its matching NVT endpoint and conserves the energy it inherits, so its mean temperature
reflects where that endpoint stopped and is not thermostatted. All values measured, none
extrapolated.}
\label{tab:temperature}
\resizebox{\linewidth}{!}{%
\begin{tabular}{llrrrrr}
\toprule
System & Model & Ensemble & Mean (K) & SEM & Leg range (K) & Frame range (K) \\
\midrule
water & Standard & NVT & 300.06 & 1.34 & 298.18--302.66 & 274.00--333.31 \\
water & Standard & NPT & 297.68 & 0.36 & 297.02--298.26 & 270.52--320.71 \\
water & Standard & NVE & 299.47 & 4.32 & 290.83--303.84 & 270.49--325.17 \\
water & Flash & NVT & 298.24 & 0.52 & 297.20--298.82 & 272.51--325.03 \\
water & Flash & NPT & 299.83 & 0.48 & 299.28--300.79 & 269.62--326.18 \\
water & Flash & NVE & 303.34 & 3.59 & 298.38--310.33 & 281.49--329.67 \\
NaCl & Standard & NVT & 298.54 & 0.29 & 298.00--298.99 & 292.38--304.31 \\
NaCl & Standard & NPT & 298.75 & 0.26 & 298.23--299.02 & 291.89--305.28 \\
NaCl & Standard & NVE & 298.60 & 0.93 & 296.93--300.16 & 292.96--304.98 \\
NaCl & Flash & NVT & 298.99 & 0.14 & 298.72--299.20 & 292.81--305.44 \\
NaCl & Flash & NPT & 299.28 & 0.14 & 299.10--299.57 & 292.54--309.01 \\
NaCl & Flash & NVE & 299.18 & 0.47 & 298.33--299.97 & 294.34--304.11 \\
DEKSAN & Standard & NVT & 298.40 & 0.21 & 298.08--298.81 & 288.90--307.34 \\
DEKSAN & Standard & NPT & 298.94 & 0.04 & 298.86--299.00 & 289.17--307.67 \\
DEKSAN & Standard & NVE & 299.26 & 0.76 & 297.97--300.59 & 291.30--307.00 \\
DEKSAN & Flash & NVT & 298.84 & 0.57 & 298.22--299.97 & 288.24--313.28 \\
DEKSAN & Flash & NPT & 298.44 & 0.73 & 297.38--299.85 & 288.04--307.55 \\
DEKSAN & Flash & NVE & 298.26 & 1.07 & 297.09--300.40 & 288.68--306.44 \\
\midrule
\multicolumn{3}{l}{All 54 legs} & 299.12 & --- & 290.83--310.33 & --- \\
\bottomrule
\end{tabular}}
\end{table}

\textbf{Chemical integrity.} Table~\ref{tab:covalent} reports the covalent-bond audit. Only
the DEKSAN peptide carries it: the water and NaCl boxes are held together by the same
inferred covalent graph, but the audit's per-leg record stores the solute bond statistics
only for the peptide, so for those two systems chemical integrity rests on the weaker
zero-overlap and zero-fusion evidence below rather than on a bond-length measurement. Across
the 18 peptide legs the inferred graph is 196 bonds in every single leg, matching the
reference exactly, and the worst stretch of any bond in any frame of any leg is a ratio of
1.17548 against a gate limit of 1.5. The peptide stays folded, with mean RMSD 0.91 to
1.20\,\AA{} and mean radius of gyration 5.89 to 6.02\,\AA{} in all 18 legs.

The weaker evidence does cover all 54 legs: no leg recorded any anomaly, and specifically
none recorded an oxygen-oxygen contact below 0.65\,\AA{} or an oxygen-fusion event below
1.8\,\AA{}. That is absence of atomic overlap, not bond preservation, and the distinction
matters for the water-only boxes, where no intramolecular coordinate was tracked at all.

\begin{table}[h]
\centering
\small
\caption{DEKSAN covalent-bond audit over the 18 peptide legs, gate limit 1.5 on the stretch
ratio. The bond count is inferred from covalent radii and is 196 in every leg, matching the
reference graph. ``Max stretch'' is the worst ratio over all bonds and all frames of all legs
of that model, and ``Max bond'' the longest absolute bond. RMSD and $R_g$ are leg means, so
the ranges are across legs rather than across frames. No value is reported for water or NaCl
because the per-leg record stores solute bond statistics for the peptide only.}
\label{tab:covalent}
\resizebox{\linewidth}{!}{%
\begin{tabular}{llrrrrr}
\toprule
Model & Covalent bonds & Max stretch & Max bond (\AA{}) & RMSD mean (\AA{}) & $R_g$ mean (\AA{}) \\
\midrule
Standard & 196 & 1.17148 & 1.6819 & 0.931--1.198 & 5.888--5.998 \\
Flash & 196 & 1.17548 & 1.7116 & 0.913--1.203 & 5.912--6.024 \\
\midrule
Reference & 196 & 1.00000 & --- & --- & --- \\
\bottomrule
\end{tabular}}
\end{table}

\textbf{Energy conservation.} Table~\ref{tab:nvedrift} gives the constant-energy drift per
leg. Flash sits below Standard on all three systems, but the difference is resolvable only on
the NaCl box. On water and DEKSAN the Flash drift is indistinguishable from zero and changes sign
between replicas ($-0.00101$, $-0.00020$, $+0.00016$ and $-0.00012$, $-0.00011$, $+0.00058$\,meV/atom/ps),
so those two ratios are ratios of noise and the ranking there is not resolvable at this caliber. The drift is small in absolute
terms either way, at most $+0.0117$\,meV/atom/ps.

\begin{table}[h]
\centering
\small
\caption{Constant-energy drift in meV/atom/ps, signed, one row per leg. Negative values
indicate the energy falling. The three-replica means are $+0.00612$, $+0.00900$ and
$+0.00535$ for Standard water, NaCl and DEKSAN, against $-0.00035$, $+0.00081$ and
$+0.00011$ for Flash. Measured over an 18\,ps window (t = 2 to 20\,ps), so a drift this small bounds the
short-time behaviour only and is not a long-time conservation claim.}
\label{tab:nvedrift}
\resizebox{\linewidth}{!}{%
\begin{tabular}{llrrrr}
\toprule
System & Model & Seed 104729 & Seed 130363 & Seed 155921 & Mean \\
\midrule
water & Standard & $+0.00545$ & $+0.00529$ & $+0.00761$ & $+0.00612$ \\
water & Flash & $-0.00101$ & $-0.00020$ & $+0.00016$ & $-0.00035$ \\
NaCl & Standard & $+0.01174$ & $+0.00767$ & $+0.00761$ & $+0.00900$ \\
NaCl & Flash & $+0.00129$ & $+0.00065$ & $+0.00050$ & $+0.00081$ \\
DEKSAN & Standard & $+0.00483$ & $+0.00647$ & $+0.00475$ & $+0.00535$ \\
DEKSAN & Flash & $-0.00012$ & $-0.00011$ & $+0.00058$ & $+0.00011$ \\
\bottomrule
\end{tabular}}
\end{table}

\subsection[1 ns NVT campaign]{Duration Beyond the Matrix: The \texorpdfstring{1\,ns}{1 ns} NVT Campaign}
\label{sec:app_ns1}

\textbf{Duration beyond 20\,ps.} The matrix above stops at 20\,ps by design, shorter than
every failure this work records elsewhere. The separate 1\,ns run (protocol above;
Table~\ref{tab:ns1main}) answers the duration question for one system and nothing else. Its
bounds are load-bearing: one ensemble, one seed and a fixed cell, so no density, RDF or
thermodynamic quantity follows; and it does not retire the 40, 58, 60, 511.9 and 680\,ps onsets on
other systems and configurations. Both checkpoints carry this system past the latest onset
without a structural failure.

\textbf{Classical control.} The GROMACS control described above was propagated from the same
state for 1000\,ps. Its C$\alpha$ displacement from its
own start is 1.0799\,\AA{}; Flash's is 1.964\,\AA{}, and the Flash and classical 1000\,ps
endpoints are separated by 1.8273\,\AA{} after a common C$\alpha$ alignment. The classical
protein's mass-weighted $R_g$, evaluated by \texttt{gmx gyrate} over the Protein group, spans
16.881--17.268\,\AA{} and ends at 17.043\,\AA{}. Figure~\ref{fig:physical} reports both full
classical displacement time series alongside Standard and Flash in panel (e). Panel (f) instead
reports C$\alpha$ RMSD to the CASP15 T1123-D1 experimental structure
\citep{kryshtafovych2023critical}, using the 214 resolved C$\alpha$ atoms shared by the target
and an AlphaFold2-based CASP15 submission (group TS270; PDB entry T1123TS270\_1). That
submission is shown as a horizontal reference at 5.102\,\AA{}; it is an external reference, not
a DeepMind CASP entry. Over 1\,ns, the final
ground-truth RMSDs are 5.064\,\AA{} (Standard), 4.809\,\AA{} (Flash), and 4.889\,\AA{}
(classical). The comparison covers one system, one ensemble and one seed, so these curves do not
establish a general accuracy ranking.

\FloatBarrier

\section{Capacity Variants and Warm-Start Accuracy}
\label{sec:app_capacity}

Table~\ref{tab:accuracyext} carries three retired non-production configurations evaluated
on the full split at the training-time batch-mean caliber; they were not re-scored at the
exact component-weighted caliber.

\begin{table}[h]
\centering
\footnotesize
\caption{Retired non-production configurations of this work, on the full 2762021-sample
validation split at the training-time batch-mean caliber. Force MAE is per-component
and energy MAE is at equal prominence. The Flash configuration is worse on force than
every Standard configuration. Single-seed.}
\label{tab:accuracyext}
\resizebox{\linewidth}{!}{%
\begin{tabular}{llrrccl}
\toprule
Model & Routing & Force MAE & Energy MAE & Val.\ split & Force mode & Provenance \\
 & & (meV/\AA{}) & (meV/atom) & & & \\
\midrule
E2Former-V3-Standard, 6L, 5.5\,\AA{} & Standard & 7.200 & 1.999 & full & conserv. & this work \\
E2Former-V3-Standard, 6L, 6\,\AA{} & Standard & 7.651 & 1.475 & full & conserv. & this work \\
E2Former-V3-Flash, 6L, 32\,ch, 5.5\,\AA{} & Flash & 10.506 & 3.921 & full & conserv. & this work \\
\bottomrule
\end{tabular}}
\end{table}

Table~\ref{tab:capacity} is the evidence behind the statement in
Appendix~\ref{sec:app_bounds} that averaged error does not rank dynamical stability: the
spreads are 0.4\% on validation force, 0.4\% on held-out force, 1.2\% on validation
energy and 1.8\% on the held-out energy metric, while the dynamics outcomes are not
comparable at all. Table~\ref{tab:mdonset} gives the barrier profile of the same variants
and controls on one rigid proton-transfer coordinate; for the Flash rows, pure water is
the all-low-resolution routing limit, for which there is no absolute reference, so only
the ordering across checkpoints is meaningful; a barrier of 0.0 means the two basins
merged rather than that the measurement failed. Table~\ref{tab:warmstart} records the
direct-force warm-start checkpoints, which are validated on the 1M subsample and are
therefore not comparable to the full-split rows of Table~\ref{tab:accuracy}.

A separate ensemble-and-box-size scan of one model recorded failure onsets from 16.2 to
189.0\,ps, an 11.7-fold range. Those runs are not the row-wise capacity comparison in
Table~\ref{tab:capacity}. Likewise, the 680\,ps capacity-sweep trajectory and the separately
located 511.9\,ps onset below are distinct trajectories and are not competing timestamps for
one run.

\begin{table}[h]
\centering
\small
\caption{Low-resolution-capacity variants: indistinguishable averaged error, divergent dynamics.
Trained 200k-step checkpoints. Both $E/F$ columns report energy MAE in meV/atom and
per-component force MAE in meV/\AA{}. Validation and held-out columns use different splits and
calibers; compare within a column only. Cells marked ``not measured'' are not zero and no
value is substituted. The variant with the lowest held-out energy and force MAEs
nevertheless developed a frozen ion pair from 60\,ps.}
\label{tab:capacity}
\resizebox{\linewidth}{!}{%
\begin{tabular}{lllp{5.2cm}}
\toprule
Variant & Validation $E$ / $F$ & Held-out $E$ / $F$ & Dynamics outcome \\
\midrule
64 scalar $+$ 64 vector low-resolution & 13.91 / 18.91 & 0.406 / 8.354 & clean 1\,ns; clean 200\,ps on the molecular set \\
128 scalar $+$ 128 vector low-resolution & 14.08 / 18.93 & 0.401 / 8.382 & fused water pairs from 40\,ps, 968 of 1000 pairs fused, minimum O-O 2.033\,\AA{}; diverged at 2.5 to 3.0\,ps on the molecular set \\
64 scalar $+$ 64 vector, two tail blocks & 14.05 / 18.99 & \textbf{0.399 / 8.346} & frozen ion pair from 60\,ps; clean 200\,ps on the molecular set \\
64 scalar low-resolution (Flash reference) & not measured & not measured & fused at 680\,ps with two-proton defects \\
Full-resolution and Standard controls & not measured & not measured & clean \\
\bottomrule
\end{tabular}}
\end{table}

\begin{table}[h]
\centering
\small
\caption{Barrier profile of the same variants and controls on one rigid proton-transfer
coordinate, in kcal/mol. The Flash rows are evaluated in pure water, its
all-low-resolution routing limit; LR denotes the low-resolution role. There is no absolute
reference for this coordinate, so only the ordering across
checkpoints is meaningful, and a barrier of 0.0 means the two basins merged rather than
that the measurement failed.}
\label{tab:mdonset}
\begin{tabular}{@{}llrrrl@{}}
\toprule
Checkpoint & Routing & At 2.80\,\AA{} & At 2.31\,\AA{} & Best coupled & Observed MD outcome \\
\midrule
64 scalar $+$ 64 vector & Flash limit & 23.9 & 0.1 & 10.5 & survived 1\,ns \\
64 scalar & Flash limit & 26.2 & 4.6 & 9.8 & failed at 511.9\,ps \\
64 scalar, two tail & Flash limit & 24.2 & 2.5 & 9.7 & not run \\
128 scalar $+$ 128 vector & Flash limit & 27.5 & \textbf{0.0} & 7.2 & failed at 40\,ps \\
Standard, 4 layers & Standard & 48.6 & 26.3 & 36.0 & never failed \\
Full-resolution uniform & uniform & 46.0 & 25.9 & 34.5 & never failed \\
\bottomrule
\end{tabular}
\end{table}

\begin{table}[h]
\centering
\small
\caption{Direct-force warm-start checkpoints, validated on the \textbf{1M subsample} and
therefore not comparable with the full-split rows of Table~\ref{tab:accuracy}; the
subsample reads about 6.5\% optimistic. Force mode is direct except where stated, and
direct-force rows must not be read as conservative-force accuracy. The last two rows are
the published direct-force variants of the baselines in Table~\ref{tab:accuracy}, on their
own unstated split. Single-seed.}
\label{tab:warmstart}
\resizebox{\linewidth}{!}{%
\begin{tabular}{llrrl}
\toprule
Model & Routing & Force MAE (meV/\AA{}) & Energy MAE (meV/atom) & Force mode \\
\midrule
Standard, 4L, 6\,\AA{} & Standard & 10.310 & 1.390 & direct \\
Standard, 6L, 5.5\,\AA{} & Standard & 8.668 & 1.250 & direct \\
Flash, 6L, 32\,ch, 5\,\AA{} & Flash & 12.808 & 1.783 & direct \\
Standard, 4L, 6\,\AA{} (earlier run) & Standard & 8.147 & 1.200 & conservative \\
\midrule
E2Former-V2 \citep{huang2026formerv} & uniform & 9.55 & 1.24 & direct \\
eSEN \citep{fu2025learning} & uniform & 9.92 & 1.49 & direct \\
\bottomrule
\end{tabular}}
\end{table}

\FloatBarrier

\section[Limitations and bounds]{Limitations, Threats to Validity, and Outstanding Measurements}
\label{sec:app_bounds}

This appendix records the boundary values and the outstanding measurements behind the scope
statements in Section~\ref{sec:discussion}.

\subsection{Boundaries by claim}

\textbf{Accuracy.} The force claim is per-component MAE. Energy MAE is 2.190\,meV/atom for
Standard and 3.941 for Flash, against E2Former-V2's published 1.16; the energy column is not
argued in our favour. Standard's 8.146\,meV/\AA{} is lower than the published values compared
here, but those values use unstated conventions, so this is a level comparison rather than a
like-for-like ranking. Flash is less accurate than E2Former-V2 (11.272 against 8.38). Every
OMol25 trained result is single-seed, so no statistical claim attaches to an OMol25 accuracy
difference; the SPICE seed check in Section~\ref{sec:seedvar} adds two seeds at one recipe but
does not extend to OMol25. The same-seed rerun in Section~\ref{sec:ablation} bounds
implementation variance, not seed variance.

\textbf{Attribution.} The two production configurations differ in depth, cutoff, routing,
persistent-state layout, attention parameterisation and FFN at once. The 16-cell ablation separates
depth, cutoff, FFN, attention parameterisation, light width and the light-head partition; the
mixed-resolution family has a cumulative estimate only. The block schedule and water-oxygen
routing move with that family, and no accuracy-valid state-layout variant was measured. No
production-recipe-matched full-resolution control was trained;
the SPICE cell A is matched only at the 200k ablation recipe. The numerical gates used random
weights and were not re-run on the final conservative checkpoints.
The role-stratified MAE is absolute and unnormalised by force magnitude.

\textbf{Dynamics.} The 20\,ps matrix is shorter than the water-box failure onsets recorded
elsewhere: near 40, 58 and 60\,ps, one located onset at 511.9\,ps, and one at 680\,ps. The
constant-pressure densities are 1.0720--1.0762\,g/cm$^3$ against an experimental 0.997,
and water compressibilities are 2.66--3.42$\times10^{-5}$/bar against
4.52$\times10^{-5}$ (roughly 25--40\% too stiff; Section~\ref{sec:observables},
Table~\ref{tab:observables}). The separate 1\,ns campaign covers one system, one ensemble and
one seed (Table~\ref{tab:ns1main}) and does not transfer to the three systems of the 20\,ps
matrix. The 100\,ps NVE check in Section~\ref{sec:nve100} covers only a 512-molecule water box
(Table~\ref{tab:nve100}). Neither that check nor the 1\,ns campaign establishes general
long-time stability or thermodynamic accuracy; the density and compressibility biases above are
not resolved by either.

\subsection{Outstanding measurements}

Hydrogen-pair observables, ion residence and exchange
rates, the $\geq 200$\,ps screening tier and the 1\,ns tier of the 54-leg matrix were not run.
The study has no production-recipe-matched full-resolution control; the SPICE cell A is a
same-recipe control at the ablation recipe only. We claim no reduced angular resolution, no
learned or dynamic routing, no optimal partition, no zero-shot transfer of the routing policy
to another solvent, and no demonstrated validity in reactive regimes.

\subsection{Random-routing controls}

No trained random-routing control at the same low-resolution
fraction exists. An assignment can be resampled per training sample, resampled per MD frame, or
fixed for a trajectory. Per-sample resampling sends the same chemical environment through
different pathways across examples and makes the workload less predictable. Per-frame
resampling lets an atom change pathway as it moves, which changes the evaluation independently
of the physical motion and can introduce artificial discontinuities in energy and forces. A
fixed mask avoids temporal switching but must be used in both training and inference; applying a
random mask only at inference to a checkpoint trained with chemical routing is not a valid
control. Chemical identity is an application-driven prior, motivated by precedents such as
SHAKE, MARTINI and QM/MM region selection \citep{ryckaert1977numerical, marrink2007martini,
senn2009methods}; those methods do not establish that this partition outperforms a properly
trained random-routing model. The claim that atom-role-aware allocation beats a
chemistry-agnostic allocation of the same low-resolution fraction therefore remains a
conjecture.

\subsection{Threats to validity}

\begin{table}[h]
\centering
\small
\caption{The main threats to the claims and the measurement that would remove each.}
\label{tab:threats}
\begin{tabular}{@{}>{\raggedright\arraybackslash}p{0.28\linewidth}>{\raggedright\arraybackslash}p{0.34\linewidth}>{\raggedright\arraybackslash}p{0.28\linewidth}@{}}
\toprule
Threat & Consequence for the claim & What would change it \\
\midrule
OMol25 single seed & No statistical claim attaches to an OMol25 accuracy difference; the SPICE three-seed check is separate (Section~\ref{sec:seedvar}). & Multi-seed OMol25 retraining at the same recipe. \\
One device per number & Ratios are device-specific and not universal. & Same-device rerun of numerator and denominator. \\
No production-recipe full-resolution control & Allocation and training budget are confounded at the production schedule; the SPICE cell A is a same-recipe control at the ablation recipe. & A full-resolution retrain at the production budget. \\
No random-routing control & The chemistry mechanism remains a conjecture; an inference-only random mask would not test it. & A trained random-routing model at the same low-resolution fraction. \\
Two configurations are not a single-variable ablation & Differences cannot be assigned to one axis. & The single-variable ablation, which covers the separable axes only. \\
State-layout accuracy is unmeasured & No claim about persistent-state layout. & An accuracy-valid state-layout ablation. \\
20\,ps horizon & Short-time stability and local structure only. & A 200\,ps / 1\,ns matrix on all systems. \\
Role MAE is unnormalised & A small LR value is not proof of adequacy. & Force-magnitude-normalised and per-species error. \\
Published baselines have unstated conventions & Cross-convention comparisons are levels, not ties. & Re-scored baselines or explicit conventions. \\
No reactive regimes & The scope is non-reactive, explicitly solvated systems. & Transition-path and reactive tests. \\
\bottomrule
\end{tabular}
\end{table}

\end{document}